\documentclass[a4paper,11pt]{article}
\pdfoutput=1 

\usepackage{jheppub} 

\usepackage[T1]{fontenc} 
\usepackage{soul}
\usepackage{color}

\usepackage{empheq}
\usepackage{mathrsfs} 
\usepackage{url}
\usepackage[utf8]{inputenc}
\usepackage[toc,page]{appendix}
\usepackage[autostyle]{csquotes}
\usepackage{slashed}
\usepackage{amssymb}
\usepackage{graphicx}
\usepackage{longtable}
\usepackage{array}
\usepackage{bm} 
\usepackage{amsmath}
\usepackage{nccmath}
\usepackage{cancel}
\usepackage{mathtools}
\usepackage{amsfonts}
\usepackage{amssymb}
\usepackage[labelfont=bf]{caption}
\usepackage{paralist} 
\usepackage{hyperref}
\usepackage{cleveref}
\DeclareMathAlphabet{\mathpzc}{OT1}{pzc}{m}{it} 
\usepackage{tensor}

\title{\boldmath Logarithmic correction to black hole entropy in universal low-energy string theory models}

\author[1]{Sudip Karan}\note{Alternative e-mail: \tt{sudip.karaan@gmail.com}.}
\author{and Gurmeet Singh Punia}
\affiliation{Indian Institute of Science Education \& Research Bhopal,\\ Bhopal Bypass Road, Bhauri, Bhopal 420066, India}

\emailAdd{skaran@iiserb.ac.in}
\emailAdd{gurmeet17@iiserb.ac.in}

\abstract{We calculate the logarithmic correction to the entropy of asymptotically flat and AdS black holes (rotating, non-rotating, charged, and uncharged) embedded in Einstein-Maxwell-dilaton (EMD) theories with $U(1)$-charged. The leading quantum gravitational corrections are achieved in both extremal and non-extremal limits of black hole temperature by designing a common Euclidean gravity setup that evaluates the ``logarithmic term'' from one-loop effective actions via heat kernel method-based calculations. EMD theories are universal building blocks of compactified string theory or supergravity models in 4D. For a concrete example, we generalize the entire setup and calculate logarithmic corrections for black holes in $U(1)^2$-charged EMD models intersecting with $\mathcal{N}=4$ ungauged and gauged bosonic supergravity. In contrast to flat backgrounds, all the AdS$_4$ results are found to be {non-topological}, providing a wider ``infrared window'' into the microscopic degrees of freedom of black holes in string theory.}

\keywords{Black Holes in String Theory and Supergravity, AdS-CFT Correspondence}

\arxivnumber{2210.16230}

\dedicated{\href{https://doi.org/10.1007/JHEP03(2023)028}{https://doi.org/10.1007/JHEP03(2023)028}}

\begin{document} 
\maketitle
\flushbottom

\section{Introduction}\label{intro}

Black holes are fascinating as well as peculiar gravitational objects for which the quantum gravity effects are significant. A proper understanding of the microscopic viability of quantum gravity candidates by interpreting black hole entropy has been one of the major attractions of fundamental physics over the past several decades. Inside the framework of Einstein's general relativity, the entropy of a black hole is described by the seminal \emph{Bekenstein-Hawking area law} (BHAL) \cite{Bekenstein:1973jb,Hawking:1975sh}, which is equal to one-quarter of the area of the event horizon. String theory has already gained remarkable success by providing a counting of microstates underlying the entropy of various classes of flat and AdS black holes to establish the BHAL (e.g., see \cite{Strominger:1996sh,Maldacena:1996gb,Horowitz:1996fn,Emparan:2006it,Mandal:2010cj,Sen:2014aja,Belin:2016knb,Benini:2019dyp,Gang:2019uay,PandoZayas:2020iqr,Liu:2017vll,Benini:2015eyy,Liu:2018bac}). For a non-trivial consistency check, there has been trendy progress on the macroscopic front (the low energy or IR limit), i.e., in the description of Einstein gravity by incorporating possible quantum gravitational correction to BHAL describing black hole entropy (semi-)classically and approximately at tree level. This paper aims to explore the same direction.
 
It is well-known that the leading and fundamental quantum gravitational correction to black hole entropies is a \emph{logarithmic term}. However, the total quantum corrected black hole entropy has the following generic form (with horizon area $\mathcal{A}_H$ and Planck length $\ell_P$)
\begin{align}\label{int1}
	S_{\text{bh}}(\mathcal{A}_H) =  \frac{\mathcal{A}_H}{4\ell_P^2}+ \mathcal{C}\ln \left(\frac{\mathcal{A}_{H}}{4\ell_P^2}\right) + \sum_{n=1}^\infty \kappa_n {\left(\frac{\mathcal{A}_H}{4\ell_P^2}\right)}^{-n+1} + \eta \exp \left(-\frac{\mathcal{A}_H}{4\ell_P^2}\right),
\end{align}
where the leading term $ \frac{\mathcal{A}_H}{4\ell_P^2}$ is BHAL, the second sub-leading term proportional to $\ln \mathcal{A}_{H}$ is the logarithmic correction (e.g., see \cite{Solodukhin:1995na,Solodukhin:1995nb,Kaul:2000rk,Carlip:2000nv,Banerjee:2008cf,Banerjee:2009fz,Majhi:2009gi,Cai:2010ua,Banerjee:2011oo,Banerjee:2011pp,Sen:2012qq,Sen:2012rr,Bhattacharyya:2012ss,Chowdhury:2014np,Gupta:2014ns,Jeon:2017ij,Karan:2019sk,Sen:2013ns,Keeler:2014nn,Charles:2015nn,Larsen:2015nx,Castro:2018tg,Banerjee:2020wbr,Karan:2020sk,Karan:2021teq,Banerjee:2021pdy,David:2021eoq,Xiao:2021zly,Calmet:2021lny,Delgado:2022pcc,Pourhassan:2022auo}) and the renaming terms are the other power-law or quantum-loop\footnote{The n-loop quantum corrections are proportional to $\mathcal{A}_H^{-n+1}$, where $\ln \mathcal{A}_H$ term is also of one-loop.} (perturbative) and exponential (non-perturbative) corrections \cite{Chatterjee:2020iuf,Dabholkar:2014ema}. In the large-charge limit\footnote{In the large-charge limit, black hole parameters like its charge, angular momentum, mass, etc., are scaled (keeping their dimensionless ratios fixed) so that the horizon area appears much larger than the Planck area, i.e., $\mathcal{A}_H \gg \ell_P^2$. Throughout this paper, we will work on this limit.} of black holes, the logarithmic correction is fully dominant over others. These logarithmic entropy corrections also turn out to be \emph{universal} since they are inescapable in the structure of every quantum gravity, even via many different approaches like -- Euclidean effective action method \cite{Banerjee:2011oo,Banerjee:2011pp,Sen:2012qq,Sen:2012rr,Bhattacharyya:2012ss,Chowdhury:2014np,Gupta:2014ns,Jeon:2017ij,Karan:2019sk,Sen:2013ns,Keeler:2014nn,Charles:2015nn,Larsen:2015nx,Castro:2018tg,Banerjee:2020wbr,Karan:2020sk,Karan:2021teq,Banerjee:2021pdy,David:2021eoq}, quantum tunneling \cite{Banerjee:2008cf,Banerjee:2009fz,Majhi:2009gi}, conical singularity \cite{Solodukhin:1995na,Solodukhin:1995nb}, Cardy formula \cite{Carlip:2000nv}, conformal anomaly \cite{Cai:2010ua}, quantum geometry \cite{Kaul:2000rk}, non-local quantum gravity \cite{Xiao:2021zly,Calmet:2021lny,Delgado:2022pcc,Pourhassan:2022auo}, etc. {For a gravity model} coupled to the higher-curvature terms beyond two-derivative, the expansion \eqref{int1} in principle holds a similar form, except the BHAL gets modified into the Bekenstein-Hawking-Wald formula \cite{Wald:1993rw} by capturing the classical higher-derivative corrections to black hole entropy. Technically, the loop contributions of the higher-curvature terms give rise to a distinct class of power-law corrections with the relevant constant prefactors depending on both the coupling constants associated with the higher-curvature terms and the quantum fluctuation data of the concerned theory. But the universality status of logarithmic corrections is so robust or fundamental that they are insensitive to the higher-derivative or power-law corrections \cite{Banerjee:2011oo} and entirely determined by the one-loop contribution of the two-derivative sector of the theory. Note that the logarithmic prefactors $(\mathcal{C}, \kappa, \eta)$ in the formula \eqref{int1} control the relative strengths of corresponding quantum corrections, which generally depend on the details of UV completion of the concerned low-energy gravity theory. Surprisingly, the logarithmic corrections and their prefactor $\mathcal{C}$ are {special} since they are entirely computable from the knowledge of only low-energy modes (IR data), i.e., massless fluctuations\footnote{The account of massive fluctuations leads to the corrections to black hole entropy that are suppressed by the inverse powers of $m^2\mathcal{A}_H$ but cannot give $\ln \mathcal{A}_H$ terms. Also, see \Cref{massfoot}.} running in one-loop \cite{Banerjee:2011oo,Banerjee:2011pp,Sen:2012qq,Sen:2012rr,Bhattacharyya:2012ss,Chowdhury:2014np,Gupta:2014ns,Jeon:2017ij,Karan:2019sk,Sen:2013ns,Keeler:2014nn,Charles:2015nn,Larsen:2015nx,Castro:2018tg,Banerjee:2020wbr,Karan:2020sk,Karan:2021teq,Banerjee:2021pdy,David:2021eoq}. This fundamental feature makes them a strong infrared laboratory for the most active litmus test, i.e., any enumeration of black hole microstates inside the structure of string theory must agree with the logarithmic corrected entropy. However, there are often huge technical challenges to overcome in evaluating them \cite{Mandal:2010cj,Sen:2014aja}. In this paper, we will explain how to compute logarithmic corrections for all rotating (and non-rotating) as well as charged (and uncharged) black holes in the low-energy model of Einstein-Maxwell-dilaton theory by structuring a common and efficient setup.

Our central objective is to address the specific question: how to obtain the logarithmic correction to black hole entropy in the most ubiquitous building blocks of effective gravity models that are an IR limit of compactified string theories in 4D spacetimes? One such popular model is the Einstein-Maxwell-dilaton (\emph{abbreviated to} EMD) theory that universally structures the 4D description of various higher-dimensional GR-inspired theories and supergravity \cite{Gibbons:1982ih,Gibbons:1987ps,Garfinkle:1990qj,Horowitz:1991cd,Kallosh:1992ii,Khuri:1995xk,Gao:2004tv,Becker:2006dvp,Cvetic:2014vsa,GoulartSantos:2017dun,Guo:2021zed,Zhang:2021edm,Myung:2020dqt}. Especially the supergravities are well-known compactified string vacuums (typically type-II and type-I on a Calabi–Yau three-fold \cite{Grana:2006mg,Freedman:2012xp}) that already have well-established microscopic counterparts. EMD theories are toy models for studying the string-loop effects from the macroscopic gravity side, which recently attracted some serious attention and motivated the current paper. These are nothing but the Einstein gravity model coupled to the Maxwell sector via the non-minimal coupling function of a dilaton (as a fixed scalar field\footnote{The scalar field, dubbed as \emph{dilaton}, controls how the extra dimension(s) dilates along the compactified 4D spacetimes via the low-energy EMD models \cite{Becker:2006dvp,Freedman:2012xp}.}), describing the central bosonic sector of \emph{true} supergravity theories for some specific choices of the Maxwell-dilaton coupling constant.  Obviously, the EMD models are the  natural but non-trivial generalization of a simple Einstein-Maxwell (EM) theory. However, the presence of non-minimal Maxwell-dilaton coupling precludes the EM theory from being a consistent truncation of this EMD class. In other words, any EM background (at least the charged black holes) does not solve the EMD theories.\footnote{An EM theory minimally coupled to a scalar field never gives rise to new black hole solutions beyond the limit of the Kerr-Newman family.} But quite surprisingly, there exists an exceptional case where general EM backgrounds like Schwarzschild, Reissner-Nordstr\"om, and Kerr black holes can be uplifted or embedded into the EMD models as scalar hair (or dilaton)-free black hole solutions (please refer to \cref{embd}). Similarly, it is possible to embed the Schwarzschild-AdS, Reissner-Nordstr\"om-AdS and Kerr-AdS black holes in a consistent EMD theory with a negative cosmological constant (\emph{abbreviated to} EMD-AdS theory) and their intersecting sectors in gauged supergravity \cite{Gao:2004tv,GoulartSantos:2017dun,Guo:2021zed,Zhang:2021edm}. These embedding choices of black hole backgrounds effectively intensify the prospect of microscopic consistency of calculated quantum correction results inside string theory \cite{Strominger:1996sh,Maldacena:1996gb,Horowitz:1996fn,Emparan:2006it,Mandal:2010cj,Sen:2014aja,Belin:2016knb,Benini:2019dyp,Gang:2019uay,PandoZayas:2020iqr,Liu:2017vll}. To date, pioneered by Ashoke Sen and collaborators and then followed by many other groups, the logarithmic corrections are mostly reported for the full Kerr-Newman family of black holes in EM theory \cite{Bhattacharyya:2012ss,Sen:2012rr,Sen:2013ns,Karan:2021teq} and all $\mathcal{N} \geq 1$ ungauged supergravity \cite{Banerjee:2011oo,Banerjee:2011pp,Sen:2012qq,Gupta:2014ns,Keeler:2014nn,Larsen:2015nx,Karan:2019sk,Banerjee:2020wbr,Karan:2020sk,Banerjee:2021pdy,Charles:2015nn,Castro:2018tg}. Few results are also available for AdS$_4$ black holes by Jeon et al. \cite{Jeon:2017ij} and David et al. \cite{David:2021eoq}. All this motivated us to the particular objective of this paper, i.e., computation of the logarithmic correction for all flat and AdS scalar-free black holes in the EMD and embedded supergravity theories. 

We plan to employ the standard and most successful Euclidean quantum gravity approach \cite{Gibbons:1977ta,Hawking:1978td} to address the question of logarithmic corrections in this paper. In this process, we explicitly test Sen's quantum entropy function formalism \cite{Sen:2008wa,Sen:2009wb,Sen:2009wc} and the Euclidean gravity treatment \cite{Sen:2013ns} for extremal and non-extremal black holes,  respectively. The underlying framework is computing the Euclidean path integral of any gravity theory perturbatively via the saddle-point expansion, considering the black hole solutions as a classical saddle-point. The Bekenstein-Hawking formula (or Wald entropy if higher-derivative terms are incorporated into gravity action) arises from the entropy evaluated on the on-shell saddle-point, while the quantum corrections to black hole entropy are different order loop contributions to the Euclidean path integral. For the logarithmic correction, one needs to extract and evaluate the exclusive ``logarithmic term'' from the one-loop quantum effective action part of massless fluctuations. To fulfill this purpose, the heat kernel method \cite{Hawking:1977te,Denardo:1982tb,Avramidi:1994th,Barvinsky:2015} is a practical and effective tool that has successfully reproduced correct results in all available cases \cite{Banerjee:2011oo,Banerjee:2011pp,Sen:2012qq,Sen:2012rr,Bhattacharyya:2012ss,Chowdhury:2014np,Gupta:2014ns,Jeon:2017ij,Karan:2019sk,Sen:2013ns,Keeler:2014nn,Charles:2015nn,Larsen:2015nx,Castro:2018tg,Banerjee:2020wbr,Karan:2020sk,Karan:2021teq,Banerjee:2021pdy,David:2021eoq}. Here the one-loop quantum effective actions are estimated by computing expansion coefficients of the heat kernel operator controlling all quadratic or one-loop fluctuations around a concerned black hole background. To achieve the ambitious goal of exploring all uncharged, non-rotating, charged and rotating quantum black holes from a single platform, we will cast the Seeley-DeWitt expansion \cite{Seeley:1966tt,Seeley:1969uu,DeWitt:1965ff,DeWitt:1967gg,DeWitt:1967hh,DeWitt:1967ii} of heat kernel,\footnote{Technically, the Seeley-DeWitt coefficients capture only the local part of the heat kernel, while a global contribution also exists due to zero modes of the heat kernel operator. Please refer to \cref{setup}.} followed by Gilkey's approach \cite{Vassilevich:2003ll} of computing the relevant coefficients that are only invariants induced from the background curvature. Most other acknowledged approaches, e.g., the eigenfunction expansion method \cite{Banerjee:2011oo,Banerjee:2011pp,Sen:2012qq,Gupta:2014ns} and its related avatars \cite{Keeler:2014nn,Larsen:2015nx}, are strictly limited to non-rotating extremal black hole backgrounds having a near-horizon geometry with rotational symmetry (i.e., Bertotti-Robinson or AdS$_2\times S^2$ type). In contrast, Gilkey’s Seeley-DeWitt approach \cite{Vassilevich:2003ll} gained immense success by providing logarithmic corrections for the full Kerr-Newman family (i.e., Schwarzschild, Reissner-Nordstr\"om, Kerr and Kerr-Newman) of black holes in both the extremal \cite{Sen:2012rr,Bhattacharyya:2012ss,Karan:2019sk,Banerjee:2020wbr,Karan:2020sk,Karan:2021teq} and non-extremal \cite{Sen:2013ns,Charles:2015nn,Castro:2018tg,Karan:2021teq,Banerjee:2021pdy} limits, irrespective of being non-supersymmetric or supersymmetric. So far, this success is chiefly for asymptotically-flat backgrounds, but we have overcome the challenges of extending the setup for asymptotically-AdS$_4$ black holes in this paper.

Let us highlight the prime technical findings and the remaining content of this paper. In \cref{setup}, by fusing the mentioned heat kernel treatment \cite{Vassilevich:2003ll} and Euclidean quantum gravity setups \cite{Sen:2008wa,Sen:2009wb,Sen:2009wc,Sen:2013ns}, we structure an efficient handbook that can compute logarithmic corrections to the entropy of all flat and AdS black holes via a common platform in 4D. Here we will see how a particular Seeley-DeWitt coefficient $a_4(x)$ (defined in \cref{set14}), encoding all the trace anomaly and central charge data, evaluates the logarithmic corrections to the entropy of 4D black holes in both extremal and non-extremal limits of their temperature. In \cref{EMD1a}, we consider a typical $U(1)$-charged EMD-AdS theory with a single Einstein-dilaton coupling of arbitrary strength. We then embed possible rotating and non-rotating EM backgrounds by setting appropriate constraints on the Maxwell or $U(1)$ charges that vanish the background dilaton. We finally calculate the first three Seeley-DeWitt coefficients (i.e., up to $a_4(x)$) by fluctuating the entire content EMD-AdS theory around the embedded backgrounds for the quadratic order. The specific heat kernel results, depicted in \cref{sdc29} for an arbitrary dilaton coupling, are obtained entirely on-shell and found to be manageable in terms of the background invariants preserving the electromagnetic duality. At any point, by setting the flat-space limit of vanishing cosmological constant, we can retrieve the same Seeley-DeWitt results for the $U(1)$-charged EMD theory. Next, \cref{EMD1b} utilizes the heat kernel data of \cref{EMD1a} and computes the logarithmic correction formulas for the Schwarzschild-AdS, Reissner-Nordstr\"om-AdS and Kerr-AdS black holes embedded in the EMD-AdS theory as well as for the Schwarzschild, Reissner-Nordstr\"om and Kerr black holes embedded in the EMD theory. The non-extremal and extremal relations are separately derived for the three choices of the dilaton coupling constant as per string theory and recorded as in \Cref{logs}. We analyzed the universal status of all the results and found that the logarithmic corrections for AdS$_4$ black holes are non-topological. In contrast, the corrections are topological (i.e., pure number and independent of black hole parameters) for the case of flat black holes, except for the non-extremal charged Reissner-Nordstr\"om background.
{
	\renewcommand{\arraystretch}{1.3}
	\begin{table}[t]
		\centering
		\hspace{-0.2in}
		\begin{tabular}{|>{\centering}p{1.65in}|>{\centering}p{2.4in}|>{\centering}p{1.35in}|}
			\hline
			\textbf{Logarithmic Correction Results} & \textbf{$\bm{U(1)}$-charged EMD and EMD-AdS Theories} & \textbf{$\bm{U(1)}^2$ EMD embedded $\mathcal{N}=4$ Supergravity} \tabularnewline \hline \hline
			Schwarzschild & \eqref{flat1a}, \eqref{flat2a}, \eqref{flat3a}  & \eqref{sdclog10a} \tabularnewline \hline
			Schwarzschild-AdS & \eqref{ads1a}, \eqref{ads2a}, \eqref{ads3a}  & \eqref{sdclog9a} \tabularnewline \hline
			Kerr & \eqref{flat1b}, \eqref{flat2b}, \eqref{flat3b} & \eqref{sdclog10b}, \eqref{sdclog10d} \tabularnewline \hline
			Kerr-AdS & (\ref{ads1c}, \ref{ads1e}), (\ref{ads2c}, \ref{ads2e}), (\ref{ads3c}, \ref{ads3e})  & \eqref{sdclog9c}, \eqref{sdclog9e} \tabularnewline \hline
			Reissner-Nordstr\"om & \eqref{flat1c}, \eqref{flat2c}, \eqref{flat3c} & \eqref{sdclog10c}, \eqref{sdclog10e} \tabularnewline \hline
			Reissner-Nordstr\"om-AdS & (\ref{ads1b}, \ref{ads1d}), (\ref{ads2b}, \ref{ads2d}), (\ref{ads3b}, \ref{ads3d})  & \eqref{sdclog9b}, \eqref{sdclog9d} \tabularnewline \hline
		\end{tabular}
\caption{The list of logarithmic entropy correction results for extremal and non-extremal black holes in ${U(1)}$ EMD theories and ${U(1)^2}$ EMD embedded $\mathcal{N}=4$ bosonic supergravity.}\label{logs}
	\end{table}
} 

\Cref{EMD2} is another novel and central part of this paper where we upgrade the whole Euclidean quantum gravity and heat kernel setup for a more generalized $U(1)^2$-charged EMD theory with two Maxwell-dilaton couplings that can be explicitly embedded into $\mathcal{N}=4$ supergravity (e.g., see \cite{Kallosh:1992ii,Myung:2020dqt,Cvetic:2014vsa} and citations therein). The embedding condition is determined by a fixed choice of the two dilaton coupling constants that exactly describes the bosonic sector of the SO(4) version of $\mathcal{N}=4$ ungauged and gauged supergravity in 4D. Similarly, we computed the necessary Seeley-DeWitt coefficient $a_4(x)$ (see \cref{sdclog7}) and utilized it to explore logarithmic corrections for the flat and AdS black holes in the special class of $U(1)^2$-charged EMD embedded $\mathcal{N}=4$ ungauged and gauged supergravity. As recorded in \Cref{logs}, the correction results depict a similar kind of universal or topological profile as we witnessed for the black holes in $U(1)$-charged models. All these calculated leading quantum corrections and their observed ``topological {vs.} non-topological'' status provide a wider ``infrared window'' to probe into microscopic degrees of freedom underlying black hole entropy in string theory.

Finally, in \cref{diss}, we conclude this paper with a summary, discussing the novelty of the results and making some remarks on their future implications and outlook. \Cref{calcul} encodes a handful of details about the complicated Seeley-DeWitt trace calculations in the $U(1)^2$-charged EMD model embedded into $\mathcal{N}=4$ supergravity for interested readers. \Cref{holo} includes a brief note on the holographic renormalization process used in integrating the necessary Seeley-DeWitt invariants around the concerned AdS$_4$ black hole backgrounds. In \Cref{enhci}, we list the explicit forms of the similar curvature invariants and their regulated integrations around the near-horizon geometry of extremal black holes for proceeding via the QEF formalism. \Cref{genlog} {includes the general logarithmic correction formulas for the black holes embedded in the class of $U(1)$-charged EMD-AdS and EMD models having a common dilaton coupling constant parameter.}


\section{Effective action and heat kernel ``recipe'' for logarithmic correction}\label{setup}

This section sets up the central working formula for computing the one-loop effective action for finding the logarithmic correction to black hole entropy using the Euclidean quantum gravity approach \cite{Gibbons:1977ta,Hawking:1978td}, followed by the heat kernel method \cite{Hawking:1977te,Denardo:1982tb,Avramidi:1994th,Barvinsky:2015,Vassilevich:2003ll,Sen:2013ns}. Here the calculation treatments for the interested 4D black holes are explicitly highlighted.  

\subsection{The setup}\label{setups}
Let us consider the generic rotating and charged black hole solution in a $D$-dimensional gravity theory described by the following path integral\footnote{Throughout, Boltzmann constant ($k_{B}$), Planck-Dirac constant ($\hbar$) and speed of light ($c$) are set as unity.}
\begin{align}\label{set1}
	\begin{gathered}
	\mathcal{Z}(\beta,\vec{\omega},\vec{\mu}) = \int \mathscr{D}[g,\varphi]\exp\left(-\mathcal{S}_E[g,\varphi]\right),\\[3pt]
	\mathcal{S}_E[g,\varphi] = \int \mathrm{d}^Dx\sqrt{\det g} \thinspace\mathscr{L}\left[g_{\mu\nu}, \varphi\right],
	\end{gathered}
\end{align}
where $\mathcal{S}_E[g,\varphi]$ is the Euclideanized action and $\mathscr{D}[g,\varphi]$ is a measure over all fields $\varphi$ and space-time metric $g_{\mu\nu}$ with the asymptotic boundary conditions controlled by the fixed inverse temperature $\beta$, angular velocities $\vec{\omega}$, and chemical potentials $\vec{\mu}$, which are thermodynamically dual to black hole mass $M$, charges $\vec{Q}$, and angular momenta $\vec{J}$, respectively. Then the grand canonical partition function \eqref{set1} can be related to the microcanonical entropy of the black hole via the Legendre transformation \cite{Gibbons:1977ta},\footnote{
	$M$, $\vec{J}$, and $\vec{Q}$ of a black hole are determined by the potentials $\beta$, $\vec{\omega}$, and $\vec{\mu}$ and vice versa \cite{Sen:2013ns}:
	\begin{align*}
		\beta = \frac{\partial S_{\text{BH}}}{\partial M},\thinspace \vec{\omega} = \frac{\partial S_{\text{BH}}}{\partial\vec{J}},\thinspace \vec{\mu} = \frac{\partial S_{\text{BH}}}{\partial\vec{Q}} \Longleftrightarrow M = - \frac{\partial \ln {\mathcal{Z}}}{\partial \beta},\thinspace \vec{J} = - \frac{\partial \ln {\mathcal{Z}}}{\partial \vec{\omega}},\thinspace \vec{Q} = - \frac{\partial \ln {\mathcal{Z}}}{\partial \vec{\mu}}.	
	\end{align*}
}
\begin{align}\label{set3}
	S_{\text{bh}}(M, \vec{J}, \vec{Q}) = \ln {\mathcal{Z}}(\beta,\vec{\omega},\vec{\mu}) + \beta M + \vec{\omega}\cdot\vec{J} + \vec{\mu}\cdot \vec{Q}.
\end{align}
Considering the black hole as a classical saddle-point $(\bar{g},\bar{\varphi})$, we can fluctuate the entire field content (including the metric) for small quantum fluctuations ${\phi}_m =\lbrace\tilde{g},\tilde{\varphi}\rbrace$,
\begin{equation}\label{set2}
	g = \bar{g} + \tilde{g}, \enspace \varphi = \bar\varphi + \tilde{\varphi},
\end{equation}
and then compute the effective action perturbatively via the saddle point or loop expansion up to the one-loop as
\begin{align}\label{set100}
\Gamma[g,\varphi]=	- \ln \mathcal{Z}(\beta,\vec{\omega},\vec{\mu}) \simeq \mathcal{S}_E[\bar{g},\bar{\varphi}] + \mathpzc{W}.
\end{align}
Here the leading contribution $\mathcal{S}_E[\bar{g},\bar{\varphi}]$ is the Euclideanized on-shell action which always gives rise to the classical Bekenstein-Hawking entropy \cite{Gibbons:1977ta,Hawking:1978td} or the Bekenstein-Hawking-Wald entropy \cite{Wald:1993rw} when the action incorporates higher curvature terms beyond the two-derivative limit of Einstein's gravity,
\begin{align}
	S_{\text{BH}}(M, \vec{J}, \vec{Q}) = \frac{\mathcal{A}_{H}(M, \vec{J}, \vec{Q})}{4G_D} + \cdots.
\end{align}
On the other hand, the sub-leading (next to the saddle-point) contribution $\mathpzc{W}$ is the effective action at one-loop, described by the functional determinant of the kinetic operator $\mathcal{H}= \frac{\delta^2\mathcal{S}_E}{\delta {\phi}_m^2}$ controlling all the quadratic fluctuations as
\begin{align}\label{set4}
	\mathpzc{W} = -\ln \int \mathscr{D}[\phi_m]\exp\left(-\int \mathrm{d}^Dx\sqrt{\det \bar{g}}\thinspace\phi_m \mathcal{H}^m_n\phi^n\right) = \frac{\chi}{2}\ln \det \mathcal{H},
\end{align}   
where $\chi = \pm 1$ stands for bosons and fermions, respectively. Evidently, the effective action \eqref{set4} determines the one-loop correction part in the total quantum black hole entropy \eqref{set3} when one utilizes the expansion \eqref{set100}, i.e., 
\begin{align}\label{set5}
	S_{\text{bh}}(M, \vec{J}, \vec{Q}) = S_{\text{BH}}(M, \vec{J}, \vec{Q}) - \mathpzc{W}  + \beta M + \vec{\omega}\cdot\vec{J} + \vec{\mu}\cdot \vec{Q},
\end{align}
where all the potential parameters $\left(\beta,\thinspace \vec{\omega},\thinspace \vec{\mu}\right)$ are set to be fixed on their classical or saddle-point values \cite{Sen:2013ns}. Next, the task is to evaluate as well as regulate the UV divergences of the one-loop effective action $\mathpzc{W}$ by the \emph{heat kernel method}, i.e., expressing the one-loop functional determinant $\det \mathcal{H}$ in terms of the diagonal elements of the operator $e^{-\tau \mathcal{H}}$ for a proper time $\tau$ \cite{Vassilevich:2003ll}. Interestingly, our desired logarithmic correction to black hole entropy emerged from the UV-independent part of $\mathpzc{W}$ when only massless fluctuations are turned on. However, the heat kernel method can only compute the local part of the logarithmic correction. In contrast, there exists a global contribution from two distinct sources: 
\begin{inparaenum}[(i)]
	\item the zero-modes inside the eigenvalue spectrum of the kinetic operator $\mathcal{H}$, and
	\item  the explicit correction due to upgrading the black hole from a grand canonical to a microcanonical ensemble during the process \eqref{set3}.	
\end{inparaenum}
These global or zero-mode contributions are incompatible with the heat kernel setup and must be treated separately (see \cref{zeromode}). Inside the heat kernel treatment, one must interpret $\mathpzc{W}$ as the explicit effective action of the black hole by subtracting the thermal gas contribution of the theory with which the related black hole is in equilibrium. The whole setup is depicted as follows. 

Suppose $ f_i^{m}$ are the orthonormal eigenfunctions of the one-loop kinetic operator $\mathcal{H}$ with eigenvalues $ h_i$, then the related heat kernel is introduced as 
\begin{align}\label{set6}
	K^{mn}(x,x^\prime;\tau) = \sum_i e^{-h_i\tau}f_i^m(x)f_i^n(y),
\end{align}
where $m$ and $n$ are indices labeling the particular fluctuations along with their tensor indices at two distinct spacetime points, $x$ and $y$. Here the proper time $\tau$ acts as the auxiliary heat kernel parameter having a dimension of length-square. Then, we can express the heat trace $K(\tau)$ into the following form where its local and zero mode parts are separated
\begin{align}\label{set7}
K(\tau) = \int \mathrm{d}^Dx\sqrt{\det \bar{g}}\thinspace I_{mn}K^{mn}(x,x;\tau) = \sideset{}{'}\sum_{i\atop (h_i \neq 0)} e^{-\tau h_i} + n_{\text{zm}},
\end{align}
where $I_{mn}$ is an effective metric or projection operator in the space of individual fluctuations. The primed part in the heat trace form \eqref{set7} depicts the removal of the total number of zero-modes $n_{\text{zm}}$ associated with the eigenfunctions $ f^0_{i}$ inducing zero eigenvalues (i.e., $\mathcal{H}f^0_{i} = 0$) so that,
\begin{align}\label{set8}
n_{\text{zm}} = \sum_i\int \mathrm{d}^Dx \sqrt{\text{det}\thinspace \bar{g}}\thinspace f^0_{i}(x)f^0_{i}(x).
\end{align} 
With the heat kernel setup mentioned above, the Schwinger-DeWitt proper time representation \cite{Schwinger:1951sp,DeWitt:1975ps} allows us to express the local or non-zero mode part of one-loop effective action into the following renormalized form
\begin{align}\label{set9}
	\mathpzc{W}^\prime = -\frac{\chi}{2}\int_\epsilon^\infty \frac{\mathrm{d}\tau}{\tau} \Big(K(\tau) - n_{\text{zm}} \Big),
\end{align}
where $\epsilon$ is a UV cut-off that is of the order of Planck area $\ell_P^2$ (or equivalently $G_D$ in the choice of units of this paper) and independent of the size of the concerned black hole. At this point, it is necessary to emphasize that simply calculating the effective action \eqref{set9} is not sufficient to find the logarithmic correction to black hole entropy. This is mainly because the saddle-point solution or black hole is always in equilibrium with a thermal gas of all particles present in theory. But Ashoke Sen in \cite{Sen:2013ns} showed a treatment to eliminate the thermal gas contribution and isolate the exact piece of effective action that is exclusively related to the quantum black hole entropy or underlying microstates. In this progress, one first needs to confine the original Euclidean black hole of radius $\mathfrak{R}$ (such that its horizon area scales as $\mathcal{A}_H \sim \mathfrak{R}^{D-2}$) inside a thermal box of size $\zeta$. The logarithmic entropy correction we are interested in this paper is highly sensitive to the scaling regime of black hole backgrounds and will be calculated under the scaling limits \cite{Sen:2013ns},
\begin{align}\label{set10}
\bar{g}_{\mu\nu} \to L^2 \bar{g}_{\mu\nu}, \enspace M \to L^{D-3}M, \enspace  \vec{Q} \to L^{D-3}\vec{Q},\enspace  \vec{J} \to L^{D-2}\vec{J},
\end{align}
which further leads to the following relations for a common and large length scale ${L}$
\begin{align}\label{set11}
	\begin{gathered}
		S_{\text{BH}} (M,\vec{J},\vec{Q}) = L^{D-2}S_{\text{BH}} (M,\vec{J},\vec{Q}),\\[3pt]
		\mathfrak{R} (M,\vec{J},\vec{Q}) = L\mathfrak{R} (M,\vec{J},\vec{Q}).
	\end{gathered}
\end{align}
Next, we should consider a similar but rescaled black hole solution of radius $\mathfrak{R}_0$ and confine it inside an identical thermal box of size $\zeta^{(0)} = (\mathfrak{R}_0/\mathfrak{R})\zeta$. This choice essentially allows the new black hole system to satisfy the same fixed scaling choices \eqref{set10} and \eqref{set11} for $L = \mathfrak{R}_0/\mathfrak{R}$. Consequently, the eigenvalues $h^{(0)}_i$ and $h_i$ of the kinetic operator $\mathcal{H}$ inside the new and original systems are related via
\begin{align}
	h^{(0)}_i = h_i/L^2= h_i\left(\mathfrak{R}_0/\mathfrak{R}\right)^2 ,
\end{align}\label{set12}
which immediately resets the heat kernel parameter in the new black hole system as $\tau \to \tau/L^2$ for satisfying the one-loop effective action form \eqref{set9}. At this stage, it has been ascertained that the leading thermal gas contributions to the effective actions of the two systems with the original and rescaled black holes are exactly identical \cite{Sen:2013ns}. This feature is crucial and suggests that the subtraction of the effective action forms \eqref{set9} for the two systems will completely eliminate the equal thermal gas contributions and provide the exact difference in the black hole quantum effective actions, i.e.,
 \begin{align}\label{set13}
 	\Delta \mathpzc{W}^\prime =\left(\mathpzc{W}^\prime_\mathfrak{R} + \mathpzc{W}^\prime_{\text{gas}}\right)-\left(\mathpzc{W}^\prime_{\mathfrak{R}_0} + \mathpzc{W}^\prime_{\text{gas}}\right)= -\frac{\chi}{2}\int_\epsilon^{\epsilon/L^2} \frac{\mathrm{d}\tau}{\tau} \Big(K(\tau) - n_{\text{zm}} \Big).
 \end{align}
We can see that the above effective action integral has a dominant contribution within the IR integration range $\epsilon/\mathfrak{R}^2 < \tau/\mathfrak{R}^2 < \epsilon/{\mathfrak{R}_0}^2 $. In this regime, we must set the large-charge limit on the black holes, i.e., $\mathfrak{R} \gg \sqrt{\epsilon}$ and ${\mathfrak{R}_0} \gg \sqrt{\epsilon}$ (since $\epsilon\sim {\ell_P}^2\sim G_D$), which will enforce a small proper time variable $\tau$ and allow us to cast the following short-time heat trace expansion \cite{Seeley:1966tt,Seeley:1969uu,DeWitt:1965ff,DeWitt:1967gg,DeWitt:1967hh,DeWitt:1967ii}
\begin{align}\label{set14}
	K(\tau) \overset{\tau \to 0}{=} \int \mathrm{d}^Dx \sqrt{\text{det}\thinspace \bar{g}}\thinspace \sum_{n=0}^\infty \tau^{n-\frac{D}{2}}a_{2n}(x),
\end{align}
where $a_{2n}(x)$ are the well-known \emph{Seeley-DeWitt coefficients}. Interestingly, it is found that only the $\tau$-independent term of the expansion \eqref{set14}, i.e., the term involving $a_{2n}(x)$ coefficient with $n=\dfrac{D}{2}$, extracts the desired ``logarithmic term'' within the effective action form \eqref{set13},   
\begin{align}\label{set15}
	\Delta \mathpzc{W}^\prime = \int \mathrm{d}^Dx \sqrt{\text{det}\thinspace \bar{g}}\thinspace \Bigg[& \frac{a_0(x)}{D\epsilon^{\frac{D}{2}}}\left(\frac{\mathfrak{R_0}^D}{\mathfrak{R}^D}-1\right)+ \frac{a_2(x)}{(D-2)\epsilon^{\frac{D}{2}-1}}\left(\frac{\mathfrak{R_0}^{D-2}}{\mathfrak{R}^{D-2}}-1\right) \nonumber \\
	& + \mathcal{O}(\epsilon^{-1}) + a_D(x)\ln \left(\frac{\mathfrak{R_0}}{\mathfrak{R}}\right) + \mathcal{O}(\epsilon)\Bigg]-\chi n_{\text{zm}}\ln \left(\frac{\mathfrak{R_0}}{\mathfrak{R}}\right).
\end{align}
Notice that the parameter $\chi$ controlling the spin-type signatures is absorbed inside the terms involving Seeley-DeWitt coefficients, which is further readjusted appropriately in the formula \eqref{comp5}. Here the first two terms related to $a_0(x)$ and $a_2(x)$ coefficients capture the one-loop renormalization of cosmological and gravitational constants, respectively.\footnote{The one-loop renormalization of cosmological and gravitational constants via $a_0(x)$ and $a_2(x)$ are ambiguous and scheme-dependent. For an example, refer to Appendix D of \cite{David:2021eoq}.} Since the UV cutoff $\epsilon$ is set as small, we can always neglect the higher-order divergent terms of \eqref{set15} in the limit $\epsilon \to 0$ and identify the explicit form of logarithmic correction to the local part of one-loop effective action associated with the original black hole of radius $\mathfrak{R}$ as\footnote{\label{massfoot} Note that the effective action \eqref{set4} with massive $m$ fluctuations turned on will read as $\mathpzc{W} = \mp\frac{1}{2}\ln \det (\mathcal{H} + m^2)$, setting the heat trace expansion $\sim e^{-\tau m^2}\sum \tau^{n-\frac{D}{2}}a_{2n}(x) $ which never leads to $\ln \mathcal{A}_H$ in the one-loop correction terms via the integration \eqref{set13}. Thus, as argued earlier, the logarithmic correction to black hole entropy is always insensitive to massive modes.} 
\begin{align}\label{set16}
\mathpzc{W}^\prime_{\text{log}} = - \frac{1}{(D-2)}\left[ \int \mathrm{d}^Dx \sqrt{\text{det}\thinspace \bar{g}}\thinspace a_D(x)-\chi n_{\text{zm}}\right] \ln \left(\frac{\mathcal{A}_H}{G_D}\right).
\end{align} 
We can see that the above relation sets a fixed integration range $\epsilon \ll \tau \ll \mathcal{A}_H$ for any generic black hole of horizon area $\mathcal{A}_H$ in the integral \eqref{set9}. Apart from the local part \eqref{set16}, there is also zero-mode contribution $\mathpzc{W}_{\text{zm}}$ to the one-loop effective action, i.e., 
\begin{align}\label{set17}
		\mathpzc{W}_{\text{zm}} = -\ln\int \mathscr{D}[\phi_m]\Big\rvert_{\mathcal{H}\phi_m = 0},
\end{align}
for which $\exp\left(-\int \mathrm{d}^Dx\sqrt{\det \bar{g}}\thinspace\phi_m \mathcal{H}^m_n\phi^n\right) = 1$. These zero modes originated due to different asymptotic symmetry transformations (e.g., gauge transformations, diffeomorphism invariance, etc.) that do not disappear at infinity. A typical way to analyze the zero modes is by changing the integration variables from fields to the parameters labeling the underlying asymptotic symmetries in the volume integral \eqref{set17} so that the related Jacobian assigns a $\mathfrak{R}^{\beta_{\phi}}$ factor to every zero mode of the original black hole \cite{Banerjee:2011oo,Banerjee:2011pp,Sen:2012rr,Sen:2012qq}, i.e., 
\begin{align}\label{set18}
		e^{-\mathpzc{W}_{\text{zm}}} \simeq {\mathfrak{R}}^{ \sum\limits_{\lbrace\phi\rbrace}\chi \beta_{\phi} n^{0}_{\phi}},
\end{align}
where $n^{0}_\phi$ is the number of zero modes related to each fluctuation such that the total zero-modes $n_{\text{zm}} = \sum\limits_{\lbrace\phi\rbrace} n^{0}_\phi$ satisfy the definition \eqref{set8}. The parameter $\beta_{\phi}$ is a constant number that depends on the scaling dimensions of zero-mode integrals and can be computed by normalizing the path integral for each fluctuation via
\begin{align}\label{set18x}
	\int \mathscr{D}[\phi_m]\exp \left[- \mathfrak{R}^{\beta_{\phi}}\int \mathrm{d}^Dx \sqrt{\det \bar{g}^{(0)}}\,I^{(0)mn}\phi_m\phi_n\right] = 1,
\end{align}
where the background metric is scaled as $\bar{g}_{\mu\nu} = \mathfrak{R}^2\bar{g}^{(0)}_{\mu\nu}$ which further rescales the effective metric $I^{mn}$ of the fluctuations $\phi_m$. Here $\bar{g}^{(0)}_{\mu\nu}$ and $I^{(0)mn}$ are $\mathfrak{R}$-independent. In $D$-dimensional spacetimes, the treatment in \eqref{set18x} computes $\beta_2 = \frac{D}{2}$, $\beta_{1}=\frac{D-2}{2}$ and $\beta_{3/2}=D-1$ respectively for the graviton, vector or gauge field and spin-3/2 Rarita-Schwinger fluctuations \cite{Banerjee:2011oo,Banerjee:2011pp,Sen:2012rr,Sen:2012qq,Bhattacharyya:2012ye,Gupta:2014ns}. Finally, we recombine the local \eqref{set16} and zero-mode \eqref{set18} contributions and express the net logarithmic contribution in the one-loop effective action for black hole quantum entropy as
\begin{align}\label{set19}
\mathpzc{W}_{\text{log}} = - \frac{1}{(D-2)}\left[ \int_{\text{BH geometry}} \mathrm{d}^Dx \sqrt{\text{det}\thinspace \bar{g}}\thinspace a_D(x) + \sum_{\lbrace \phi\rbrace}\chi(\beta_{\phi}- 1)n^0_{\phi}\right] \ln \left(\frac{\mathcal{A}_H}{G_D}\right).
\end{align} 
It is important to mention that the Seeley-DeWitt coefficients with odd index, i.e., $a_{2n+1}(x)$ always vanish over the manifolds without boundary due to the lack of diffeomorphism invariant scalar functions connected to the background metric \cite{Vassilevich:2003ll}. This, in turn, explains the absence of the local part involving $a_D(x)$ in the relation \eqref{set19} for all quantum black holes in all odd $D$-dimensional spacetimes. 
In those cases, only the zero-mode part will contribute to the logarithmic correction.

\subsection{Treatment for black holes in the extremal limit}\label{extlim}
The analysis of any extremal black hole via the Euclidean quantum gravity setup of \cref{setups} is a bit tricky. In this line, a naive way is to directly set the extremal or zero-temperature limit $\beta \to \infty$ (or $T_{\mathrm{bh}}= (\frac{\partial S_{\text{bh}}}{\partial M})^{-1} \to 0$) on the general non-extremal or finite-temperature setup. But in this extremal limit, the relevant infinite thermal circle of radius $\dfrac{1}{\beta}$ will make the on-shell Euclideanized action and one-loop effective action in \eqref{set100} divergent. This divergence can be viewed as a correction to the respective actions, which is actually an infinite shift in the ground state energy induced due to the extremal limit. We will bypass this issue of analyzing the extremal black holes via the two different treatments described below.
 
The most efficient and precise treatment to regulate the divergence due to extremality is Sen's quantum entropy function (QEF) formalism \cite{Sen:2008wa,Sen:2009wb,Sen:2009wc}. To date, this treatment is highly successful in computing the logarithmic and other quantum corrections to the extremal black hole entropy \cite{Banerjee:2011oo,Banerjee:2011pp,Sen:2012rr,Sen:2012qq,Chowdhury:2014np,Gupta:2014ns,Bhattacharyya:2012ss,Karan:2019sk,Karan:2020sk,Banerjee:2020wbr,Karan:2021teq,Jeon:2017ij,Larsen:2015nx}. Based on the AdS/CFT correspondence and the fact that any extremal-near-horizon black hole background always accommodates an $\mathrm{AdS}_2$ part, the QEF formalism proved that only the ``finite'' piece of the near-horizon partition function contributes to the quantum degeneracy of extremal black holes. This, in turn, leads to a revised local part in the one-loop effective action form \eqref{set19} as 
\begin{align}\label{set20}
	\int_{\text{BH geometry}} \mathrm{d}^Dx \sqrt{\text{det}\thinspace \bar{g}}\thinspace a_D(x)\Big\rvert_{\beta \to \infty} \equiv {\left\langle \int_{\text{near-horizon}} \mathrm{d}^Dx \sqrt{\text{det}\thinspace \bar{g}}\thinspace a_D(x) \right\rangle}_{AdS_2}^{\text{finite}}.
\end{align} 
Here $\langle \rangle$ denotes the integration of $a_D(x)$ coefficient over only the finite extremal near-horizon geometry (i.e., by dropping all regulated $\mathrm{AdS}_2$-boundary independent terms) structured as $\mathrm{AdS}_2 \times \sum_{\bar{g}}^{D-2}$ where $\sum_{\bar{g}}$ is a $(D-2)$ dimensional space of all the compact and angular coordinates fibered over the $\mathrm{AdS}_2$ part. For more technical details, readers are referred to \cite{Banerjee:2011pp,Sen:2012rr,Sen:2012qq,Bhattacharyya:2012ss,Karan:2019sk,Karan:2020sk,Banerjee:2020wbr,Karan:2021teq}.

Another alternative of the QEF formalism is recently used by David et al. in \cite{David:2021eoq}, which has successfully resolved the problem of handling the divergent terms arising due to employing the extremal limit directly on the Euclidean gravity structure of non-extremal black holes. This prescription involves managing the outer horizon geometry (e.g., see the relation \eqref{ebh1}) as an expansion of $\beta$ in the limit $\beta\to\infty$ (\textit{small-temperature expansion}) such that we can impose the extremal limit on various charges by fixing their values and then express the local part of one-loop effective action \eqref{set19} up to a finite constant as
\begin{align}\label{set21}
	\int_{\text{BH geometry}} \mathrm{d}^Dx \sqrt{\text{det}\thinspace \bar{g}}\thinspace a_D(x)\Big\rvert_{\beta \to \infty} \equiv \mathcal{C}_0 + \mathcal{C}_1\beta + \mathcal{O}(\beta^{-1}).
\end{align}
We can see that the second term, linear in $\beta$ with a constant prefactor $\mathcal{C}_1$, is divergent. This divergence is nothing but an infinite shift of the ground state energy due to quadratic fluctuations and hence can be removed via a proper renormalization procedure. The terms with inverse order of must $\beta$ vanish in the extremal limit. Thus, after neglecting the divergent term and setting $\beta \to \infty$, it is convenient to identify the $\beta$-independent constant term $\mathcal{C}_0$ as a finite and unambiguous contribution to the one-loop quantum entropy for the extremal black holes. This extremal treatment via the small-temperature expansion seems almost equivalent to the principle of the QEF formalism.  In the current paper, we cast both the treatments \eqref{set20} and \eqref{set21} for calculating logarithmic correction to the entropy of extremal flat and AdS black holes in 4D and check the consistencies of the obtained results (e.g., please refer to \cref{ebh}).

\subsection{Computation for four-dimensional black holes} \label{comp}

In the present Euclidean quantum gravity setup, the central working formula for calculating logarithmic correction to the entropy of any four-dimensional black hole is
\begin{subequations}\label{comp1}
	\begin{align}\label{comp1a}
	\Delta S_{\text{BH}} &= \frac{1}{2}(\mathcal{C}_{\text{local}}+\mathcal{C}_{\text{zm}})\ln\left(\frac{\mathcal{A}_{H}}{G_D}\right),
	\end{align}
with the local ($\mathcal{C}_{\text{local}}$) and global or zero-mode ($\mathcal{C}_{\text{zm}}$) contributions given by the relations, 
	\begin{align}
		\mathcal{C}_{\text{local}} &= \int_{\text{BH geometry}} \mathrm{d}^4x \sqrt{\text{det}\thinspace \bar{g}}\thinspace a_4(x),\label{comp1b}\\
		\mathcal{C}_{\text{zm}}&= \sum_{\lbrace \phi\rbrace}\chi(\beta_{\phi}- 1)n^0_{\phi}.\label{comp1c}
	\end{align}
\end{subequations}
\subsubsection{Local contribution}\label{local}
We will follow the heat kernel manual \cite{Vassilevich:2003ll} to compute the third-order Seeley-DeWitt coefficient $a_4(x)$ only in terms of different curvature invariants of 4D black hole backgrounds. In this progress, we first need to adjust the quadratic fluctuated action, i.e.,
\begin{align}\label{comp2}
	\delta^2 \mathcal{S}[\phi_m]= \int \mathrm{d}^4x \sqrt{\text{det}\thinspace \bar{g}}\thinspace \phi_m \mathcal{H}^m_n\phi^n,
\end{align} 
such that the relevant matrix structure of the kinetic operator $\mathcal{H}^m_n$ becomes Laplace-type of the following schematic
\begin{align}\label{comp3}
	\mathcal{H}^m_n = -D_\rho D^\rho I^m_n - (N_\rho D^\rho)^m_n - P^m_n,
\end{align}
where $D_\rho$ is the space-time covariant derivative incorporating the spin and Christoffel connections, $P$ and $N_\rho$ are arbitrary matrices that combinedly determine the potential part, and $I$ is a unit matrix in the space of each fluctuation.\footnote{Generally, $I_0 = 1$, $I_1= \bar{g}^{\mu\nu}$, $I_{1/2}= \mathbb{I}_4$, and $I_{3/2}= \mathbb{I}_4 \bar{g}^{\mu\nu}$ for the scalar, vector, spin-1/2 and spin-3/2 Rarita-Schwinger fluctuations, respectively. $\mathbb{I}_4$ is the identity matrix of Clifford algebra describing 4D spinors. In this paper, $I$ for the spin-2 graviton will be structured as the DeWitt metric \eqref{sdc9}. The trace of $I$ will depict the effective off-shell degrees of freedom of any specific fluctuation in the concerned theory.}  Next, we need to absorb the linear derivative term of \eqref{comp3} into a more compact form,
\begin{subequations}\label{comp4}
\begin{align}\label{comp4a}
	\mathcal{H}^m_n = -\mathcal{D}_\rho\mathcal{D}^\rho I^m_n - E^m_n,
\end{align}  
where the covariant derivative $D_\rho$ is redefined by incorporating a new parameter $\omega_\rho$ controlling the gauge connection between fluctuations,
\begin{align}\label{comp4b}
	\mathcal{D}_\rho \phi_m= D_\rho \phi_m + (\omega_\rho)^m_n \phi^n, \thinspace (\omega_\rho)^m_n = \frac{1}{2}(N_\rho)^m_n \quad\forall\, m\neq n, 
\end{align}
so that the matrix-valued effective potential $E$ is expressed as
\begin{align}
	\phi_m E^m_n \phi^n &= \phi_m P^m_n \phi^n -\phi_m(D_\rho\omega^\rho)^m_n \phi^n -\phi_m(\omega_\rho)^{mp}(\omega^\rho)_{pn}\phi^n. \label{comp4c}
\end{align}
Similarly, we can express the commutator curvature $\Omega_{\rho\sigma}= [\mathcal{D}_\rho,\mathcal{D}_\sigma]$ associated with the new covariant derivative $\mathcal{D}_\rho$ as
\begin{align}
	\phi_m\left(\Omega_{\rho\sigma}\right)^m_n \phi^n  = \phi_m [D_\rho,D_\sigma] \phi^m + \phi_m {D_{[\rho}\omega_{\sigma]}} ^m_n \phi^n + \phi_m[\omega_\rho,\omega_\sigma]^m_n \phi^n.\label{comp4d}
\end{align}
\end{subequations}
With the help of the above setup, the Seeley-DeWitt coefficient $a_4(x)$ can capture all the matrix-structure data of the kinetic operator $\mathcal{H}$ in terms of all the background curvatures $R_{\mu\nu\rho\sigma}$, $R_{\mu\nu}$, $R$, $\Omega_{\rho\sigma}$ and $E$ via the following formula \cite{Vassilevich:2003ll}
\begin{align}\label{comp5}
	a_4(x) &= \frac{\chi}{16\pi^2} \bigg\lbrace\frac{1}{6}\mathrm{Tr}(D_\rho D^\rho E)+ \frac{1}{30}\mathrm{Tr}(D_\rho D^\rho R) +\frac{1}{2}\mathrm{Tr}(E^2) +\frac{1}{6}R\mathrm{Tr}(E)  \\ \nonumber
	&\qquad \quad + \frac{1}{12}\mathrm{Tr}(\Omega_{\rho\sigma}\Omega^{\rho\sigma})+\frac{1}{180}\bigg(R_{\mu\nu\rho\sigma}R^{\mu\nu\rho\sigma}-R_{\mu\nu}R^{\mu\nu}+ \frac{5}{2} R^2\bigg)\mathrm{Tr}(I)\bigg\rbrace,
\end{align}  
where ``Tr'' is the trace operation over the index $m$ labeling all fluctuations including their tensor indices $\phi_m$. To define these traces, the identity matrix $I$ plays a crucial role by acting as a projection operator or effective metric for individual fluctuations (e.g., see the definitions \eqref{calcul44}). Note that, we will always neglect the total derivative terms (e.g., the first two terms in formula \eqref{comp5}) for the remaining analysis of this paper. When one integrates $a_4(x)$ around the appropriate part of extremal and non-extremal asymptotic black hole geometries via the integral \eqref{comp1b}, all the total derivative terms will appear as non-contributing boundary terms. 
Since our primary motive is to find the logarithmic correction for 4D black holes, it is sufficient to calculate Seeley-DeWitt coefficients only up to the $a_4(x)$ order. The relevant formulas for the coefficients $a_1(x)$ and $a_2(x)$ are \cite{Vassilevich:2003ll}
\begin{align}
	a_0(x) &= \frac{\chi}{16\pi^2}\mathrm{Tr}(I),\label{comp9a} \\
	a_2(x) &=\frac{\chi}{16\pi^2}\bigg\lbrace\mathrm{Tr}(E) + \frac{1}{6}R\,\mathrm{Tr}(I) \bigg\rbrace.\label{comp9b}
\end{align}
The above-mentioned heat kernel approach strictly demands the kinetic operator $\mathcal{H}$ in a quadratic form. But the quadratic action for the case of fermionic fluctuations $\Psi$ is always structured by a first-order operator $\slashed{D}$,   
\begin{align}\label{comp6}
	\delta^2 \mathcal{S}[{\Psi}_m,\bar{\Psi}_m]= \int \mathrm{d}^4x \sqrt{\text{det}\thinspace \bar{g}}\thinspace \bar{\Psi}_m \slashed{D}^m_n{\Psi}^n.
\end{align}
By following the technique used in \cite{Sen:2012qq}, we can upgrade the fermionic operator $\slashed{D}$ into a second-order operator $\mathcal{H}^m_n = \left(\slashed{D}^\dagger\right)^{mp} \slashed{D}_{pn}$ controlling the one-loop action \eqref{set4} via   
\begin{align}\label{comp7}
	\text{ln det}\thinspace \slashed{D}=\text{ln det}\thinspace \slashed{D}^\dagger = \frac{1}{2}\thinspace \text{ln det}\thinspace\slashed{D}^\dagger\slashed{D}, 
\end{align}
which essentially sets $\chi = 1$, $-1$ and $-\frac{1}{2}$ respectively for the bosons, Dirac fermions and Majorana fermions into the heat kernel setup via the formula \eqref{comp5}. These signatures will be the opposite for the ghost fields whenever the theory is gauge-fixed. For interested readers, we want to refer to \cite{Karan:2018ac} for more details about the Seeley-DeWitt calculations of elementary and minimal
spin-1/2 and spin-3/2 fermionic fields around a general background.
\subsubsection{Zero-mode contribution}\label{zeromode}
The parameters $(\beta_{\phi}, n^0_{\phi})$, controlling the zero-mode or global contribution \eqref{comp1c} of different field fluctuations around the flat and AdS black hole backgrounds, are well-known and reported in many works \cite{Banerjee:2011oo,Banerjee:2011pp,Sen:2012rr,Sen:2012qq,Sen:2013ns,Bhattacharyya:2012ye,Liu:2017vbl,Camporesi:1994ye}. For 4D backgrounds, the results  are listed in \Cref{tab:zeromode} which allows structuring the following compact formula for the global part of logarithmic corrections
\begin{table}[t]
	\setlength{\tabcolsep}{4pt} 
	\renewcommand{\arraystretch}{1.0}
	\captionsetup{font=small}
	\centering 
	\begin{tabular}{|l| c c c c c|} 
		\hline 
		 & \textbf{Metric} & \textbf{Scalar} & \textbf{Vector} & \textbf{Spin-1/2} & \textbf{Spin-3/2}
		\\[-0.9ex]
		\raisebox{1.5ex}{\textbf{Black hole backgrounds}} & $(\beta_{\phi}= 2)$ & $(\beta_{\phi}= 0)$ & $(\beta_{\phi}= 1)$ & $(\beta_{\phi}= 1)$ & $(\beta_{\phi}= 3)$
		\\ [0.7ex]
		\hline\hline 
		
		Non-extremal non-rotating & -3 & 0 & -1 & 0 & 0  \\[1ex]
		Non-extremal rotating     & -1  & 0 & -1 & 0 & 0  \\[1ex]
		
		Extremal (non-BPS) non-rotating &-6 & 0 & -1 & 0 & 0  \\[1ex]
		Extremal (non-BPS) rotating &-4 & 0 & -1 & 0 & 0  \\[1ex]
		Extremal (BPS) non-rotating & 2 & 0 & -1 & 0 & -4  \\[1ex]
		
		Extremal (BPS) rotating     & 4 & 0 & -1 & 0 & -4  \\[1ex]
		\hline 
	\end{tabular}
	\caption{{List of zero-mode number $n^0_{\phi}$ for different fluctuations around the 4D black hole backgrounds. The values of the scaling dimension parameter $\beta_{\phi}$ are mentioned in the corresponding columns. The negative values for the AdS$_4$ and extremal near-horizon backgrounds arise from the renormalized finite part of the zero-mode integral \eqref{set8} \cite{Banerjee:2011oo,Banerjee:2011pp,Sen:2012rr,Sen:2012qq,Bhattacharyya:2012ye,Liu:2017vbl}.}} \label{tab:zeromode}
\end{table}
\begin{align}\label{comp8}
		\mathcal{C}_{\text{zm}} = -(3+\mathbb{K})+ 3\delta_{\text{non-ext}} +  2\mathbb{N}_{\text{BPS}},
\end{align} 
where only the metric and spin-3/2 gravitino fluctuations contribute. For an extremal black bole ($\beta \to \infty$), the related analysis fully relies on the near-horizon geometry and is controlled by a dominant AdS$_2$ part which provides $-3$ zero-modes corresponding to unbroken translational symmetry (i.e., SL($2,\mathbb{R}$) symmetry of AdS$_2$ spaces). However, this contribution will be eliminated for non-extremal black holes where one needs to shift from the near-horizon to full geometry analysis by incorporating a $3\delta_{\text{non-ext}}$ contribution to account for the additional IR volume integration due to a finite $\beta$ (inverse temperature). Thus, $\delta_{\text{non-ext}}$ is 1 for non-extremal black holes, otherwise 0. Indeed the same fact also holds for AdS$_4$ black holes, which generally admit a 2-form zero-mode that vanishes when embedded in 4D theories \cite{Bhattacharyya:2012ye,Liu:2017vbl,Camporesi:1994ye}. In addition, there will be a contribution of $-K$ zero-modes associated with the rotational symmetries of both extremal and non-extremal black holes, which are induced due to change in the ensemble from grand canonical to microcanonical in step \eqref{set3} by fixing black hole mass $M$, charge $\vec{Q}$, and angular momentum $\vec{J}$. It is found that $K$ is 3 for spherically symmetric non-rotating ($J_3 = \vec{J^2}= 0$) backgrounds, otherwise 1 for any rotating black hole. Apart from these zero modes of the metric fluctuation (or graviton), the spin-3/2 gravitino fluctuation around a BPS black hole gives rise to a $2\mathbb{N}_{\text{BPS}}$ contribution, where  $\mathbb{N}_{\text{BPS}}=4$ corresponds to the number of preserved supersymmetry (i.e., the number of generators of PSU(1, 1|2) near-horizon symmetry) of BPS solutions in supergravity. However, around the backgrounds of any non-supersymmetric black hole, there will be no spin-3/2 zero-modes, i.e., $\mathbb{N}_{\text{BPS}}=0$. 
In this paper, we will directly use the formula \eqref{comp8} or the data recorded in \Cref{czero} to obtain the zero-mode contributions of black holes embedded into four-dimensional EMD and EMD-AdS theories.



\section{Heat kernel coefficients in EMD and EMD-AdS theories}\label{EMD1a}
This section aims to quantize the four-dimensional Einstein-Maxwell-dilaton (EMD) gravity models by fluctuating all the content and then analyze the quadratic fluctuation spectrum to calculate the Seeley-DeWitt coefficients for finding the logarithmic correction to the entropy of embedded black holes. Throughout, we will consider a generic $U(1)$-charged EMD theory with a negative cosmological constant $\Lambda$, i.e., evolving in AdS$_4$ space of boundary $\ell$ so that $\Lambda = -\dfrac{3}{\ell^2}$. In the end, the account of flat-space limit $\ell \to \infty$ will serve the similar heat kernel results for the simple EMD theory with a vanishing cosmological constant (i.e., $\Lambda=0$). 

The dynamics of a general four-dimensional $U(1)$-charged EMD model is given by the action,\footnote{The strength gravitational interaction is set as $8\pi G_4 = \frac{1}{2}$.}
\begin{align}\label{hk1}
\mathcal{S}[g_{\mu\nu}, A_\mu, \Phi]= \int \mathrm{d}^4x \sqrt{\det g} \left(\mathcal{R}-2\Lambda- 2D_\mu\Phi
	D^\mu\Phi- f(\Phi){F}_{\mu\nu}{F}^{\mu\nu} \right), f(\Phi)=e^{-2\kappa\Phi},
\end{align}
where $\mathcal{R}= g^{\mu\nu}\mathcal{R}\indices{^\alpha_\mu_\alpha_\nu}$ is the Ricci curvature scalar related to the spacetime metric $g_{\mu\nu}$, ${F}_{\mu\nu}={D_{[\mu}A_{\nu]}}$ is the 2-form field strength of the $U(1)$ Maxwell field $A_\mu$ and $\Phi$ is a dilaton field that is minimally coupled to Einstein's gravity but non-minimally coupled to the Maxwell sector. The ``Maxwell-Dilaton'' interaction is controlled by an exponential coupling function $f(\Phi)=e^{-2\kappa\Phi}$ with the dilaton coupling constant $\kappa$. As already mentioned, the EMD families of theories are of special interest in string theory and appear as a universal toy model of the low-energy effective theories of superstring or supergravities. Generally, the dilaton coupling function $f(\Phi)$ is determined by the tree-level string interactions where the string couplings are related via the vacuum expectation value of dilaton, $g_s = e^{\left\langle \Phi\right\rangle}$. The dilaton coupling constant $\kappa$ is fixed via the dimensional reduction or compactification of consistent string theories in 4D. 
It is found that the specific values $\kappa=\frac{1}{\sqrt{3}}$, $\kappa=1$ and $\kappa=\sqrt{3}$ particularly embed the EMD models \eqref{hk1} into various supergravity and  Kaluza-Klein theories \cite{Gibbons:1982ih,Gibbons:1987ps,Garfinkle:1990qj,Horowitz:1991cd,Kallosh:1992ii,Khuri:1995xk,Gao:2004tv,Becker:2006dvp,Cvetic:2014vsa,GoulartSantos:2017dun,Guo:2021zed,Zhang:2021edm,Myung:2020dqt}. Note that the exponential nature of the dilaton coupling function $f(\Phi)$ is so that its expansion can mimic the string loop expansion. In this paper, we will also execute a similar expansion while fluctuating the EMD action \eqref{hk1} around classical backgrounds to find the one-loop quantum correction.



\subsection{Equations of motion and embedding of Einstein-Maxwell backgrounds}\label{embd}
Evolution of the action \eqref{hk1} with respect to the metric, Maxwell field, and dilaton steams out the EMD equations of motion for an arbitrary classical background $(\bar{g}_{\mu\nu}, \bar{A}_\mu, \bar{\Phi})$. The gravitational field equations are obtained as
\begin{subequations}\label{embd1}
\begin{align}
R_{\mu\nu}-\frac{1}{2}\bar{g}_{\mu\nu}R + \Lambda \bar{g}_{\mu\nu} = T_{\mu\nu}^{\mathrm{(dilaton)}} +  T_{\mu\nu}^{U(1)},\label{embd1a}
\end{align}
with the following dilaton and $U(1)$ or Maxwell parts of the total stress-energy tensor
\begin{align}
T_{\mu\nu}^{\mathrm{(dilaton)}} &= 2 \left( D_\mu\bar{\Phi}D_\nu \bar{\Phi} - \frac{1}{2} \bar{g}_{\mu\nu}D_\rho\bar{\Phi}D^\rho \bar{\Phi}\right),\label{embd1b}\\
T_{\mu\nu}^{U(1)} &= e^{-2\kappa \bar{\Phi}}T_{\mu\nu}^{\mathrm{(Maxwell)}} = 2e^{-2\kappa \bar{\Phi}}\left ( \bar{F}_{\mu\rho}\bar{F}\indices{_\nu^\rho}-\frac{1}{4}\bar{g}_{\mu\nu}\bar{F}_{\rho\sigma}\bar{F}^{\rho\sigma}\right),\label{embd1c}
\end{align}
\end{subequations}
where $\bar{F}_{\mu\nu}= {\partial_{[\mu}\bar{A}_{\nu]}}$ is the background Maxwell field strength. The Maxwell and dilaton evolution equations take the forms, 
\begin{align}
	& D_\mu \left(e^{-2\kappa \bar{\Phi}}\bar{F}^{\mu\nu}\right) = 0, \thinspace D_{[\mu}\bar{F}_{\rho\sigma]}=0,\label{embd2}\\
	& D_\mu D^\mu \bar{\Phi} + \frac{1}{2}\kappa  e^{-2\kappa \bar{\Phi}}\bar{F}_{\mu\nu}\bar{F}^{\mu\nu} =0.\label{embd3}
\end{align}
The above equations of motion are invariant under both the electromagnetic duality rotation and dilaton transformation, 
\begin{align*}
	\left\lbrace Q \to P,\thinspace  P \to Q \right\rbrace \enspace \text{and} \enspace f(\bar{\Phi}) \to \dfrac{1}{f(\bar{\Phi})},
\end{align*}
where $Q$ and $P$ are respectively the electric and magnetic charges related to the background Maxwell field $\bar{A}_{\mu}$. Indeed, the EMD background solutions remain invariant for the change $(\kappa, \bar{\Phi}) \to (-\kappa,-\bar{\Phi})$, where any alteration of the signature of $\kappa$ must imply a flip in the signature of $\bar{\Phi}$ and vice-versa. 

The EMD theory \eqref{hk1} is a natural but nontrivial generalization of the standard Einstein-Maxwell (EM) of gravity that is minimally coupled to the Maxwell sector,
\begin{align}\label{embd4}
\mathcal{S}[g_{\mu\nu}, A_\mu]= \int \mathrm{d}^4x \sqrt{\det g} \left(\mathcal{R}-2\Lambda- {F}_{\mu\nu}{F}^{\mu\nu} \right),
\end{align} 
where the background solutions are well-known Kerr-Newman, Kerr, Reissner-Nordstr\"om and Schwarzschild black holes satisfying the following equations of motion  
\begin{align}\label{embd5}
\begin{gathered}
	R_{\mu\nu}- \bar{g}_{\mu\nu}\Lambda = 2 \bar{F}_{\mu\rho}\bar{F}\indices{_\nu^\rho}- \frac{1}{2}\bar{g}_{\mu\nu}\bar{F}_{\rho\sigma}\bar{F}^{\rho\sigma} , \thinspace R = 4\Lambda, \\
	 D_\mu \bar{F}^{\mu\nu} = 0, \thinspace D_{[\mu}\bar{F}_{\rho\sigma]}=0. 
\end{gathered}
\end{align}
Generally, the EM system can not be obtained as a consistent truncation of EMD theories due to the presence of the Maxwell-dilaton coupling term $e^{-2\kappa\Phi}{F}_{\mu\nu}{F}^{\mu\nu}$ in the action \eqref{hk1}. But, it is possible to recover the EM backgrounds from the EMD equations of motion \eqref{embd1}-\eqref{embd3} for two specific cases: 
\begin{inparaenum}[(i)]
	\item $\bar{\Phi}= \mathrm{constant}$ and $\kappa=0$,  
	\item $\bar{\Phi}= \mathrm{constant}$ and $\bar{F}_{\mu\nu}\bar{F}^{\mu\nu}=0$.\footnote{Note that the constant dilaton explicitly can not recover an EM system from the EMD theory but reduces it into the Brans-Dicke-Maxwell system with a Brans-Dicke coupling constant $\omega=-1$ \cite{Cai:1996pj}.}
\end{inparaenum}
The former case is nothing but a natural limit describing an EM theory minimally coupled to the kinetic part of a constant scalar field (not necessarily a dilaton). In contrast, the latter case can embed the EM backgrounds into the EMD theory and is of particular interest to this paper. For $\kappa \neq 0$, it is possible to interpret $\bar{\Phi}=0$ as a solution of EMD equations if $\bar{F}_{\mu\nu}\bar{F}^{\mu\nu}=0$. This is the particular EM embedding we will restrict ourselves, which includes an equal-charged Reissner-Nordstr\"om (RN), but not in general an equal-charged Kerr-Newman black hole. In other words, a non-rotating dyonic black hole with equal electric and magnetic charges always satisfies the EM embedding condition,
\begin{align}\label{embd6}
	\begin{gathered}
		Q=P, \thinspace J=0 \longrightarrow \bar{F}^{\text{(EMD)}}_{\mu\nu}\bar{F}^{\text{(EMD)}\mu\nu}=0,\thinspace \bar{\Phi}^{\text{(EMD)}} =0.
	\end{gathered}  
\end{align}
The above condition is naturally satisfied by the uncharged Schwarzschild and Kerr backgrounds and is also common for both the asymptotically flat and AdS limits of the embedded black holes. The whole fact can be checked by setting $Q=P$ and $(Q,P)=0$  in the relation \eqref{nbh6} for both $J=0$ and $J \neq 0$. Here we want to emphasize an important remark -- a genuine RN solution of the EM theory \eqref{embd4} can contain any combination of charges $(Q,P)$ with an effective value, $Q_e= \sqrt{Q^2 + P^2}$, which is different from the RN black holes embedded in the EMD theory \eqref{hk1} that must support $Q=P$. In summary, the constraint \eqref{embd6} on EMD background effectively decouples the non-minimal Maxwell-Dilaton interaction from the dilaton equation \eqref{embd3} and thus guarantees that the scalar-free or bald Reissner-Nordstr\"om, Kerr and Schwarzschild as solutions of the EMD theory satisfying the EM background equations \eqref{embd5}. This entire setup of EM embedding will essentially boost the prospect of microscopic visibility of the black holes when we compute logarithmic corrections to their entropy by fluctuating the EMD content. In this line, one should not worry about the setting of vanishing dilaton background (i.e., $\bar{\Phi} =0$) since, while quantizing the EMD theory, there will always be the Maxwell-dilaton interactions and related contributions via the perturbative expansion of the coupling function $f(\Phi)$ for any small dilaton fluctuation $\tilde{\Phi}$ around $\bar{\Phi}=0)$,  
\begin{align}\label{embd7}
	f(\bar{\Phi}+ \tilde{\Phi})\big\vert_{\bar{\Phi}=0} \approx f(\bar{\Phi})+ \tilde{\Phi}\frac{\mathrm{d}f(\Phi)}{\mathrm{d}\Phi}\bigg\vert_{\bar{\Phi}=0} + \frac{\tilde{\Phi}^2}{2}\frac{\mathrm{d}^2f(\Phi)}{\mathrm{d}\Phi^2}\bigg\vert_{\bar{\Phi}=0}+ \cdots\cdotp.
\end{align}

\subsection{Quadratic fluctuations and computation of Seeley-DeWitt coefficients}\label{sdc}

To determine quantum entropy corrections for black holes in EMD theory, one needs to analyze the spectrum of quadratic fluctuations of action \eqref{hk1}. We aim to do this via computing the Seeley-DeWitt coefficients up to the first three orders by fluctuating the field content of EMD theory around the EM background, where dilaton is assumed to vanish at the background. In particular, we consider the following fluctuations
\begin{align}\label{sdc1}
	\begin{gathered}
			g_{\mu\nu} = \bar{g}_{\mu\nu} + \sqrt{2}h_{\mu\nu}, \enspace A_\mu = \bar{A}_\mu + \frac{1}{2}a_\mu,\enspace \Phi = \cancelto{0}{\bar{\Phi}} + \tilde{\Phi},\\
		F_{\mu\nu} = \bar{F}_{\mu\nu}+\frac{1}{2}f_{\mu\nu}, \enspace {f}_{\mu\nu}={\partial_{[\mu}a_{\nu]}}={D_{[\mu}a_{\nu]}},
	\end{gathered}
\end{align} 
where $h_{\mu\nu}$ and $a_\mu$ are the thermal fluctuations of metric (i.e., graviton) and Maxwell field (i.e., graviphoton) with the field strength ${f}_{\mu\nu}$. Inside the present setup, dilaton actually behaves as its own fluctuation while sharing the common EM background $(\bar{g}_{\mu\nu}, \bar{A}_\mu)$ with graviton and graviphoton. This will, in turn, induce several non-minimal couplings between the dilaton, graviton and graviphoton via the background field strength $\bar{F}_{\mu\nu}$. The normalization factors of the graviton and graviphoton in \eqref{sdc1} are set according to the convention of \cite{Bhattacharyya:2012ss}, which will be essential in the following heat kernel treatment. With the fluctuations \eqref{sdc1} and \eqref{embd7}, we obtain the quadratic-order variation of the action \eqref{hk1} up to a total derivative and express the minimal blocks of ``cosmological Einstein gravity'' and ``Maxwell-dilaton'' as
\begin{subequations}
\begin{align}
		\delta^2 \left(\sqrt{\det g}\,(\mathcal{R}-2\Lambda)\right) &= \frac{1}{2}\sqrt{\det \bar{g}} \bigg[h_{\mu\nu}D_\rho D^\rho h^{\mu\nu}- h\indices{^\mu_\mu}D_\rho D^\rho h\indices{^\nu_\nu}-2 h^{\nu\rho}D_\mu D_\nu h\indices{^\mu_\rho} \nonumber\\
		&\qquad  +2h^{\mu\nu}D_\mu D_\nu h\indices{^\alpha_\alpha}+2 {R}_{\mu\nu}\big(2h^{\mu\rho}h\indices{^\nu_\rho}-h\indices{^\alpha_\alpha} h^{\mu\nu}\big)\nonumber\\
		&\qquad  -\left(R-2\Lambda\right)h_{\mu\nu}h^{\mu\nu}  + \bigg(\frac{R}{2}-\Lambda\bigg)(h\indices{^\alpha_\alpha})^2\bigg],\label{sdc2}
\end{align}
\begin{align}
&-\delta^2 \left(\sqrt{\det g}\,(2D_\mu\Phi
D^\mu\Phi+ f(\Phi){F}_{\mu\nu}{F}^{\mu\nu})\right) = \frac{1}{2}\sqrt{\det \bar{g}}  \bigg[4\Phi D_\rho D^\rho \Phi -\frac{1}{2}f_{\mu\nu}f^{\mu\nu} \nonumber \\
&\hspace{1.9in} - 4 \bar{F}_{\mu\nu}\bar{F}_{\alpha\beta}h^{\mu\alpha}h^{\nu\beta} - 8 \bar{F}_{\mu\nu}\bar{F}^{\mu\alpha}h^{\nu\beta}h_{\alpha\beta}
 + 4\bar{F}_{\mu\nu}\bar{F}\indices{_\alpha^\nu}h\indices{^\rho_\rho}h^{\mu\alpha}    \nonumber \\ 
&\hspace{1.9in} + \bar{F}_{\mu\nu}\bar{F}^{\mu\nu}\Big(h_{\alpha\beta}h^{\alpha\beta}-\frac{1}{2}(h\indices{^\rho_\rho})^2\Big) - 4 \kappa^2 \bar{F}_{\mu\nu}\bar{F}^{\mu\nu} \Phi^2  \nonumber \\ 
&\hspace{1.9in} -8\sqrt{2}\kappa \bar{F}_{\mu\nu}\bar{F}\indices{_\alpha^\nu}\Phi h^{\mu\alpha} + 2\sqrt{2}\kappa \bar{F}_{\mu\nu}\bar{F}^{\mu\nu} \Phi h\indices{^\alpha_\alpha}    \nonumber\\
&\hspace{1.9in}  + 4 \kappa \bar{F}_{\mu\nu}\Phi f^{\mu\nu} + 4\sqrt{2}\bar{F}_{\mu\nu}h^{\mu\alpha}f\indices{_\alpha^\nu} - \sqrt{2}\bar{F}_{\mu\nu}h\indices{^\rho_\rho} f^{\mu\nu} \bigg].\label{sdc3}
\end{align}
\end{subequations}
Before proceeding further, we find it necessary to gauge-fix the theory. One can execute the gauge-fixing exactly the same way as in the asymptotically-flat case (i.e., $\Lambda=0$) \cite{Bhattacharyya:2012ss} since incorporating a cosmological term into the action never affects the gauge invariance of a concerned theory. The most convenient choice is to set the harmonic gauge $D_\mu h^{\mu\rho}-\frac{1}{2}D^\rho h\indices{^\alpha_\alpha}=0$ and Lorentz gauge $D_\mu a^\mu=0$ via adding the gauge-fixing term,   
\begin{align}\label{sdc4}
\mathcal{S}_{\text{gauge}}= -\int \mathrm{d}^4x \sqrt{\det \bar{g}}\left[\left(D_\mu h^{\mu\rho}-\frac{1}{2}D^\rho h\indices{^\alpha_\alpha}\right)\left(D^\nu h_{\nu\rho}-\frac{1}{2}D_\rho h\indices{^\beta_\beta}\right)  + \frac{1}{2}(D_\mu a^\mu)^2\right],
\end{align} 
followed by the compensating ghost term \cite{Banerjee:2011oo},
\begin{align}\label{sdc5}
	\begin{split}
		\mathcal{S}_{\text{ghost}}[b_\mu,c_\mu,b,c] &=\int \mathrm{d}^4x \sqrt{\det \bar{g}}\big[ b_\mu D_\rho D^\rho c^\mu +b D_\rho D^\rho c +  b_\mu R^{\mu\nu} c_\nu -2 b \bar{F}^{\rho\nu} D_\rho c_\nu\big],
	\end{split}
\end{align}
where the vector fields $(b_\mu, c_\mu)$ are diffeomorphism ghosts related to the graviton and the scalar fields $(b, c)$ are ghosts induced due to the gauge invariance of graviphoton. These ghost fields are all minimally-coupled and never interact with any of the graviton, graviphoton, and dilaton fluctuations. So we have the freedom to evaluate their contribution separately (see \cref{sdc28}). Then, without accounting for the ghosts, the  gauge fixed quadratic fluctuated EMD action is obtained as
{\allowdisplaybreaks
\begin{align}\label{sdc6}
	{\delta^2\mathcal{S}}[h_{\mu\nu}, a_\mu, \Phi]&= \frac{1}{2}\int \mathrm{d}^4x \sqrt{\det \bar{g}}\bigg[h_{\mu\nu}D_\rho D^\rho h^{\mu\nu}- \frac{1}{2}h\indices{^\mu_\mu}D_\rho D^\rho h\indices{^\nu_\nu}+ a_\mu D_\rho D^\rho a^\mu \nonumber\\
	&\qquad\quad + 4\Phi D_\rho D^\rho \Phi + 2R_{\mu\alpha\nu\beta}h^{\mu\nu}h^{\alpha\beta} -2R_{\mu\nu}h^{\mu\rho}h\indices{^\nu_\rho}   \nonumber \\
	&\qquad\quad  - 4 \bar{F}_{\mu\nu}\bar{F}_{\alpha\beta}h^{\mu\alpha}h^{\nu\beta} +2\Lambda \Big(h_{\mu\nu}h^{\mu\nu}-\frac{1}{2}(h\indices{^\mu_\mu})^2\Big) -a_\mu R^{\mu\nu} a_\nu  \nonumber \\
	&\qquad\quad - 4\sqrt{2}\kappa R_{\mu\alpha}\Phi h^{\mu\alpha} + 4\sqrt{2}\kappa \Lambda \Phi h\indices{^\mu_\mu} + 8\kappa \Phi \bar{F}^{\rho\mu}D_\rho a_\mu \nonumber\\
	&\qquad\quad +2\sqrt{2}h_{\mu\nu}\big(2\bar{g}^{\nu\rho} \bar{F}^{\mu\alpha}- 2\bar{g}^{\alpha\nu} \bar{F}^{\mu\rho}-\bar{g}^{\mu\nu}\bar{F}^{\rho\alpha}\big) (D_\rho a_\alpha) \bigg],
\end{align}}
where the EM embedding condition \eqref{embd6} and background equations of motion \eqref{embd5} are utilized to achieve the above intricate form. Now, we are in a position to carry out the heat kernel method depicted in \cref{local}. But before that, we must resolve a few issues to extract the kinetic operator $\mathcal{H}$ out of the quadratic action \eqref{sdc6} in the desired Laplace-type form \eqref{comp3}. The kinetic terms of each fluctuation in \eqref{sdc6} must be adjusted into the same state of normalization. But notice that the kinetic part of graviton consists of two terms with a total of ten degrees of freedom. In particular, the appearance of an additional kinetic term associated with the graviton trace, i.e., $h\indices{^\mu_\mu}D_\rho D^\rho h\indices{^\nu_\nu}$ is problematic. As a resolution, there exists a convenient method of transforming the explicit graviton kinetic part $h_{\mu\nu}D_\rho D^\rho h^{\mu\nu}$ into the appropriate irreducible representation of $SL(2, \mathbb{C})$ \cite{Christensen:1979iy,Christensen:1978md,Gibbons:1978ji}. In this progress, we need to split the graviton $h_{\mu\nu}$ into its trace-free (as well as symmetric) component $\hat{h}_{\mu\nu}$ and trace component $h$,
\begin{align}\label{sdc7}
	\begin{split}
		\hat{h}_{\mu\nu} &= h_{\mu\nu} - \frac{1}{4}\bar{g}_{\mu\nu}h, \\
		h &= h\indices{^\mu_\mu} = \bar{g}^{\mu\nu}h_{\mu\nu},
	\end{split}
\end{align}    
and then express the fluctuated action as
{
	\allowdisplaybreaks
\begin{align}\label{sdc8}
	{\delta^2\mathcal{S}}[h_{\mu\nu}, a_\mu, \Phi]&= \frac{1}{2}\int \mathrm{d}^4x \sqrt{\det \bar{g}}\bigg[I^{\hat{h}_{\mu\nu}\hat{h}_{\alpha\beta}}	\hat{h}_{\mu\nu}D_\rho D^\rho \hat{h}_{\alpha\beta}- \frac{1}{4} h D_\rho D^\rho h  \nonumber\\
	&\qquad + a_\mu D_\rho D^\rho a^\mu + \Phi D_\rho D^\rho \Phi + 2R^{\mu\alpha\nu\beta}\hat{h}_{\mu\nu}\hat{h}_{\alpha\beta} -2R^{\mu\nu}\hat{h}_{\mu\rho}\hat{h}\indices{_\nu^\rho}   \nonumber \\
	&\qquad  - 4 \bar{F}^{\mu\nu}\bar{F}^{\alpha\beta}\hat{h}_{\mu\alpha}\hat{h}_{\nu\beta} - 2 \bar{F}^{\mu\alpha}\bar{F}\indices{^\nu_\alpha}\hat{h}_{\mu\nu}h +2\Lambda \Big(\hat{h}_{\mu\nu}\hat{h}^{\mu\nu}-\frac{1}{4}h^2\Big)   \nonumber \\
	&\qquad -a_\mu R^{\mu\nu} a_\nu - 2\sqrt{2}\kappa R^{\mu\nu}\Phi \hat{h}_{\mu\nu}  + 4\kappa \Phi \bar{F}^{\rho\mu}D_\rho a_\mu \nonumber\\
	&\qquad + 4\sqrt{2}\hat{h}_{\mu\nu}\big(\bar{g}^{\nu\rho} \bar{F}^{\mu\alpha}- \bar{g}^{\alpha\nu} \bar{F}^{\mu\rho}\big) (D_\rho a_\alpha) \bigg],
\end{align}}
where the dilaton has been rescaled as $\Phi \to \dfrac{1}{2}\Phi$. Notice that the kinetic term of the trace-free graviton is now modulated via a newly introduced operator $I^{\hat{h}_{\mu\nu}\hat{h}_{\alpha\beta}}$. This operator is a projection operator onto the trace-free graviton, which should contain $10-1 =9$ independent off-shell degrees of freedom and can be structured via the following combination of background metrices, 
\begin{align}\label{sdc9}
I^{\hat{h}_{\mu\nu}\hat{h}_{\alpha\beta}} = \frac{1}{2}\left(\bar{g}^{\mu\alpha} \bar{g}^{\nu\beta} + \bar{g}^{\mu\beta} \bar{g}^{\nu\alpha}-\frac{1}{2}\bar{g}^{\mu\nu}\bar{g}^{\alpha\beta}\right).
\end{align}
The covariant partner $I_{\hat{h}_{\mu\nu}\hat{h}_{\alpha\beta}}$ also follows exactly a similar form. The projection operator form \eqref{sdc9} actually acts as an effective background metric (called DeWitt metric) to contract the indices of arbitrary matrices $\mathbb{M}$ acting on the trace-free graviton fluctuations,
\begin{align}\label{sdc10}
	\phi_m \mathbb{M}^m_n\phi^n = \hat{h}_{\mu\nu} {\mathbb{M}}^{\hat{h}_{\mu\nu}\hat{h}_{\alpha\beta}}\hat{h}_{\alpha\beta},
\end{align}
 and define their (as well as of its own) traces as
 \begin{align}\label{sdc11}
 	\begin{split}
 			\text{Tr}(\mathbb{M}) &= \mathbb{M}\indices{^{\hat{h}_{\mu\nu}}_{\hat{h}_{\mu\nu}}} = {I}_{\hat{h}_{\mu\nu}\hat{h}_{\alpha\beta}}{\mathbb{M}}^{\hat{h}_{\mu\nu}\hat{h}_{\alpha\beta}},\\
 			\text{Tr}(\mathbb{M}^2) &= \mathbb{M}\indices{^{\hat{h}_{\mu\nu}}_{\hat{h}_{\alpha\beta}}}\mathbb{M}\indices{^{\hat{h}_{\alpha\beta}}_{\hat{h}_{\mu\nu}}} = {I}_{\hat{h}_{\alpha\beta}\hat{h}_{\rho\sigma}}{I}_{\hat{h}_{\mu\nu}\hat{h}_{\gamma\delta}}{\mathbb{M}}^{\hat{h}_{\mu\nu}\hat{h}_{\rho\sigma}}{\mathbb{M}}^{\hat{h}_{\alpha\beta}\hat{h}_{\gamma\delta}}.
 	\end{split}
 \end{align}
The quadratic action form \eqref{sdc8} still suffers the ``conformal factor problem'' due to the negative signature of graviton trace kinetic term, making the Euclidean one-loop path integral \eqref{set4} unbounded, divergent and ill-defined. However, this problem can be resolved by performing the standard treatment \cite{Gibbons:1978ac,Mazur:1989by} of conformal rotation along the imaginary axis, i.e., $h \to i\hat{h} $ with a new real graviton trace $\hat{h}$. In addition, we should rescale the redefined $\hat{h}$ so that the normalization state of its kinetic term matches that of $\hat{h}_{\mu\nu}$, $a_\mu$ and $\Phi$. Altogether, we set
\begin{align}
h = 2i\hat{h},
\end{align}
 and express the quadratic action of EMD theory as
\begin{align}\label{sdc13}
	\delta^2 \mathcal{S}[\hat{h}_{\mu\nu}, \hat{h}, a_\mu, \Phi]= \frac{1}{2}\int \mathrm{d}^4x \sqrt{\det \bar{g}}\, \phi_m \mathcal{H}^m_n\phi^n,
\end{align}
where the kinetic operator $\mathcal{H}$ operating on the effective fluctuations $\phi_m = \big\lbrace \hat{h}_{\mu\nu}, \hat{h}, a_\mu, \Phi \big\rbrace$ is obtained in the following Laplace-type form
{
\allowdisplaybreaks
\begin{align}\label{sdc14}
&\phi_m \mathcal{H}^m_n\phi^n = \hat{h}_{\mu\nu}\bigg\lbrace I^{\hat{h}_{\mu\nu}\hat{h}_{\alpha\beta}}D_\rho D^\rho +R^{\mu\alpha\nu\beta}+R^{\mu\beta\nu\alpha}-2\left(\bar{F}^{\mu\alpha}\bar{F}^{\nu\beta}+\bar{F}^{\mu\beta}\bar{F}^{\nu\alpha}\right) \nonumber \\
 &\qquad\qquad -\frac{1}{2}\left(\bar{g}^{\mu\alpha}R^{\nu\beta}+\bar{g}^{\nu\alpha}R^{\mu\beta}+ \bar{g}^{\mu\beta}R^{\nu\alpha}+\bar{g}^{\nu\beta}R^{\mu\alpha}\right) + \Lambda\left(\bar{g}^{\mu\alpha} \bar{g}^{\nu\beta} + \bar{g}^{\mu\beta} \bar{g}^{\nu\alpha}\right)\bigg\rbrace \hat{h}_{\alpha\beta} \nonumber \\
 & \qquad\qquad +\hat{h} \big(D_\rho D^\rho + 2\Lambda\big) \hat{h} + a_\mu \big(\bar{g}^{\mu\nu} D_\rho D^\rho - R^{\mu\nu}\big) a_\nu + \Phi D_\rho D^\rho \Phi - i \hat{h}_{\mu\nu} R^{\mu\nu} \hat{h}     \nonumber \\
 &\qquad\qquad  - i \hat{h} R^{\mu\nu} \hat{h}_{\mu\nu} - \sqrt{2}\kappa \hat{h}_{\mu\nu} R^{\mu\nu} \Phi - \sqrt{2}\kappa \Phi R^{\mu\nu}\hat{h}_{\mu\nu}+ 2\kappa a_\mu \bar{F}^{\mu\rho} D_\rho\Phi - 2\kappa  \Phi \bar{F}^{\mu\rho} D_\rho a_\mu \nonumber \\
 &\qquad\qquad  +  \frac{\sqrt{2}}{2}\hat{h}_{\mu\nu}\bigg\lbrace \left( D^\mu \bar{F}^{\alpha\nu} + D^\nu \bar{F}^{\alpha\mu} \right)+ 2 \left(\bar{g}^{\mu\rho}\bar{F}^{\nu\alpha}+\bar{g}^{\nu\rho}\bar{F}^{\mu\alpha}-\bar{g}^{\mu\alpha}\bar{F}^{\nu\rho}-\bar{g}^{\nu\alpha}\bar{F}^{\mu\rho}\right) D_\rho\bigg\rbrace a_\alpha \nonumber\\
 &\qquad\qquad +  \frac{\sqrt{2}}{2}a_\alpha\bigg\lbrace \left( D^\mu \bar{F}^{\alpha\nu} + D^\nu \bar{F}^{\alpha\mu} \right)- 2 \left(\bar{g}^{\mu\rho}\bar{F}^{\nu\alpha}+\bar{g}^{\nu\rho}\bar{F}^{\mu\alpha}-\bar{g}^{\mu\alpha}\bar{F}^{\nu\rho}-\bar{g}^{\nu\alpha}\bar{F}^{\mu\rho}\right) D_\rho\bigg\rbrace \hat{h}_{\mu\nu},
\end{align}
}
where each component is made symmetric or anti-symmetric with respect to the corresponding fluctuations and also associated with their Hermitian pair by adjusting up to a total derivative.\footnote{For example, one can use the following schematic treatment for all commuting fluctuations $\phi_m$
	\begin{align}
		\phi_m \mathbb{K}^\rho (D_\rho\phi_n) = \frac{1}{2} \phi_m \Big(\mathbb{K}^\rho D_\rho - \frac{1}{2} (D_\rho \mathbb{K}^\rho) \Big) \phi_n - \frac{1}{2} \phi_n \Big(\mathbb{K}^\rho D_\rho + \frac{1}{2} (D_\rho \mathbb{K}^\rho) \Big) \phi_m.
\end{align}}   
The above operator form may look complicated but is now structured exactly as the schematic \eqref{comp3} from where one can explicitly read off the necessary matrices $I$, $P$ and $N_\rho$. Components of the identity operator or effective metric $I$ that controls the kinetic part of fluctuations and their projections are expressed as
\begin{align}\label{sdc15}
	\phi_m I^{mn}\phi_n &= \hat{h}_{\mu\nu} I^{\hat{h}_{\mu\nu}\hat{h}_{\alpha\beta}} \hat{h}_{\alpha\beta} + \hat{h}\hat{h} + a_\mu \bar{g}^{\mu\nu} a_\nu + \Phi\Phi.
\end{align}
Taking a trace over the above components will count how many off-shell degrees of freedom are effectively acting in the quadratic fluctuation loop of the theory. On the other hand, the matrices $P$ and $N^\rho$ that combinedly control all the minimal and non-minimal interactions between fluctuations in the kinetic operator \eqref{sdc14} have the following components 
{
\allowdisplaybreaks
\begin{align}
\phi_m (N^\rho)^{mn}\phi_n &= \sqrt{2} \hat{h}_{\mu\nu}\bigg(\bar{g}^{\mu\rho}\bar{F}^{\nu\alpha}+\bar{g}^{\nu\rho}\bar{F}^{\mu\alpha}-\bar{g}^{\mu\alpha}\bar{F}^{\nu\rho}-\bar{g}^{\nu\alpha}\bar{F}^{\mu\rho}\bigg) a_\alpha \nonumber \\
& \qquad - \sqrt{2} a_\alpha\bigg(\bar{g}^{\mu\rho}\bar{F}^{\nu\alpha}+\bar{g}^{\nu\rho}\bar{F}^{\mu\alpha}-\bar{g}^{\mu\alpha}\bar{F}^{\nu\rho}-\bar{g}^{\nu\alpha}\bar{F}^{\mu\rho}\bigg) \hat{h}_{\mu\nu} \nonumber \\
&\qquad + a_\mu \Big(2\kappa \bar{F}^{\mu\rho} \Big)\Phi +  \Phi\Big(- 2\kappa \bar{F}^{\mu\rho}\Big) a_\mu, \label{sdc16}\\[5pt]
\phi_m P^{mn}\phi_n &= \hat{h}_{\mu\nu}\bigg( R^{\mu\alpha\nu\beta}+R^{\mu\beta\nu\alpha}-2\left(\bar{F}^{\mu\alpha}\bar{F}^{\nu\beta}+\bar{F}^{\mu\beta}\bar{F}^{\nu\alpha}\right)\nonumber\\
&\qquad\quad -\frac{1}{2}\left(\bar{g}^{\mu\alpha}R^{\nu\beta}+\bar{g}^{\nu\alpha}R^{\mu\beta}+ \bar{g}^{\mu\beta}R^{\nu\alpha}+\bar{g}^{\nu\beta}R^{\mu\alpha}\right) \nonumber\\
& \qquad\quad + \Lambda\left(\bar{g}^{\mu\alpha} \bar{g}^{\nu\beta} + \bar{g}^{\mu\beta} \bar{g}^{\nu\alpha}\right)\bigg) \hat{h}_{\alpha\beta} + 2\hat{h}\Lambda\hat{h} - a_\mu R^{\mu\nu}a_\nu \nonumber \\
& \quad  - i \hat{h}_{\mu\nu} R^{\mu\nu} \hat{h} - i \hat{h} R^{\mu\nu} \hat{h}_{\mu\nu} + \hat{h}_{\mu\nu}\Big(-\sqrt{2}\kappa R^{\mu\nu}\Big) \Phi + \Phi\Big(- \sqrt{2}\kappa R^{\mu\nu}\Big)\hat{h}_{\mu\nu}  \nonumber\\
& \quad + \frac{\sqrt{2}}{2}\hat{h}_{\mu\nu}\bigg( D^\mu \bar{F}^{\alpha\nu} + D^\nu \bar{F}^{\alpha\mu} \bigg)a_\alpha + \frac{\sqrt{2}}{2}a_\alpha\bigg( D^\mu \bar{F}^{\alpha\nu} + D^\nu \bar{F}^{\alpha\mu} \bigg)\hat{h}_{\mu\nu}.	\label{sdc17}
\end{align}
}
The above relations encode all information one needs to know about the quadratic fluctuation of EMD theory, which can be utilized further to write the operator form \eqref{comp4} in terms of the more generic matrices describing the gauge connection $\omega_\rho$, effective matrix-valued potential $E$ and curvature commutator $\Omega_{\rho\sigma}$. In this progress, one first needs to derive the components of the contraction $\omega^\rho\omega_\rho$,
\begin{align}\label{sdc18}
\phi_m (\omega^\rho\omega_\rho)^{mn}\phi_n &= \hat{h}_{\mu\nu}\bigg( \bar{g}^{\mu\alpha}\bar{F}^{\rho\nu}\bar{F}\indices{^\beta_\rho}+ \bar{g}^{\mu\beta}\bar{F}^{\rho\nu}\bar{F}\indices{^\alpha_\rho}+ \bar{g}^{\nu\alpha}\bar{F}^{\rho\mu}\bar{F}\indices{^\beta_\rho}+ \bar{g}^{\nu\beta}\bar{F}^{\rho\mu}\bar{F}\indices{^\alpha_\rho} \nonumber \\
&\quad  -2\left(\bar{F}^{\mu\alpha}\bar{F}^{\nu\beta}+\bar{F}^{\mu\beta}\bar{F}^{\nu\alpha}\right)\bigg)\hat{h}_{\alpha\beta} + a_\mu\bigg(-\left(2+ \kappa^2\right)\bar{F}^{\mu\rho}\bar{F}\indices{^\nu_\rho}\bigg)a_\nu \nonumber \\
& \quad +\hat{h}_{\mu\nu}\bigg(-2\sqrt{2}\kappa\bar{F}^{\mu\rho}\bar{F}\indices{^\nu_\alpha}\bigg)\Phi + \Phi\bigg(-2\sqrt{2}\kappa\bar{F}^{\mu\rho}\bar{F}\indices{^\nu_\alpha}\bigg)\hat{h}_{\mu\nu},
\end{align}
followed by components of the commutation between two connections $\omega_\rho$ and $\omega_\sigma$ as
{\allowdisplaybreaks
\begin{align}\label{sdc19}
\phi_m [\omega_\rho,\omega_\sigma]^{mn}\phi_n &=\phi_m(\omega_\rho)^{mp}(\omega_\sigma)\indices{_p^n}\phi_n - \big(\rho \leftrightarrow \sigma\big), \nonumber \\ 
&=\frac{1}{2}\hat{h}_{\mu\nu}\bigg(\left(\bar{g}^\mu_\rho \bar{F}^{\nu\theta}+ \bar{g}^\nu_\rho \bar{F}^{\mu\theta}\right)\left(\bar{g}^\alpha_\theta\bar{F}\indices{^\beta_\sigma}+ \bar{g}^\beta_\theta\bar{F}\indices{^\alpha_\sigma}\right) \nonumber \\
&\quad -\left(\bar{g}^\mu_\rho \bar{F}^{\nu\theta}+ \bar{g}^\nu_\rho \bar{F}^{\mu\theta}\right)\left(\bar{g}^\alpha_\sigma\bar{F}\indices{^\beta_\theta}+ \bar{g}^\beta_\sigma\bar{F}\indices{^\alpha_\theta}\right) \nonumber\\
&\quad + \left(\bar{g}^{\mu\theta}\bar{F}\indices{^\nu_\rho}+ \bar{g}^{\nu\theta}\bar{F}\indices{^\mu_\rho}\right)\left(\bar{g}^\alpha_\sigma \bar{F}\indices{^\beta_\theta}+ \bar{g}^\beta_\sigma\bar{F}\indices{^\alpha_\theta}\right) \nonumber \\
&\quad - \left(\bar{g}^{\mu\theta}\bar{F}\indices{^\nu_\rho}+ \bar{g}^{\nu\theta}\bar{F}\indices{^\mu_\rho}\right)\left(\bar{g}^\alpha_\theta \bar{F}\indices{^\beta_\sigma}+ \bar{g}^\beta_\theta\bar{F}\indices{^\alpha_\sigma}\right)\bigg)\hat{h}_{\alpha\beta} \nonumber \\
&\quad + a_\alpha\bigg({\bar{g}_\rho^\beta}\bar{F}^{\nu\alpha}\bar{F}_{\nu\sigma}- \bar{g}_{\rho\sigma}\bar{F}^{\nu\alpha}\bar{F}\indices{_\nu^\beta}+ \bar{g}^\alpha_\sigma\bar{F}\indices{^\nu_\rho}\bar{F}\indices{_\nu^\beta}\nonumber \\
&\quad -\bar{g}^{\alpha\beta}\bar{F}\indices{^\nu_\rho}\bar{F}_{\nu\sigma}-2 \bar{F}\indices{_\sigma^\alpha}\bar{F}\indices{_\rho^\beta}-2\bar{F}\indices{^\alpha^\beta}\bar{F}_{\rho\sigma}+ (2-\kappa^2)\bar{F}\indices{_\rho^\alpha}\bar{F}\indices{_\sigma^\beta}\bigg)a_\beta \nonumber \\
&\quad + \frac{\sqrt{2}}{2}\kappa\hat{h}_{\mu\nu}\bigg(\bar{F}\indices{^\mu_\rho}\bar{F}\indices{_\sigma^\nu}+ \bar{F}\indices{^\nu_\rho}\bar{F}\indices{_\sigma^\mu}- \bar{g}^\mu_\rho \bar{F}^{\nu\alpha}\bar{F}_{\sigma_\alpha}- \bar{g}^\nu_\rho \bar{F}^{\mu\alpha}\bar{F}_{\sigma_\alpha}\bigg) \Phi \nonumber \\
&\quad + \frac{\sqrt{2}}{2}\kappa\Phi\bigg(\bar{F}\indices{^\mu_\rho}\bar{F}\indices{_\sigma^\nu}+ \bar{F}\indices{^\nu_\rho}\bar{F}\indices{_\sigma^\mu}- \bar{g}^\mu_\rho \bar{F}^{\nu\alpha}\bar{F}_{\sigma_\alpha}- \bar{g}^\nu_\rho \bar{F}^{\mu\alpha}\bar{F}_{\sigma_\alpha}\bigg) \hat{h}_{\mu\nu} \nonumber \\
&\quad + \Phi\Big(-\kappa^2 \bar{F}\indices{_\rho^\alpha}\bar{F}_{\sigma_\alpha}\Big)\Phi - \big(\rho \leftrightarrow \sigma\big).
\end{align}
} 
The above equalities rejected all the terms involving $\bar{F}_{\mu\nu}\bar{F}^{\mu\nu}$ for satisfying the embedding condition \eqref{embd6}. Also, during these derivations, one must remember to make use of the appropriate projection operators \eqref{sdc15} while raising or lowering indices for the contraction operations. Note that the $(\hat{h}_{\mu\nu}\Phi)$ and $(\Phi\hat{h}_{\mu\nu})$ components originated since the trace-free graviton and dilaton fluctuation create a virtual connection through the contraction of connections $(\omega_\rho)^{\hat{h}_{\mu\nu}a_\alpha}$ and $(\omega_\sigma)^{a_\alpha\Phi}$. Next, we require the covariant derivative commutation $[D_\rho,D_\sigma]$ operating on each fluctuation. Recognizing that the covariant derivative $D_\rho$ commutes while operating on scalars (e.g., the graviton trace $\hat{h}$ and dilaton $\Phi$) but not when acting on any other vector, tensor, or spinor fields, we obtain
\begin{align}\label{sdc20}
\phi_m [D_\rho,D_\sigma] \phi^m = \frac{1}{2} \hat{h}_{\mu\nu}\bigg(\bar{g}^{\mu\alpha} R\indices{^{\nu\beta}_{\rho\sigma}} +\bar{g}^{\mu\beta}R\indices{^{\nu\alpha}_{\rho\sigma}}+\bar{g}^{\nu\alpha}R\indices{^{\mu\beta}_{\rho\sigma}}+\bar{g}^{\nu\beta}R\indices{^{\mu\alpha}_{\rho\sigma}}\bigg)\hat{h}_{\alpha\beta} + a_\alpha R\indices{^\alpha^\beta_\rho_\sigma} a_\beta.
\end{align}   
Finally, it is necessary to achieve components of $(D_\rho\omega^\rho)$ and $ {D_{[\rho}\omega_{\sigma]}}$ by operating the covariant derivative on the field connections $\omega_\rho$, yielding
{\allowdisplaybreaks 
\begin{align}
	\phi_m(D_\rho\omega^\rho)^{mn}\phi_n &= \frac{\sqrt{2}}{2}\hat{h}_{\mu\nu}\bigg(D^\mu \bar{F}^{\nu\alpha}+ D^\nu \bar{F}^{\mu\alpha}\bigg)a_\alpha + \frac{\sqrt{2}}{2}a_\alpha\bigg(D^\mu \bar{F}^{\alpha\nu}+ D^\nu \bar{F}^{\alpha\mu}\bigg)\hat{h}_{\mu\nu}, \label{sdc21} \\
	\phi_m{D_{[\rho}\omega_{\sigma]}}^{mn}\phi_n &= \phi_m(D_\rho \omega_\sigma)^{mn}\phi_n - \big(\rho \leftrightarrow \sigma\big),    \nonumber\\
	&= \frac{\sqrt{2}}{2} \hat{h}_{\mu\nu}\bigg(\bar{g}^\mu_\sigma D_\rho \bar{F}^{\nu\alpha}+ \bar{g}^\nu_\sigma D_\rho \bar{F}^{\mu\alpha} - \bar{g}^{\mu\alpha}D_\rho \bar{F}\indices{^\nu_\sigma}- \bar{g}^{\nu\alpha}D_\rho \bar{F}\indices{^\mu_\sigma}\bigg) a_\alpha \nonumber  \\
	&\quad + \frac{\sqrt{2}}{2} a_\alpha\bigg(\bar{g}^\mu_\sigma D_\rho \bar{F}^{\alpha\nu}+ \bar{g}^\nu_\sigma D_\rho \bar{F}^{\alpha\mu} - \bar{g}^{\mu\alpha}D_\rho \bar{F}\indices{_\sigma^\nu}- \bar{g}^{\nu\alpha}D_\rho \bar{F}\indices{_\sigma^\mu}\bigg) \hat{h}_{\mu\nu} \nonumber \\
	& \quad + a_\alpha \Big(\kappa D_\rho \bar{F}\indices{^\alpha_\sigma}\Big)\Phi +  \Phi\Big(\kappa D_\rho \bar{F}\indices{_\sigma^\alpha}\Big) a_\alpha - \big(\rho \leftrightarrow \sigma\big), \label{sdc22}
\end{align}
}
where the covariant derivatives are only acting on the background field strengths, not over the related fluctuations. Technically, $(a_\mu \Phi)$ and $(\Phi a_\mu)$ components are also possible in \eqref{sdc21}, but they eventually diapered due to the constraint of Maxwell equations. At this stage, one needs to substitute all the matrix-valued background data of \cref{sdc18,sdc19,sdc20,sdc21,sdc22} into the formulas \eqref{comp4c} and \eqref{comp4d}, and extract the most simplified forms of all $E$ and $\Omega_{\rho\sigma}$ components. Due to our specific choice of background, one may initially end up with a vanishing $E^{a_\alpha \hat{h}_{\mu\nu}}$ component. But in that case, we are always allowed to adjust the Hermitian pair $E^{\hat{h}_{\mu\nu} a_\alpha} $ until both the components of the commuting fluctuations share a non-vanishing result. Finally, with the help of all $I$, $E$, $\Omega_{\rho\sigma}$ matrix components, we calculate the traces required in the Seeley-DeWitt formulas \eqref{comp5} to \eqref{comp9b}. These traces are defined and executed via a similar treatment as mentioned in \eqref{sdc11}, where the use of an appropriate projection or identity operator is found to be extremely crucial. We have used all classical equations of motion \cref{embd5} with the embedding condition \cref{embd6} so that the final trace results are simplified only in terms of background invariants. 
The trace calculations are tedious but manageable via a systematic approach. Thus, without providing any intermediate technical details,\footnote{The readers are referred  to \cref{calcul} for a more general but complex heat kernel trace calculations in the $U(1)^2$-charged EMD theory embedded into $\mathcal{N}=4$ supergravity.} we shall quote only the final results,
{\allowdisplaybreaks
\begin{align}\label{sdc23}
		\text{Tr}(I) &= 9 + 1 + 4 + 1 = 15, \nonumber\\
		\text{Tr}(E) &= -8\Lambda, \nonumber\\
		\text{Tr}(E^2) &= 3 R_{\mu\nu\rho\sigma}R^{\mu\nu\rho\sigma}+ \left(\frac{\kappa^4}{4}-7\right)R_{\mu\nu}R^{\mu\nu} \nonumber\\
		&\quad\enspace  - \left(\kappa^4 - 32\right)\Lambda^2 + 3R_{\mu\nu\rho\sigma}\bar{F}^{\mu\nu}\bar{F}^{\rho\sigma}, \nonumber\\[5pt]
		\text{Tr}\left(\Omega_{\rho\sigma}\Omega^{\rho\sigma}\right) &= -7 R_{\mu\nu\rho\sigma}R^{\mu\nu\rho\sigma} + \left(\frac{\kappa^4}{2} - 12 \kappa^2 + 56\right)R_{\mu\nu}R^{\mu\nu} \nonumber\\
		& \quad\enspace  - \left(2\kappa^4 - 48\kappa^2 + 224\right)\Lambda^2-18 R_{\mu\nu\rho\sigma}\bar{F}^{\mu\nu}\bar{F}^{\rho\sigma}. 
\end{align}}
Similarly, we will now evaluate the contributions from ghost fields that were excluded so far, including the trace data \eqref{sdc23}. Evidently, the operator describing the ghost action \eqref{sdc5} is not in the prescribed Laplace-type form since the kinetic terms are not diagonalized. However, at any point, it is convenient to choose the following redefinitions
\begin{align}\label{sdc24}
	\begin{split}
		b_\mu \to \frac{\sqrt{2}}{2}\left(c_\mu - ib_\mu\right), \enspace b \to \frac{\sqrt{2}}{2}\left(c - ib\right), \\
	c_\mu \to \frac{\sqrt{2}}{2}\left(c_\mu + ib_\mu\right), \enspace c \to \frac{\sqrt{2}}{2}\left(c + ib\right), 
	\end{split}
\end{align}
and express the kinetic operator acting on ghost fields $\phi_m = \big\lbrace b_\mu, c_\mu, b, c \big\rbrace$ as 
\begin{align}\label{sdc25}
	\phi_m \mathcal{H}^m_n\phi^n &= b_\mu\left(\bar{g}^{\mu\nu}\Box + R^{\mu\nu}\right)b_\nu + c_\mu\left(\bar{g}^{\mu\nu}\Box + R^{\mu\nu}\right)c_\nu + b \Box b + c \Box c \nonumber\\
	&\quad + b_\mu \bar{F}^{\rho\mu}\left( D_\rho b + i D_\rho c \right) +  c_\mu \bar{F}^{\rho\mu}\left(D_\rho c- i D_\rho b  \right) \nonumber\\
	& \quad - b \bar{F}^{\rho\mu}\left(D_\rho b_\mu -i D_\rho c_\mu\right) - c \bar{F}^{\rho\mu}\left(D_\rho c_\mu + i D_\rho b_\mu\right),
\end{align}
where $\Box = D_\rho D^\rho$. It is now straightforward to extract the matrices defined in \eqref{comp3} and \eqref{comp4},
\begin{align}\label{sdc26}
\begin{split}
	\phi_m I^{mn}\phi_n &=b_\mu\bar{g}^{\mu\nu}b_\nu + c_\mu\bar{g}^{\mu\nu}c_\nu + b b + c c,\\
	\phi_m P^{mn}\phi_n &=b_\mu R^{\mu\nu}c_\nu + c_\mu R^{\mu\nu}b_\nu,\\
	\phi_m (\omega^\rho)^{mn}\phi_n &=  \frac{1}{2} b_\mu \big( \bar{F}^{\rho\mu}\big) b - \frac{1}{2} b \big(\bar{F}^{\rho\mu}\big) b_\mu  + \frac{1}{2} b_\mu \big(i \bar{F}^{\rho\mu}\big) c - \frac{1}{2} c \big(i\bar{F}^{\rho\mu}\big) b_\mu \\
	& \quad + \frac{1}{2} c_\mu \big( \bar{F}^{\rho\mu}\big) c - \frac{1}{2} c \big(\bar{F}^{\rho\mu}\big) c_\mu  + \frac{1}{2} c_\mu \big(i \bar{F}^{\rho\mu}\big) b - \frac{1}{2} b \big(i\bar{F}^{\rho\mu}\big) c_\mu, 
\end{split}
\end{align}    
followed by,
\begin{align}\label{sdc27}
\begin{split}
	\phi_m E^{mn}\phi_n &=b_\mu R^{\mu\nu}c_\nu + c_\mu R^{\mu\nu}b_\nu,\\
	\phi_m \left(\Omega_{\rho\sigma}\right)^{mn}\phi_n &=  b_\mu \Big(R\indices{^\mu^\nu_\rho_\sigma}\Big)b_\nu + c_\mu \Big(R\indices{^\mu^\nu_\rho_\sigma}\Big)c_\nu - \frac{1}{2}b_\mu \big(D^\mu \bar{F}_{\rho\sigma}\big)b \\
	& \quad + \frac{1}{2} b \big(D^\mu \bar{F}_{\rho\sigma}\big)b_\mu - \frac{1}{2}b_\mu \big(iD^\mu \bar{F}_{\rho\sigma}\big)c + \frac{1}{2} c \big(i D^\mu \bar{F}_{\rho\sigma}\big)b_\mu \\
	& \quad + \frac{1}{2}c_\mu \big(iD^\mu \bar{F}_{\rho\sigma}\big)b - \frac{1}{2} b \big(iD^\mu \bar{F}_{\rho\sigma}\big)c_\mu - \frac{1}{2}c_\mu \big(D^\mu \bar{F}_{\rho\sigma}\big)c\\
	&\quad  + \frac{1}{2} c \big( D^\mu \bar{F}_{\rho\sigma}\big)c_\mu,
\end{split}	
\end{align}
where all other valid components are completely dissolved due to our specific background, equations of motion and Maxwell-Bianchi identities \eqref{embd5}. The calculated traces of $I$, $E$, $E^2$ and $\Omega_{\rho\sigma} \Omega^{\rho\sigma}$ for ghost fields are
\begin{align}\label{sdc28}
	\begin{split}
		\text{Tr}(I) &= 4 + 4 + 1 + 1 = 10, \\
		\text{Tr}(E) &= 8\Lambda, \enspace \text{Tr}(E^2) = 2 R_{\mu\nu}R^{\mu\nu}, \\
		\text{Tr}\left(\Omega_{\rho\sigma}\Omega^{\rho\sigma}\right) &= -2 R_{\mu\nu\rho\sigma}R^{\mu\nu\rho\sigma}.
	\end{split}
\end{align}
Finally, the Seeley-DeWitt coefficients for the $U(1)$-charged EMD-AdS theory can be achieved by appropriately utilizing both the gauge-fixed \eqref{sdc23} and ghost \eqref{sdc28} trace data into the formulae \eqref{comp5} to \eqref{comp9b}. The ghost fields in \eqref{sdc5} are bosons, hence their contribution must be associated with an overall minus signature, i.e., by setting $\chi =-1$ because of their reverse spin-statistics. Up to the third order, the Seeley-DeWitt coefficient results are 
\begin{align}\label{sdc29}
	\begin{split}
		(4\pi)^2 {a_0}^{\text{$U(1)$-EMD}}(x) &= 5,\\[5pt]
		(4\pi)^2 {a_2}^{\text{$U(1)$-EMD}}(x) &= -\frac{38}{3} \Lambda,\\[5pt]
		(4\pi)^2 {a_4}^{\text{$U(1)$-EMD}}(x) &= \frac{10}{9}R_{\mu\nu\rho\sigma}R^{\mu\nu\rho\sigma} + \left(\frac{\kappa^4}{6}-\kappa^2 + \frac{5}{36}\right)R_{\mu\nu}R^{\mu\nu}\\
		&\qquad - \left(\frac{\kappa^4}{24} - \frac{\kappa^2}{4} + \frac{55}{72} \right)\Lambda^2.
	\end{split}
\end{align}
The same results for the $U(1)$-charged EMD theory in flat space are achieved by simply setting $\Lambda=0$ into the above formulas, or one can proceed without the cosmological constant from the beginning. The ${a_4}^{\text{$U(1)$-EMD}}$ relation is going to be one of the central results of this paper. Before proceeding further into its implication in logarithmic correction to black hole entropy, we find it worth making a few remarks:

\begin{enumerate}
	
\item Even though the trace data \eqref{sdc23} includes the $R_{\mu\nu\rho\sigma}\bar{F}^{\mu\nu}\bar{F}^{\rho\sigma}$ terms, ${a_4}^{\text{$U(1)$-EMD}}$ is found to be free from any such term. They are exactly canceled out inside the final formula due to our specific choice of embedding EM backgrounds into the EMD system. This fact suggests that ${a_4}(x)$ for EMD fluctuations is always invariant under the electro-magnetic duality transformation. 

\item The ${a_4}^{\text{$U(1)$-EMD}}$ form in $\eqref{sdc29}$ is expressed only in terms of the background metric invariants. However, it is never advisable to ignore the terms involving background strength $\bar{F}_{\mu\nu}$ from the beginning or in any intermediate steps of trace calculations. This is because some part of the coefficient associated with $R_{\mu\nu}R^{\mu\nu}$ and $\Lambda^2$ (or $R^2$) contributions are induced from the $\bar{F}_{\mu\rho}\bar{F}\indices{_\nu^\rho}$ terms through the Einstein equation \eqref{embd5}. In this context, one should note that the $R_{\mu\nu\rho\sigma}R^{\mu\nu\rho\sigma}$ contribution is fundamental for a theory with fixed degrees of freedom. Thus, the coefficient of $R_{\mu\nu\rho\sigma}R^{\mu\nu\rho\sigma}$ contribution is fully insensitive to the non-minimal couplings and gauge interactions between fields.	
		
\item It is not trivial to recover the results of an Einstein-Maxwell theory \cite{Bhattacharyya:2012ss} from \eqref{sdc29} even after setting the dilaton coupling constant $\kappa=0$. This is because the $\kappa$-independent terms of Seeley-DeWitt data \eqref{sdc29} are encoded with all off-shell degrees of freedom of the EMD fluctuations including the dilaton.
	
\item The ${a_4}^{\text{$U(1)$-EMD}}$ value exhibits an exact match with the result of \cite{Castro:2018tg} (see eq. (A.18)) for a pure Kaluza-Klein system with the special limiting case $\kappa= \sqrt{3}$ and $\Lambda=0$. This essentially boosts our confidence regarding the consistency of the Seeley-DeWitt coefficient results (as well as the relevant logarithmic corrections in \cref{EMD1b,EMD2}) reported in this paper for more generic EMD and EMD-AdS models with arbitrary dilaton coupling $\kappa$. 
	
\end{enumerate}




\section{Logarithmic correction for black holes in EMD and EMD-AdS theories}\label{EMD1b}

In this section, we shall calculate and demonstrate the relevant treatment for the logarithmic correction to the entropy of non-extremal and extremal black holes embedded in the four-dimensional $U(1)$-charged EMD-AdS and EMD theories. 
\subsection{Trace anomalies and central charges}
Logarithmic corrections are known to be directly connected to the divergent part of trace anomalies for a gravity theory with quantum corrections.\footnote{Readers may be interested in a relevant discussion in section 7 of \cite{Keeler:2014nn}.} In their present analysis via the heat kernel treatment, the four-dimensional anomaly data are encoded inside the third-order Seeley-DeWitt coefficient $a_4(x)$. For the 4D EMD theories embedded with the Einstein-Maxwell (EM) backgrounds in AdS space (or with a negative cosmological constant $\Lambda$), we will always end up expressing,  
\begin{align}\label{emd1}
	a_4(x) = \frac{1}{16\pi^2} \Big(c_{\text{A}} W_{\mu\nu\rho\sigma}W^{\mu\nu\rho\sigma} - a_{\text{A}} E_4 + b_{\text{A}} R^2 \Big),
\end{align}
where the coefficients ($c_{\text{A}}, a_{\text{A}}$) are the central charges of the theory related to the Type-A trace anomalies -- Weyl anomaly, i.e., the Weyl tensor squared term $W_{\mu\nu\rho\sigma}W^{\mu\nu\rho\sigma}$ and four-dimensional Euler density $E_4$. In terms of the background metric, one can explicitly define,
\begin{align}\label{emd2}
	\begin{split}
	W_{\mu\nu\rho\sigma}W^{\mu\nu\rho\sigma} &= R_{\mu\nu\rho\sigma}R^{\mu\nu\rho\sigma}-2R_{\mu\nu}R^{\mu\nu} + \frac{1}{3}R^2,\\
	E_4 &= R_{\mu\nu\rho\sigma}R^{\mu\nu\rho\sigma}-4R_{\mu\nu}R^{\mu\nu} + R^2,
	\end{split}
\end{align} 
where the classical Einstein equation implies $R=4\Lambda = -\dfrac{12}{\ell^2}$, with $\ell$ being the radius of EM-AdS background solutions. Interestingly, the Seeley-DeWitt expansion method used in this paper can compute the $c_{\text{A}}$, $a_{\text{A}}$ and $b_{\text{A}}$ coefficients related to four-derivative invariants by analyzing the quadratic fluctuation data of a two-derivative action. Here one should note that in the flat case limit (i.e., $\Lambda =0$ or $\ell \to \infty$) of EM-AdS backgrounds, the Ricci scalar $R$ vanishes, and consequently, the anomaly relation \eqref{emd1} is entirely controlled by the central charges $c_{\text{A}}$ and $a_{\text{A}}$. For the four-dimensional EMD-AdS theory, the $a_4(x)$ result \eqref{sdc29} extracts the following trace anomaly data
\begin{align}\label{emd3}
	\begin{split}
		c_{\text{A}} &= \frac{1}{24}\left(2\kappa^4 - 12\kappa^2 + 55\right),\\[3pt]
		a_{\text{A}} &= \frac{1}{72}\left(6\kappa^4 - 36\kappa^2 + 85\right),\\[3pt]
		b_{\text{A}} & = \frac{1}{72}\left(\kappa^4 - 6\kappa^2 - 25\right).
	\end{split}
\end{align}  
In the flat limit of EMD-AdS theory, we will always find $b_{\text{A}}=0$ while the central charge data ($c_{\text{A}}, a_{\text{A}}$) are the same as above. Next, we aim to utilize the trace anomaly results \eqref{emd3} in the formula \eqref{comp1b} or alternatively use the following relation
\begin{align}\label{emd4}
	\mathcal{C}_{\text{local}} = \frac{1}{16\pi^2}\int_{\text{BH}} \mathrm{d}^4x \sqrt{\det\bar{g}}\left(c_{\text{A}} W_{\mu\nu\rho\sigma}W^{\mu\nu\rho\sigma} - a_{\text{A}} E_4 + b_{\text{A}} R^2 \right),
\end{align}
and then structure the explicit formulas capturing $\mathcal{C}_{\text{local}}$ contributions to calculate the logarithmic corrections for Kerr-AdS, Reissner-Nordstr\"om-AdS, and Schwarzschild-AdS as well as Kerr, Reissner-Nordstr\"om, and Schwarzschild black hole embedded in EMD-AdS and EMD theories, respectively. However, this progress demands a pressing need for the integrated $W_{\mu\nu\rho\sigma}W^{\mu\nu\rho\sigma}$, $E_4$ and $R^2$ invariants over the appropriate part of horizon or near-horizon geometries in non-extremal and extremal limits of the black hole backgrounds.  

\subsection{Non-extremal black hole backgrounds and $\mathcal{C}_{\text{local}}$ formulas} \label{nbh}

We will start with a generic background of rotating and charged asymptotically-AdS black hole in four-dimensional spacetimes. The relevant metric is given in terms of appropriate Boyer-Lindquist coordinates as   
\begin{align}\label{nbh1}
		\mathrm{d}s^2 = \bar{g}_{\mu\nu}\mathrm{d}x^\mu \mathrm{d}x^\nu &= - \frac{\Delta _r}{\rho^2} \left( \mathrm{d}t - \frac{a \sin^2 \theta }{\Xi} \mathrm{d}\phi \right)^2 + \frac{\rho^2}{\Delta_r}\mathrm{d}r^2  + \frac{\rho^2}{\Delta_\theta} \mathrm{d}\theta^2 \nonumber\\
		&\qquad + \frac{\Delta_\theta \sin^2 \theta}{\rho^2} \left(a\,\mathrm{d}t - \frac{r^2+a^2}{\Xi} \mathrm{d}\phi \right)^2,
\end{align}
where we have followed the convention of \cite{Caldarelli:1999x} and set $G_D =1$. The parameters $\Delta_r$, $\Delta_\theta$, $\rho$ and $\Xi$ are defined as
\begin{align}\label{nbh2}
	\begin{gathered}
		\Delta_r = (r^2+a^2) \bigg(1+\frac{r^2}{\ell^2}\bigg) - 2m r + q^2 + p^2,\\
		\Delta_\theta = 1 -\frac{a^2}{\ell^2} \cos^2 \theta, \quad \rho^2 = r^2 + a^2 \cos^2 \theta, \quad \Xi = 1 -\frac{a^2}{\ell^2}.
	\end{gathered}
\end{align}
The physical mass $M$, angular momentum $J$, electric charge $Q$ and magnetic charge $P$ of the black hole are characterized by the parameters $m$, $a$, $q$ and $p$ via the relations,
\begin{align}\label{nbh3}
	M = \frac{m}{\Xi^2}, \quad J= \frac{ma}{\Xi^2}, \quad Q = \frac{q}{\Xi}, \quad  P = \frac{p}{\Xi}.
\end{align}
Here the rotational parameter $a$ must satisfy the $a < \ell$ for $a \geq 0$ limit so that the metric \eqref{nbh1} represents an AdS black hole. The background metric \eqref{nbh1} also solves the Maxwell part of the field equations \eqref{embd5} for the gauge field,
\begin{align}\label{nbh5}
	\bar{A} = -\frac{qr}{\rho^2}\left(\mathrm{d}t - \frac{a \sin^2 \theta }{\Xi} \mathrm{d}\phi\right) - \frac{p\cos\theta}{\rho^2}\left(a\,\mathrm{d}t - \frac{r^2+a^2}{\Xi} \mathrm{d}\phi \right),
\end{align}
followed by its strength $\bar{F}_{\mu\nu}$ expressing,
\begin{align}\label{nbh6}
	\bar{F}_{\mu\nu}\bar{F}^{\mu\nu} = -\frac{2}{\left(r^2 + a^2 \cos ^2\theta\right)^4} &\bigg[\left(q^2-p^2\right)\left(r^4- 6a^2r^2\cos^2\theta + a^4\cos^4\theta\right) \nonumber\\
	&\enspace + 8 qpar \cos\theta\left(r^2 - a^2 \cos ^2\theta\right)\bigg].
\end{align}
In this paper, our specific choice backgrounds are those with $q=p$, $a=0$ and $q, p=0$ so that $\bar{F}_{\mu\nu}\bar{F}^{\mu\nu}$ invariant vanishes, and we can embed them into EMD theories (as discussed in \cref{embd}). Now, if $r=r_+$ is the position of the event horizon for solving $\Delta_r =0$ as the largest real root, such that
\begin{align}
	m = \frac{1}{2 {r_+} \ell ^2} \Big[{r_+}^4 + \left(\ell^2 + a^2\right){r_+}^2 + \left(a^2 + q_e^2\right)\ell^2 \Big],
\end{align}
then the Bekenstein-Hawking entropy and inverse Hawking temperature are given by
\begin{align}\label{nbh4}
	S_{\text{BH}} = \frac{4\pi \left(r_+^2 + a^2\right)}{\Xi}, \quad \beta = \frac{4\pi \left(r_+^2 + a^2\right)}{r_+ \left(1 + \frac{a^2}{\ell^2} + 3 \frac{r_+^2}{\ell^2}-\frac{a^2 + q_e^2}{r_+^2}\right)},
\end{align}
where $q_e =\sqrt{q^2+p^2}$. With the above background setup, we express the curvature invariants required in \eqref{emd1} as 
{\allowdisplaybreaks  
\begin{align}\label{nbh7}
R^2 &= \frac{144}{\ell^4},\nonumber \\
W_{\mu\nu\rho\sigma}W^{\mu\nu\rho\sigma} &= \frac{48}{\left(r^2 + a^2 \cos ^2\theta\right)^6} \bigg[ 8 r^4 \left(q_e^2 - 2mr\right)^2 - m^2 \left(r^2 + a^2 \cos ^2\theta\right)^3 \nonumber\\
&\qquad\qquad\qquad\qquad - 8r^2 \left(q_e^2 - 3mr\right)\left(q_e^2 - 2mr\right)\left(r^2 + a^2 \cos ^2\theta\right) \nonumber\\
& \qquad\qquad\qquad\qquad + \Big(q_e^4 - 10 m r q_e^2+ 18 m^2r^2\Big)\left(r^2 + a^2 \cos ^2\theta\right)^2 \bigg],\nonumber\\[5pt]
E_4 &= \frac{24}{\ell^4} + \frac{8}{\left(r^2 + a^2 \cos ^2\theta\right)^6} \bigg[ 48 r^4 \left(q_e^2 - 2mr\right)^2 - 6m^2 \left(r^2 + a^2 \cos ^2\theta\right)^3 \nonumber\\
&\qquad\qquad\qquad\qquad - 48r^2 \left(q_e^2 - 3mr\right)\left(q_e^2 - 2mr\right)\left(r^2 + a^2 \cos ^2\theta\right) \nonumber\\
& \qquad\qquad\qquad\qquad + \Big(5q_e^4 - 60 m r q_e^2+ 108 m^2r^2\Big)\left(r^2 + a^2 \cos ^2\theta\right)^2 \bigg]. 
\end{align}
}
Next, the above invariants need to be integrated over the AdS$_4$ black hole geometry \eqref{nbh1}. But, upon integration, these $a_4(x)$ invariants will diverge due to the infinite volume of AdS$_4$. In this paper, we aim to tame these divergences by following the prescription of holographic renormalization \cite{Skenderis:2002wp}. The underlying treatment is to set a large cut-off $r=r_c$ at the boundary of the geometry \eqref{nbh1} and followed by adding the holographic counterterm,
\begin{align}\label{nbh8}
	\mathcal{C}_{\text{HCT}} = \int_{\partial \mathcal{M}} \mathrm{d}^3y \sqrt{\det\gamma} \left(c_1 + c_2 \mathcal{R}\right), 
\end{align}
where $\mathcal{R}$ is the Ricci scalar related to the metric $\gamma$ describing the boundary geometry $\partial \mathcal{M}$. Then, we determine the appropriate values for $c_1$ and $c_2$ coefficients such that the combined bulk-boundary contribution $\mathcal{C}_{\text{local}} + \mathcal{C}_{\text{HCT}}$ receives a finite value for \eqref{emd4} in the limit $r_c \to \infty$. In this process, we obtain the following renormalized results for the integrated $W_{\mu\nu\rho\sigma}W^{\mu\nu\rho\sigma}$, $E_4$ and $R^2$ invariants,
\begin{subequations}\label{nbh9} 
\begin{align}\label{nbh9a}
\begin{split}	
\frac{1}{16\pi^2}\int \mathrm{d}^4x \sqrt{\det\bar{g}}\, E_4 &= 4, \\
\frac{1}{16\pi^2}\int \mathrm{d}^4x \sqrt{\det\bar{g}}\, R^2 &= R_1 + \beta R_2, \\
\frac{1}{16\pi^2}\int \mathrm{d}^4x \sqrt{\det\bar{g}}\, W_{\mu\nu\rho\sigma}W^{\mu\nu\rho\sigma} &= \frac{W_1}{\beta} + W_2 + \beta W_3,
\end{split}
\end{align}
where the $\beta$-independent parts of background Ricci and Weyl squared integrations are expressed as
{\allowdisplaybreaks
\begin{align}\label{nbh9b}
R_1 &= -\frac{24}{\Xi\ell^2}\left(r_+^2 + a^2\right), \quad R_2 = \frac{12}{\pi\Xi\ell^4}\left(\ell^2 + r_+^2\right)r_+, \nonumber\\[5pt]
		W_1 &= \frac{\pi \left(a^2+r_+^2\right)}{\Xi a^5 r_+^2 } \bigg[\left(3a^4 + 2a^2r_+^2 + 3 r_+^4\right)ar_{+} - 3 \left(r_+^4 - a^4\right) \left(a^2+r_+^2\right)\arctan\left(\frac{a}r_+\right)\bigg], \nonumber\\
		W_2 &=  \frac{\left(a^2+r_+^2\right)}{2\Xi a^5 \ell^2r_+^3 }\bigg[\Big(3 a^4 \left(\ell ^2-r_+^2\right) + 4 a^2\ell ^2 r_+^2 -3 \left(\ell ^2 + 3 r_+^2\right)r_+^4 \Big)a r_+ \nonumber\\
		&\qquad\qquad\quad\quad\enspace - 3 \left(r_+^4- a^4\right) \Big(a^2 \left(\ell ^2 - r_+^2\right) - \left(\ell ^2 + 3 r_+^2\right)r_+^2 \Big) \arctan\left(\frac{a}r_+\right)\bigg],   \nonumber\\[5pt]
		W_3 &=  \frac{1}{16\pi \Xi a^5 \ell^4\left(r_+^2 + a^2\right)r_+^4  } \bigg[3 a^9 \left(\ell ^2-r_+^2\right)^2r_+ -4 a^7 \left(3 r_+^4 + 12 \ell ^2r_+^2 +\ell ^4\right)r_+^3 \nonumber\\
		&\qquad +2 a^5 \left(5 r_+^4 -14\ell ^2 r_+^2 +\ell ^4\right)r_+^5  -4 a^3 \left(\ell ^4-9 r_+^4\right)r_+^7  +3 a \left(\ell ^2 + 3 r_+^2\right)^2r_+^9 \nonumber\\
		& \qquad - 3 \left(r_+^4 - a^4\right) \left(a^2+r_+^2\right) \Big( a^2 \left(\ell ^2 - r_+^2\right) - \left(\ell ^2 + 3 r_+^2\right)r_+^2  \Big)^2 \arctan\left(\frac{a}r_+\right)\bigg],
\end{align}
}
\end{subequations}
where $\Xi = 1 -a^2/\ell^2$. Note that the above derivations considered the analytical continuation $t \to -i\tau$ of the metric \eqref{nbh1}, where the Euclideanized time $\tau$ is identified as a periodic coordinate of period $\beta$. Also, the integrations are executed in the range $r_+ \leq r \leq r_c$ (for the bulk part), $0 \leq \theta \leq \pi$ and $0 \leq \phi \leq 2\pi$. For more details about the holographic renormalization and boundary counterterm, we refer to \Cref{holo}. The choice of holographic renormalization is natural since the logarithmic correction is an explicit correction term to the on-shell or bulk effective action. Hence, the current prescription is very convenient and always found to be providing a physically sensible, unambiguous and finite result for all the AdS$_4$ black holes. In this line, a strong validation can be found in \cite{David:2021eoq} where it has been proved that the holographic counterterm \eqref{nbh8} exactly matches the standard boundary term of the Gauss-Bonnet-Chern theorem \cite{Chern:1945wp}.\footnote{Please review Appendix E of \cite{David:2021eoq}.} Another solid verification emerges when the holographic renormalization procedure computes the correct and exact Euler characteristic value via integrating the Euler density $E_4$ around 4D black hole geometries as      
\begin{align}\label{nbh10}
	\chi = \lim_{r_c \to \infty}\left[\frac{1}{32\pi^2}\int_{\mathcal{M}} \mathrm{d}^4x \sqrt{\det\bar{g}}\, E_4 + \int_{\partial \mathcal{M}} \mathrm{d}^3y \sqrt{\det\gamma} \left(c_1 + c_2 \mathcal{R}\right)\right] = 2.
\end{align}
Finally, substitution of the relations \eqref{nbh9} into \eqref{emd4} with appropriate limits provides the $\mathcal{C}_{\text{local}}$ formulas of the non-extremal black holes embedded in EMD-AdS theory. For the Schwarzschild-AdS ($q,p,a= 0$), Reissner-Nordstr\"om-AdS ($q=p,a= 0$) and Kerr-AdS ($q,p= 0$), we successively obtain
\begin{align}\label{nbh11}
	\begin{split}
		\mathcal{C}_{\text{local}}^{\text{(Sch-AdS)}} &= -4 a_{\text{A}} + \frac{4}{\left(3 r_+^2+\ell ^2\right)\ell ^2 } \bigg[\left(c_{\text{A}}-6 b_{\text{A}}\right)r_+^4 + \left(2 c_{\text{A}} +6 b_{\text{A}}\right)\ell ^2 r_+^2  +c_{\text{A}}\ell ^4 \bigg], \\[5pt]
		\mathcal{C}_{\text{local}}^{\text{(RN-AdS)}}  &= \frac{4}{5} \left(c_{\text{A}}-5a_{\text{A}}\right) - \frac{4}{5\pi\ell ^4 r_+} \bigg[  \left(7c_{\text{A}} + 30b_{\text{A}}\right)\pi\ell^2 r_+^3 - \frac{8\pi^2c_{\text{A}} \ell^4 r_+^2}{\beta} \\
		&\qquad\quad\qquad  -\frac{\beta}{2} \Big(\left(c_{\text{A}} + 30 b_{\text{A}}\right)\ell^2 r_+^2  +  \left(4c_{\text{A}} + 30 b_{\text{A}}\right)r_+^4  + c_{\text{A}}\ell^4\Big) \bigg],\\[5pt]
		\mathcal{C}_{\text{local}}^{\text{(Kerr-AdS)}}&= -4 a_{\text{A}} + \frac{\beta }{\pi  \ell ^2 \left(r_+^2 + a^2\right) \left(\ell ^2-a^2\right)r_+ }\bigg[6b_{\text{A}} a^4\ell^2 - c_{\text{A}}a^2\ell^4     \\
		&\qquad\quad\qquad + \left(c_{\text{A}}-6 b_{\text{A}}\right)r_+^6 +\Big( \left(2c_{\text{A}} + 6 b_{\text{A}}\right)\ell ^2 -\left(c_{\text{A}} + 12 b_{\text{A}}\right)a^2\Big)r_+^4 \\
		&\qquad\quad\qquad +\Big(c_{\text{A}} \ell ^4 + \left(12 b_{\text{A}}-2c_{\text{A}}\right)a^2 \ell ^2  -6 b_{\text{A}} a^4 \Big)r_+^2   \bigg]. 
	\end{split}
\end{align}
In the flat space limit $\ell \to \infty$, the integrations \eqref{nbh9} exactly reproduce the known relations in \cite{Sen:2013ns, Charles:2015nn,Karan:2020sk}. The corresponding $\mathcal{C}_{\text{local}}$ formulas for the asymptotically-flat and non-extremal black holes embedded in EMD theory are
\begin{align}\label{nbh12}
\begin{split}
\mathcal{C}_{\text{local}}^{\text{(Sch)}} &= 4(c_{\text{A}} - a_{\text{A}}), \quad \mathcal{C}_{\text{local}}^{\text{(Kerr)}} = 4(c_{\text{A}} - a_{\text{A}}),\\[3pt]
\mathcal{C}_{\text{local}}^{\text{(RN)}} &= 4(c_{\text{A}} - a_{\text{A}}) + \frac{2c_{\text{A}} \beta q_e^4}{5\pi r_+^5},
\end{split}
\end{align}
where $\beta = 4\pi r_+^3/\left(r_+^2 - q_e^2\right)$, $r_+ = m + \sqrt{m^2 - q_e^2}$  and $q_e =  \sqrt{2}q = \sqrt{2}p$.

\subsection{Extremal limit, near-horizon backgrounds and $\mathcal{C}_{\text{local}}$ formulas}\label{ebh} 

The special treatments required to proceed with the extremal black holes are already summarized in \cref{extlim}. Here the most trivial approach is to take the extremal or zero-temperature limit $\beta \to \infty$ of the $\mathcal{C}_{\text{local}}$ contribution \eqref{emd4} through the integrated invariant relations \eqref{nbh9}. However, the direct use of extremal limit naively leads to divergences. But, we employ a systematic approach (following the work \cite{David:2021eoq}) to escape from the extremal divergence and extract a finite piece of $\mathcal{C}_{\text{local}}$ contribution. By keeping the black hole parameters $a,\, \ell,\, q,\, p$ fixed, a \textit{low-temperature expansion} of horizon radius $r_+$ in the limit $\beta \to \infty$ can be managed as     
\begin{align}\label{ebh1}
	r_+ = r_0 + \frac{2 \pi \ell_2^2}{\beta} + \mathcal{O}(\beta^{-2}),
\end{align}
where the parameter $r_0$ characterizing the `finite part' is the extremal horizon radius and $\ell_2$ is recognized as an AdS$_2$ radius. The appearance of an AdS$_2$ part inside the structure of extremal black holes is quite natural. For example, see the prescription of quantum entropy function formalism \cite{Sen:2008wa,Sen:2009wb,Sen:2009wc}. The extremal parameters ($r_0$, $\ell_2$) are related via the AdS$_4$ radius and charges in the extremal limit as
\begin{align}\label{ebh2}
	\begin{split}
		\ell^2_2 &= \frac{\ell^2\left(a^2 + r_0^2\right)}{\left(\ell^2 + a^2 + 6r_0^2\right)}, \\[3pt]
a^2+ q_e^2 & =\frac{r_0^2 \left(a^2+\ell^2+3 r_0^2\right)}{\ell^2},
	\end{split} 
\end{align}
where we must consider $r_0^2 >( \ell_2^2 - a^2)$. Now, using the typical form \eqref{ebh1} of full horizon radius $r_+$, we can systematically adjust the $\mathcal{C}_{\text{local}}$ contribution up to a finite constant as     
\begin{align}\label{ebh3}
	\lim_{\beta \to \infty} \mathcal{C}_{\text{local}} \equiv \mathcal{C}_0 + \mathcal{C}_1\beta + \mathcal{O}(\beta^{-1}).
\end{align}
The part linear in $\beta$ is divergent and can be neglected since it is nothing but an infinite shift in the ground state energy of one-loop effective action. Therefore, after setting the extremal limit $\beta \to \infty$, all the terms inverse in $\beta$ vanished and we can recognize the finite piece $\mathcal{C}_0$ as an unambiguous and effective contribution to $\mathcal{C}_{\text{local}}$. With this setup, the related extremal version of the curvature invariant integrations \eqref{nbh9} are obtained as 
{\allowdisplaybreaks
\begin{align}\label{ebh4}
		&\lim_{\beta \to \infty}\frac{1}{16\pi^2}\int \mathrm{d}^4x \sqrt{\det\bar{g}}\, E_4 = 4, \nonumber\\
	&\lim_{\beta \to \infty}	\frac{1}{16\pi^2}\int \mathrm{d}^4x \sqrt{\det\bar{g}}\, R^2 = -\frac{24 \left(a^4+4 a^2 r_0^2+3 r_0^4\right)}{\left(\ell ^2-a^2\right) \left(a^2+6 r_0^2+\ell^2\right)}, \nonumber\\[5pt]
	&\lim_{\beta \to \infty}\frac{1}{16\pi^2}\int \mathrm{d}^4x \sqrt{\det\bar{g}}\, W_{\mu\nu\rho\sigma}W^{\mu\nu\rho\sigma} = \frac{1}{8 a^4 r_0^5 \left(\ell^2-a^2\right)\left(a^2+r_0^2\right) \left(a^2+6 r_0^2+\ell ^2\right)} \bigg[ \nonumber\\
	&\qquad\qquad\qquad\qquad + a^2r_0 \Big(9a^{8}\left(2r_0^2\ell^2 - r_0^4 -\ell^4\right) +a^6 r_0^2 \left(70 r_0^2 \ell ^2-135 r_0^4+\ell ^4\right) \nonumber\\
	&\qquad\qquad\qquad\qquad +2 a^4 r_0^4 \left(106 r_0^2 \ell ^2-77 r_0^4+ 27\ell ^4\right) + 6a^2 r_0^6 \left(18 r_0^2 \ell ^2  - 13 r_0^4 + 3 \ell ^4\right)  \nonumber\\
	&\qquad\qquad\qquad\qquad - 3 r_0^8 \left(2 r_0^2 \ell ^2 + 15 r_0^4 - \ell ^4\right) \Big)-3 r_0^{11} \left(3 r_0^2+\ell ^2\right)^2 \nonumber\\
	&\qquad\qquad\qquad -12 a^3 \left(a^2+r_0^2\right){}^2 \Big(a^2 \left(r_0^2-\ell ^2\right)+r_0^2 \left(3 r_0^2+\ell ^2\right)\Big){}^2 \arctan\left(\frac{a}{r_0}\right) \bigg].
\end{align}}
We will utilize the above relations and express the $\mathcal{C}_{\text{local}}$ contribution \eqref{emd4} generally in terms of the five independent parameters $\left\lbrace r_0, \ell, a, q, p \right\rbrace$. For a Schwarzschild background ($q,p,a= 0$), extremality is not a valid limit due to having a vanishing extremal horizon $r_0$. However, for the extremal Reissner-Nordstr\"om-AdS ($q=p,a= 0$) and Kerr-AdS ($q,p= 0$) embedded in EMD-AdS theory, we derive the following explicit $\mathcal{C}_{\text{local}}$ formulas
\begin{align}\label{ebh5} 
\lim_{\beta \to \infty}\mathcal{C}_{\text{local}}^{\text{(RN-AdS)}}  & = \frac{4}{3}\left(c_{\text{A}}-3a_{\text{A}}+ 3b_{\text{A}}\right)-\frac{2}{3 r_0^2 \ell _2^2}\left(c_{\text{A}} + 3 b_{\text{A}}\right) \left(\ell _2^4 + r_0^4\right), \nonumber\\[3pt]
\lim_{\beta \to \infty}\mathcal{C}_{\text{local}}^{\text{(Kerr-AdS)}}&= -4a_{\text{A}} + \frac{2}{r_0^2 \left(\ell ^2-a^2\right) \left(a^2+r_0^2\right) \left(a^2+6 r_0^2+\ell ^2\right)} \bigg[c_{\text{A}} a^4\ell^4  \nonumber \\
&\quad + 3\left(c_{\text{A}} - 6b_{\text{A}}\right)r_0^8 + \Big(\left(4c_{\text{A}} - 42b_{\text{A}}\right)a^2 + \left(2c_{\text{A}} + 6b_{\text{A}}\right)\ell^2\Big)r_0^6 \nonumber\\
& \quad + \Big(\left(8c_{\text{A}} + 12b_{\text{A}}\right)a^2\ell^2 - \left(3c_{\text{A}} + 30b_{\text{A}}\right)a^4 - c_{\text{A}}\ell^4\Big)r_0^4 \nonumber \\
&\quad + \Big(4c_{\text{A}} a^2\ell^4 + \left(6b_{\text{A}} - 2c_{\text{A}}\right)a^4\ell^2 - 6b_{\text{A}}a^6\Big)r_0^2\bigg]              .
\end{align}
In flat space ($\ell \to \infty$), the extremal limit is acquired by setting the constraints, 
\begin{align}\label{ebh10}
	\ell_2 = \sqrt{r_0^2 + a^2},\quad  r_0 = \sqrt{a^2 + q_e^2},
\end{align}
where $r_0$ is identified as the mass parameter $m$ of asymptotically-flat black holes (see, e.g. \cite{Bhattacharyya:2012ss,Karan:2020sk}). Thus, the related $\mathcal{C}_{\text{local}}$ formulas for extremal black holes in EMD theory are
\begin{align}\label{ebh11}
	\begin{split}
		\lim_{\beta \to \infty}\mathcal{C}_{\text{local}}^{\text{(RN)}}  &= -4a_{\text{A}}, \quad \lim_{\beta \to\infty}\mathcal{C}_{\text{local}}^{\text{(Kerr)}}=  4(c_{\text{A}} -a_{\text{A}}).
	\end{split}
\end{align}  
Next, we will verify the consistency of the above formulas using the prescription of quantum entropy function (QEF) formalism \cite{Sen:2008wa,Sen:2009wb,Sen:2009wc}. QEF is the most convenient and successful approach for the quantum entropy of extremal black holes where the underlying setup entirely relies on the near-horizon geometry analysis (see the discussion in \cref{extlim}). In order to achieve the extremal near-horizon (ENH) geometry of the black hole \eqref{nbh1}, we cast the following coordinate transformations in terms of a new parameter $\lambda$ (see, e.g. \cite{David:2021eoq,Bhattacharyya:2012ss})
\begin{align}\label{ebh6}
	\begin{gathered}
		r = r_0 + \lambda \tilde{r}, \quad  t =\frac{\ell_2^2}{\lambda} \tilde{t}, \quad \phi = \tilde{\phi } + \frac{a \left(\ell ^2-a^2\right)}{\ell^2 \left(a^2+r_0^2\right)} t,\\
	\frac{r_0^2}{\ell^2}\left(3r_0^2 + a^2 + \ell^2\right) = a^2 + q_e^2 + \lambda^2,
	\end{gathered}
\end{align}  
and then set the limit $\lambda \to 0$ by keeping $\tilde{r}$ and $\tilde{t}$ fixed. Further considering the analytical continuation $\tilde{t} \to -i\tau$, the metric describing Euclideanized ENH geometry is expressed as 
\begin{align}\label{nhe}
\mathrm{d} s^2 &= (\bar{g}_{\mu\nu}\mathrm{d}x^\mu \mathrm{d}x^\nu)_{\text{ENH}} \nonumber\\
&= \frac{\ell_2^2\left(r_0^2+a^2 \cos ^2 \theta\right)}{a^2+r_0^2}\left(\tilde{r}^2 \mathrm{d}\tau^2+\frac{\mathrm{d} \tilde{r}^2}{\tilde{r}^2}\right)+\frac{\ell^2\left(r_0^2+a^2 \cos ^2 \theta\right)}{\ell^2-a^2 \cos ^2 \theta} \mathrm{d}\theta^2 \nonumber\\
&\qquad +\frac{\ell^2\left(a^2+r_0^2\right)^2\left(\ell^2-a^2 \cos ^2 \theta\right) \sin ^2 \theta}{\left(\ell^2-a^2\right)^2\left(r_0^2+a^2 \cos ^2 \theta\right)}\left(\mathrm{d} \tilde{\phi}-\frac{2 \ell_2^2 a r_0\left(\ell^2-a^2\right)}{\ell^2\left(a^2+r_0^2\right)^2} i r \mathrm{d}\tau\right)^2.
\end{align}
The above geometry form evidently accommodates an AdS$_2$ piece with coordinates $(\tau, \tilde{r})$, as required in the QEF prescription. Hence, before we proceed further and calculate $\mathcal{C}_{\text{local}}$ and the related integrated invariants, we must remove the divergence arising due to the infinite volume of AdS$_2$ geometry. By a suitable regularization, one can set an infrared cut-off at $\tilde{r} =\tilde{r}_c$ and evaluate a regularized $\mathcal{C}_{\text{local}}$ in the range, 
\begin{align}\label{ebh7}
	1 \leq \tilde{r}\leq \tilde{r}_c, \quad 0 \leq \tau \leq 2\pi, \quad 0 \leq \theta \leq \pi, \quad 0 \leq \tilde{\phi} \leq 2\pi. 
\end{align}
One may naively think that the present treatment leads to ambiguous $\mathcal{C}_{\text{local}}$ results that vary in different regularization schemes. But, we want to ensure that the QEF formalism precisely identified that only the cut-off independent finite piece of ENH geometry induces the quantum horizon degeneracy and related corrections to the extremal black hole entropies \cite{Sen:2008wa,Sen:2009wb,Sen:2009wc}. Thus, in terms of the ENH background \eqref{nhe}, the finite $\mathcal{C}_{\text{local}}$ contribution for extremal AdS$_4$ black holes is now defined as 
\begin{align}\label{ebh8}
\mathcal{C}_{\text{local}} = -2\pi \int_{\text{ENH}} \mathrm{d}\theta\,\mathrm{d}\tilde{\phi}\,\mathcal{G}(\theta)a_4(x),
\end{align}
where $-2\pi$ factor is arising from the $\tilde{r}_c$ independent part of regularized AdS$_2$ volume\footnote{The regularized AdS$_2$ volume is $\sim \int_{0}^{\tilde{r}_c} \int_{0}^{2\pi}\mathrm{d}\tilde{r} \mathrm{d}\tau = 2\pi \left(\tilde{r}_c - 1\right) $. In any arbitrary regularization scheme, the cut-off dependent part $2\pi\tilde{r}_c$ will be absorbed by either redefining the ground state energy in dual CFT$_1$ or adding boundary counter-terms \cite{Sen:2008wa,Sen:2009wb,Sen:2009wc}. Hence, only the finite and cut-off independent term proportional to $-2\pi$ contributes to the quantum entropy of extremal black holes and $\mathcal{C}_{\text{local}}$.} and $\mathcal{G}(\theta)$ is some function with the coordinates independent of the AdS$_2$ part,
\begin{align}\label{ebh9}
	\mathcal{G}(\theta) = ({\sqrt{\det\bar{g}}})_{\text{ENH}}/ \mathcal{G}(\tilde{r},\tau)= \frac{\ell ^2 \ell _2^2 }{(\ell ^2-a^2)}\left(a^2 \cos^2+r_0^2\right) \sin \theta.
\end{align}
Any regularization process never affects the cut-off independent part of the bulk contribution. Hence, the formula \eqref{ebh8} always equips an unambiguous and natural correction result for extremal black hole entropy. Finally, we employ the $a_4(x)$ form \eqref{emd1} in the formula \eqref{ebh8} where we require to integrate the $W_{\mu\nu\rho\sigma}W^{\mu\nu\rho\sigma}$, $E_4$ and $R^2$ invariants around the ENH background \eqref{nhe} modulated by the function $\mathcal{G}(\theta)$ (see \Cref{enhci} for details). The final $\mathcal{C}_{\text{local}}$ results are obtained in terms of the parameters $\lbrace\ell_2,\, r_0,\, \ell\rbrace$ and found to match exactly with the formulas \eqref{ebh5} and \eqref{ebh11} derived via taking $\beta \to \infty$ in the non-extremal full horizon geometry analysis.

{
	\renewcommand{\arraystretch}{1.3}
	\begin{table}[t]
		\centering
		\hspace{-0.2in}
		\begin{tabular}{|>{\centering}p{2.5in}|>{\centering}p{1.0in}|}
			\hline
			\textbf{Black Hole Backgrounds} & \textbf{$\bm{\mathcal{C}_{\text{zm}}}$} \tabularnewline \hline \hline
			Schwarzschild/Schwarzschild-AdS & $-3$ \tabularnewline \hline
			Kerr/Kerr-AdS & $-1$ \tabularnewline \hline
			Reissner-Nordstr\"om/Reissner-Nordstr\"om-AdS & $-3$ \tabularnewline \hline
			Extremal Kerr/Kerr-AdS & $-4$ \tabularnewline \hline
			Extremal Reissner-Nordstr\"om/Reissner-Nordstr\"om-AdS & $-6$ \tabularnewline \hline
		\end{tabular}
		\caption{$\mathcal{C}_{\text{zm}}$ contributions for the 4D black holes embedded in EMD-AdS and EMD theories. The results for extremal backgrounds are exclusively induced from their near-horizon analysis.}\label{czero}
	\end{table}
}
\subsection{Results}\label{results}
We now evaluate the logarithmic corrections to the entropy of 4D black holes embedded in the EMD-AdS and EMD theories. For the local contributions, we used the trace anomaly or central charge data \eqref{emd3} in the $\mathcal{C}_{\text{local}}$ formulas derived in \cref{nbh11,nbh12,ebh5,ebh11} for non-extremal and extremal black holes. On the other hand, the $\mathcal{C}_{\text{zm}}$ contributions are theory-independent or global. With the help of zero-mode data depicted in \cref{zeromode}, we have explicitly prepared a list in \Cref{czero} for the black hole backgrounds of this paper. Here one must note that there is no BPS solution among the embedded black holes, even in the extremal limit.\footnote{The BPS bound for AdS$_4$ black holes is $M = Q + J/\ell$ due to satisfying the constraints $p =0$ and $r_0 = \sqrt{a\ell}$ \cite{Caldarelli:1999x}. So, only an extremal Kerr-Newman-AdS black hole satisfies the BPS condition.} Finally, we computed the net logarithmic entropy corrections \eqref{comp1} for the black holes in each case of dilaton couplings $\kappa = 1$, $\kappa = \sqrt{3}$ and $\kappa = 1/\sqrt{3}$ that are associated with the supergravity or string-theory embeddings. For example, the $\kappa = 1$ case is a prototype sector that naturally arises in low-energy type I and type II superstring theories and the $\kappa = \sqrt{3}$ case directly connected to compactified or dimensionally-reduced Kaluza-Klein theory. For a more concrete example, readers are referred to \cref{EMD2}, where we will calculate the same results in a more generalized class of $U(1)^2$-charged EMD models directly embedded into $\mathcal{N}=4$ supergravity.
\subsubsection{Logarithmic corrections in $U(1)$-charged EMD-AdS theory}\label{ads}
The logarithmic correction to the entropy of Schwarzschild-AdS ($q,p,a= 0$), Reissner-Nordstr\"om-AdS ($q=p,a= 0$) and Kerr-AdS ($q,p= 0$) black holes in EMD-AdS theory are as follows. 
\begin{itemize}
\item[\scalebox{0.6}{$\blacksquare$}]{Case $\kappa=1$:} In the finite temperature or \textit{non-extremal} limit, we obtain
{\allowdisplaybreaks
\begin{align}
\Delta S_{\text{BH}}^{\text{(Sch-AdS)}} &= \bigg[-\frac{109}{36} + \frac{5}{4\left(3 r_+^2+\ell ^2\right)\ell ^2 } \left(7r_+^4 + 2\ell ^2 r_+^2  + 3\ell ^4 \right)\bigg]\ln \mathcal{A}_{H}, \label{ads1a}\\[5pt]
\Delta S_{\text{BH}}^{\text{(RN-AdS)}}  &= \bigg[-\frac{41}{18} + \frac{1}{8\pi\ell ^4 r_+} \bigg\lbrace  \frac{48\pi^2 \ell^4 r_+^2}{\beta} - 2\pi\ell^2 r_+^3  \nonumber \\
&\qquad + \beta\left(3\ell^4 -17\ell^2 r_+^2  -8 r_+^4  \right) \bigg\rbrace\bigg]\ln \mathcal{A}_{H}, \label{ads1b}\\[7pt]
\Delta S_{\text{BH}}^{\text{(Kerr-AdS)}}&= \bigg[-\frac{73}{36} + \frac{5\beta }{16\pi  \ell ^2 \left(r_+^2 + a^2\right) \left(\ell ^2-a^2\right)r_+ }\bigg\lbrace 7 r_+^6 -4 a^4\ell^2 - 3 a^2\ell^4    \nonumber \\
&\qquad +\left( 2\ell ^2 + 5 a^2\right)r_+^4 +\left(3 \ell ^4 - 14 a^2 \ell ^2  + 4 a^4 \right)r_+^2 \bigg\rbrace\bigg]\ln \mathcal{A}_{H},\label{ads1c}	
\end{align}}
followed by the \textit{extremal} limit results of Reissner-Nordstr\"om-AdS and Kerr-AdS black holes as
{\allowdisplaybreaks
\begin{align}
\Delta S_{\text{BH}}^{\text{(ext,RN-AdS)}}  & = -\left[\frac{37}{9} + \frac{5\left(\ell _2^4 + r_0^4\right)}{24 r_0^2 \ell _2^2} \right]\ln \mathcal{A}_{H}, \label{ads1d}\\[5pt]
\Delta S_{\text{BH}}^{\text{(ext,Kerr-AdS)}}&= \bigg[-\frac{127}{36} + \frac{5}{8r_0^2 \left(\ell ^2-a^2\right) \left(a^2+r_0^2\right) \left(a^2+6 r_0^2+\ell ^2\right)} \bigg\lbrace 3 a^4\ell^4  \nonumber \\
&\qquad + 21 r_0^8 + \left(40 a^2 + 2 \ell^2\right)r_0^6  + \left(16 a^2\ell^2 + 11 a^4 - 3\ell^4\right)r_0^4 \nonumber \\
&\qquad + \left(12 a^2\ell^4 - 10 a^4\ell^2 + 4 a^6\right)r_0^2\bigg\rbrace\bigg]\ln \mathcal{A}_{H},\label{ads1e}
\end{align}
}
\item[\scalebox{0.6}{$\blacksquare$}]{Case $\kappa=\sqrt{3}$:} In the finite temperature or \textit{non-extremal} limit, we calculate
	{\allowdisplaybreaks
	\begin{align}
	\Delta S_{\text{BH}}^{\text{(Sch-AdS)}} &= \bigg[-\frac{85}{36} + \frac{1}{12\left(3 r_+^2+\ell ^2\right)\ell ^2 } \left(105 r_+^4 + 6\ell ^2 r_+^2  + 37 \ell ^4 \right)\bigg]\ln \mathcal{A}_{H}, \label{ads2a}\\[7pt]
	\Delta S_{\text{BH}}^{\text{(RN-AdS)}}  &= \bigg[-\frac{157}{90} + \frac{1}{120\pi\ell ^4 r_+} \bigg\lbrace  \frac{592\pi^2 \ell^4 r_+^2}{\beta} +162\pi\ell^2 r_+^3 \nonumber\\
	&\qquad  + \beta \left(37\ell^4-303\ell^2 r_+^2 - 192 r_+^4  \right) \bigg\rbrace\bigg]\ln \mathcal{A}_{H}, \label{ads2b}\\[7pt]
	\Delta S_{\text{BH}}^{\text{(Kerr-AdS)}}&= \bigg[-\frac{49}{36} + \frac{\beta }{48\pi \ell ^2 \left(r_+^2 + a^2\right) \left(\ell ^2-a^2\right)r_+ }\bigg\lbrace 105 r_+^6 - 68 a^4\ell^2 - 37 a^2\ell^4    \nonumber \\
	& +\left( 6 \ell ^2 + 99 a^2\right)r_+^4 +\left(37 \ell ^4 - 210 a^2 \ell ^2  + 68 a^4 \right)r_+^2   \bigg\rbrace\bigg]\ln \mathcal{A}_{H},\label{ads2c}
\end{align}}
followed by the \textit{extremal} limit results obtained as
{\allowdisplaybreaks
\begin{align}
\Delta S_{\text{BH}}^{\text{(ext,RN-AdS)}}  & = -\left[\frac{34}{9} + \frac{\left(\ell _2^4 + r_0^4\right)}{24 r_0^2 \ell _2^2}\right]\ln \mathcal{A}_{H},\label{ads2d}\\[5pt]
\Delta S_{\text{BH}}^{\text{(ext,Kerr-AdS)}}&= \bigg[-\frac{103}{36} + \frac{1}{24r_0^2 \left(\ell ^2-a^2\right) \left(a^2+r_0^2\right) \left(a^2+6 r_0^2+\ell ^2\right)} \bigg\lbrace 37 a^4\ell^4  \nonumber \\
&\quad + 315 r_0^8 + \left(624 a^2 + 6\ell^2\right)r_0^6  + \left(160 a^2\ell^2 + 229 a^4 - 37\ell^4\right)r_0^4 \nonumber \\
&\quad + \left(148 a^2\ell^4 - 142 a^4\ell^2 +  68 a^6\right)r_0^2\bigg\rbrace\bigg]\ln \mathcal{A}_{H},\label{ads2e}
\end{align}
}
\item[\scalebox{0.6}{$\blacksquare$}]{Case $\kappa=1/\sqrt{3}$:} In the finite temperature or \textit{non-extremal} limit, we find
	{\allowdisplaybreaks
	\begin{align}
	\Delta S_{\text{BH}}^{\text{(Sch-AdS)}} &= \bigg[-\frac{383}{108} + \frac{1}{108\left(3 r_+^2+\ell ^2\right)\ell ^2 } \left(945 r_+^4 + 438 \ell ^2 r_+^2  + 461 \ell ^4 \right)\bigg]\ln \mathcal{A}_{H}, \label{ads3a}\\[5pt]
	\Delta S_{\text{BH}}^{\text{(RN-AdS)}}  &= \bigg[-\frac{727}{270} + \frac{1}{1080\pi\ell ^4 r_+} \bigg\lbrace \frac{7376\pi\ell^4 r_+^2}{\beta} -1614\ell^2 r_+^3 \nonumber\\
	&\qquad + \beta \left( 461\ell^4-1959\ell^2 r_+^2 - 576r_+^4  \right) \bigg\rbrace\bigg]\ln \mathcal{A}_{H},\label{ads3b}\\[10pt]
	\Delta S_{\text{BH}}^{\text{(Kerr-AdS)}}&= \bigg[-\frac{275}{108} + \frac{\beta }{432\pi  \ell ^2 \left(r_+^2 + a^2\right) \left(\ell ^2-a^2\right)r_+ }\bigg\lbrace 945 r_+^6 -484 a^4\ell^2 - 461 a^2\ell^4   \nonumber \\
	& +\left( 438\ell ^2 + 507 a^2\right)r_+^4 +\left(461 \ell ^4 - 1890 a^2 \ell ^2 + 484 a^4 \right)r_+^2 \bigg\rbrace\bigg]\ln \mathcal{A}_{H},\label{ads3c}
\end{align}}
followed by the corrections for \textit{extremal} Reissner-Nordstr\"om-AdS and Kerr-AdS black holes as
{\allowdisplaybreaks
\begin{align}
\Delta S_{\text{BH}}^{\text{(ext,RN-AdS)}}  & = -\left[\frac{118}{27}+ \frac{73\left(\ell _2^4 + r_0^4\right)}{216 r_0^2 \ell _2^2}\right]\ln \mathcal{A}_{H},\label{ads3d}\\[5pt]
\Delta S_{\text{BH}}^{\text{(ext,Kerr-AdS)}}&= \bigg[-\frac{437}{108} + \frac{1}{216 r_0^2 \left(\ell ^2-a^2\right) \left(a^2+r_0^2\right) \left(a^2+6 r_0^2+\ell ^2\right)} \bigg\lbrace 461 a^4\ell^4  \nonumber \\
&\quad + 2835 r_0^8 + \left(5232 a^2 + 438 \ell^2\right)r_0^6  + \left(2720 a^2\ell^2 + 1037 a^4 - 461 \ell^4\right)r_0^4 \nonumber \\
&\quad + \left(1844 a^2\ell^4 - 1406 a^4\ell^2 + 484 a^6\right)r_0^2\bigg\rbrace\bigg]\ln \mathcal{A}_{H},\label{ads3e}
\end{align}}
\end{itemize}
where the AdS$_2$ radius $\ell_2= \ell\sqrt{\frac{\left(a^2 + r_0^2\right)}{\left(\ell^2 + a^2 + 6r_0^2\right)}}$ and extremal radius $r_0 = \ell\sqrt{\frac{\left(a^2 + q_e^2\right)}{\left(a^2 + \ell^2 + 3r_0^2\right)}}$. The above results are novel reports and one of the central interests of this paper. Evidently, all the logarithmic corrections for AdS$_4$ black holes are entirely non-topological, i.e., expressed in terms of different black hole parameters via the dimensionless ratios $\left\lbrace a,\, r_+,\, \ell,\, \beta\right\rbrace$ and the extremal parameters $\lbrace \ell_2,\, r_0 \rbrace$. For the present cases, the non-topological nature is induced due to the typical non-vanishing forms of integrated $R^2$ and $W_{\mu\nu\rho\sigma}W^{\mu\nu\rho\sigma}$ background invariants in both the extremal and non-extremal limits. However, inside each logarithmic correction, there exists a topological piece that is free from any dependence of associated black hole parameters and necessarily induced mostly from the two kinds of global source -- the Euler characteristic $\chi =2$ (via the integrated $E_4$ invariant) and zero-mode contributions (via $\mathcal{C}_{\text{zm}}$ data in \Cref{czero}) for the 4D black hole backgrounds. In the next subsection, we further compare these non-topological AdS$_4$ logarithmic corrections with the quantum entropy correction results of asymptotically-flat black holes.    

\subsubsection{Logarithmic corrections in $U(1)$-charged EMD theory}\label{flat}

The logarithmic correction to the entropy of asymptotically flat Schwarzschild ($q,p,a= 0$), Reissner-Nordstr\"om ($q=p,a=0$) and Kerr ($q,p=0$) black holes in EMD theory for the specific cases of dilaton couplings are calculated as follows.
\begin{itemize}
\item[\scalebox{0.6}{$\blacksquare$}]{Case $\kappa=1$:} In both the \textit{non-extremal} and \textit{extremal} limits, we obtain
\begin{align}
		\Delta S_{\text{BH}}^{\text{(Sch)}} &= \frac{13}{18}\ln \mathcal{A}_{H}, \label{flat1a}\\[5pt]
		\Delta S_{\text{BH}}^{\text{(Kerr)}} &= \frac{31}{18}\ln \mathcal{A}_{H}, \quad \Delta S_{\text{BH}}^{\text{(ext,Kerr)}}= \frac{2}{9}\ln \mathcal{A}_{H}, \label{flat1b}\\[5pt]
		\Delta S_{\text{BH}}^{\text{(RN)}} &= \left[\frac{13}{18} + \frac{3\beta q_e^4}{8\pi r_+^5}\right]\ln \mathcal{A}_{H}, \enspace
		\Delta S_{\text{BH}}^{\text{(ext,RN)}} = -\frac{163}{36}\ln \mathcal{A}_{H}, \label{flat1c}
\end{align}
\item[\scalebox{0.6}{$\blacksquare$}]{Case $\kappa=\sqrt{3}$:} In both the \textit{non-extremal} and \textit{extremal} limits, we calculate
	\begin{align}
		\Delta S_{\text{BH}}^{\text{(Sch)}} &= \frac{13}{18}\ln \mathcal{A}_{H}, \label{flat2a}\\[5pt]
		\Delta S_{\text{BH}}^{\text{(Kerr)}} &= \frac{31}{18}\ln \mathcal{A}_{H}, \quad \Delta S_{\text{BH}}^{\text{(ext,Kerr)}}= \frac{2}{9}\ln \mathcal{A}_{H}, \label{flat2b}\\[5pt]
		\Delta S_{\text{BH}}^{\text{(RN)}} &= \left[\frac{13}{18} + \frac{37\beta q_e^4}{120\pi r_+^5}\right]\ln \mathcal{A}_{H}, \enspace
		\Delta S_{\text{BH}}^{\text{(ext,RN)}} = -\frac{139}{36}\ln \mathcal{A}_{H}, \label{flat2c}
	\end{align}
\item[\scalebox{0.6}{$\blacksquare$}]{Case $\kappa=1/\sqrt{3}$:} In both the \textit{non-extremal} and \textit{extremal} limits, we find
	\begin{align}
		\Delta S_{\text{BH}}^{\text{(Sch)}} &= \frac{13}{18}\ln \mathcal{A}_{H}, \label{flat3a}\\[5pt]
		\Delta S_{\text{BH}}^{\text{(Kerr)}} &= \frac{31}{18}\ln \mathcal{A}_{H}, \quad \Delta S_{\text{BH}}^{\text{(ext,Kerr)}}= \frac{2}{9}\ln \mathcal{A}_{H}, \label{flat3b}\\[5pt]
		\Delta S_{\text{BH}}^{\text{(RN)}} &= \left[\frac{13}{18} + \frac{461\beta q_e^4}{1080\pi r_+^5}\right]\ln \mathcal{A}_{H}, \enspace
		\Delta S_{\text{BH}}^{\text{(ext,RN)}} = -\frac{545}{108}\ln \mathcal{A}_{H}, \label{flat3c}
	\end{align}
\end{itemize}
where $\beta = 4\pi r_+^3/\left(r_+^2 - q_e^2\right)$, $r_+ = m + \sqrt{m^2 - q_e^2}$  and $q_e =  \sqrt{2}q = \sqrt{2}p$. The above results for asymptotically-flat black holes are significant contributions to the novel reports of this paper. In contrast to the asymptotically-AdS cases, logarithmic corrections in the EMD models are found to be simplified and less complicated. Moreover, their relevant topological nature has been transformed drastically. All the Kerr and Schwarzschild correction results are entirely topological. However, the non-extremal Reissner-Nordstr\"om black holes have a non-topological logarithmic correction in terms of the parameters $\lbrace\beta,\, r_+,\, q_e\rbrace$, which exhibits an entirely topological form in the extremal limit. This specific nature is mainly due to the particular form of $W_{\mu\nu\rho\sigma}W^{\mu\nu\rho\sigma}$ integrated around the flat backgrounds. In particular, we have found that the integrated $W_{\mu\nu\rho\sigma}W^{\mu\nu\rho\sigma}$ invariants are universal for all flat black holes in any temperature limit except for the non-extremal charged backgrounds. The whole fact is also reflected by $\mathcal{C}_{\text{local}}$ formulas derived in \cref{nbh12,ebh11} for the flat-space limit $\ell \to \infty$. Further discussion around the logarithmic correction results computed for the black holes embedded in EMD-AdS and EMD theories is included in \cref{diss}. {Also, readers are referred to} \cref{genlog} for more generic formulas in terms of a general dilaton coupling constant parameter $\kappa$.     



\section{The $U(1)^2$-charged EMD and embedded $\mathcal{N}=4$ supergravity models}\label{EMD2}
In this section, we aim to execute similar heat kernel and logarithmic entropy correction calculations in more generalized EMD and EMD-AdS models that are directly embedded into $\mathcal{N}=4$ supergravity theories. $\mathcal{N}=4$ supergravity is the most natural model for a dimensionally reduced or compactified superstring theory in four spacetimes \cite{Becker:2006dvp,Grana:2006mg,Freedman:2012xp}. In particular, there exist some special class EMD models with two Maxwell-dilaton couplings that describe the explicit bosonic solutions of the $\mathcal{N}=4$ supergravity theories. These models, dubbed as the $U(1)^2$-charged EMD theory, are known as the most common low-energy limit of bosonic superstring theories in 4D. For a detailed structure, the readers are encouraged to review \cite{Kallosh:1992ii,Myung:2020dqt,Cvetic:2014vsa} and the citations therein. 

We will begin with the action describing a generic four-dimensional $U(1)^2$-charged EMD-AdS theory, 
\begin{align}\label{emds1}
	\mathcal{S}[g_{\mu\nu}, A_{1\mu}, A_{2\mu}, \Phi]= \int \mathrm{d}^4x \sqrt{\det g} \Big(\mathcal{R}-2\Lambda- 2D_\mu\Phi
	D^\mu\Phi- f_1(\Phi){F}_{\mu\nu}{F}^{\mu\nu} - f_2(\Phi){H}_{\mu\nu}{H}^{\mu\nu} \Big),
\end{align}
where ${F}_{\mu\nu}={\partial_{[\mu}A_{1\nu]}}$ and ${H}_{\mu\nu}={\partial_{[\mu}A_{2\nu]}}$ are the two different $U(1)$ gauge or Maxwell field strengths non-minimally coupled to the same dilaton $\Phi$ but via two separate coupling functions $f_1(\Phi)=e^{-2\kappa_1\Phi}$ and $f_2(\Phi)=e^{-2\kappa_2\Phi}$. If the constant parameters $\kappa_1$ and $\kappa_2$ controlling the `Maxwell-dilaton' coupling strengths are constrained by a special choice ($\kappa_1, \kappa_2$) $\equiv$ ($1, -1$), then that explicit $U(1)^2$-charged model corresponds to the bosonic sector of the SO(4) version\footnote{As an aside from the original SO(4) version, there exists another simpler form of the $U(1)^2$-charged EMD model that corresponds to the SU(4) version of $\mathcal{N}=4$ supergravity in 4D. However, the Maxwell fields transform via the $\text{SO(4)}\,\equiv\,\text{SU(2)}\otimes\text{SU(2)}$ group in both the SO(4) and SU(4) versions. Please review \cite{Kallosh:1992ii} for more technical and structural details.} of $\mathcal{N}=4$ supergravity describing the compactified superstring theories in four spacetimes \cite{Kallosh:1992ii,Cvetic:2014vsa}. Our particular aim is to explore logarithmic correction to the entropy of black holes in the $\mathcal{N}=4$ gauged and ungauged supergravity respectively intersecting with the $U(1)^2$-charged EMD-AdS and EMD models. Like the $U(1)$-charged EMD cases (see \cref{EMD1a}), here we also plan to embed the EM black hole backgrounds satisfying the field equations \eqref{embd5}. However, the related embedding process and background constraints are found to be distinct and illustrated as follows. 

\subsection{Embedding of Reissner-Nordstr\"om, Kerr and Schwarzschild backgrounds}\label{embedd}
The Einstein equation satisfied by the classical background $(\bar{g}_{\mu\nu},\, \bar{A}_{1\mu},\, \bar{A}_{2\mu}, \bar{\Phi})$ of the $U(1)^2$-charged EMD-AdS theory with ($\kappa_1, \kappa_2$) $\equiv$ ($1, -1$) is derived as
\begin{subequations}\label{emds2}
	\begin{align}
		R_{\mu\nu}-\frac{1}{2}\bar{g}_{\mu\nu}R + \Lambda \bar{g}_{\mu\nu} = 2 D_\mu\bar{\Phi}D_\nu \bar{\Phi} - \bar{g}_{\mu\nu}D_\rho\bar{\Phi}D^\rho \bar{\Phi} + T_{\mu\nu}^{U(1)^2},\label{emds2a}
	\end{align}
where the $U(1)^2$ or Maxwell-dilaton part of the stress-energy tensor is 
\begin{align}
T_{\mu\nu}^{U(1)^2} &=e^{-2 \bar{\Phi}}\left(2 \bar{F}_{\mu\rho}\bar{F}\indices{_\nu^\rho}-\frac{1}{2}\bar{g}_{\mu\nu}\bar{F}_{\rho\sigma}\bar{F}^{\rho\sigma}\right)+ e^{2\bar{\Phi}}\left(2\bar{H}_{\mu\rho}\bar{H}\indices{_\nu^\rho}-\frac{1}{2}\bar{g}_{\mu\nu}\bar{H}_{\rho\sigma}\bar{H}^{\rho\sigma}\right).\label{emds2b}
\end{align}
\end{subequations}  
The explicit Maxwell and Maxwell-Bianchi equations are obtained as
\begin{align}\label{emds3}
	\begin{split}
	D_\mu\bar{F}^{\mu\nu} - 2\bar{F}^{\mu\nu}D_\mu\bar{\Phi} & = 0, \enspace D_{[\mu}\bar{F}_{\rho\sigma]}=0, \\
	D_\mu\bar{H}^{\mu\nu} + 2\bar{H}^{\mu\nu}D_\mu\bar{\Phi} & = 0, \enspace D_{[\mu}\bar{H}_{\rho\sigma]}=0,
	\end{split}
\end{align}
followed by the dilaton evolution equation as
\begin{align}\label{emds4}
D_\mu D^\mu \bar{\Phi} + \frac{1}{2}\left(e^{-2\bar{\Phi}}\bar{F}_{\mu\nu}\bar{F}^{\mu\nu}-e^{2\bar{\Phi}}\bar{H}_{\mu\nu}\bar{H}^{\mu\nu}\right) =0.
\end{align}
All the above field equations will satisfy the $U(1)$-charged EMD-AdS theory \eqref{hk1} with $\kappa=1$ only when the background field strength $\bar{H}_{\mu\nu}$ is absent. However, we found that the desired EM embedding condition for the present $U(1)^2$-charged case is
\begin{align}\label{emds5}
\bar{F}_{\mu\nu}\bar{F}^{\mu\nu} = \bar{H}_{\mu\nu}\bar{H}^{\mu\nu},\enspace \bar{\Phi} = 0.
\end{align}
If the two $U(1)$-charged Maxwell fields $A_{i\mu}$ have electric charges $Q_i$ and magnetic charges $P_i$, then we can set any of the following background constraints in order to satisfy the EM embedding condition 
\begin{align}\label{emds6}
	Q_1 = Q_2, \thinspace P_1 = P_2 \enspace\, \text{or} \enspace\, Q_1 = P_1, \thinspace Q_2 = P_2.
\end{align}  
In other words, the particular $U(1)^2$ EMD model admits a general class of two-charge background solutions, which will reduce to Reissner-Nordstr\"om black holes when all the charges are equal. The uncharged Kerr and Schwarzschild backgrounds always fulfill the EM embedding condition \eqref{emds5} without any constraints.  
At any point, one can also assume the Maxwell fields $A_{1\mu}$ and $A_{2\mu}$ are either electric or magnetic and can still continue with the same embedding condition. But, without any loss of generality, we want to keep the dyonic-mode of the $U(1)$ charges to become consistent with the previous parts of this paper. The combined effect of the embedding condition \eqref{emds5} and background constraints \eqref{emds6} will transform $T_{\mu\nu}^{U(1)^2}$ into a Maxwell stress-energy tensor (e.g., the $T_{\mu\nu}^{\mathrm{(Maxwell)}}$ part in \eqref{embd1c}) with an effective Maxwell field having the electric charge $Q_1 + Q_2$ and magnetic charge $P_1 + P_2$. 
As a consequence, the $U(1)^2$-charged EMD backgrounds $(\bar{g}_{\mu\nu},\, \bar{A}_{1\mu},\, \bar{A}_{2\mu}, \bar{\Phi})$ transformed into the Reissner-Nordstr\"om, Kerr and Schwarzschild (both flat and AdS) black hole solutions satisfying the equations of motion \eqref{embd5}, which in turn justifies the EM embedding. In particular, the field equations for the EM-embedded $U(1)^2$ EMD theory are expressed as
\begin{align}\label{emds7}
	\begin{split}
		&R_{\mu\nu}- \bar{g}_{\mu\nu}\Lambda = 2 \bar{\mathcal{F}}_{\mu\rho}\bar{\mathcal{F}}\indices{_\nu^\rho}- \frac{1}{2}\bar{g}_{\mu\nu}\bar{\mathcal{F}}_{\rho\sigma}\bar{\mathcal{F}}^{\rho\sigma},\quad R = 4\Lambda,  \\
		& D_\mu \bar{F}^{\mu\nu} = 0,\enspace D_\mu \bar{H}^{\mu\nu} = 0,\enspace D_{[\mu}\bar{F}_{\rho\sigma]}=0,\enspace D_{[\mu}\bar{H}_{\rho\sigma]}=0, 
	\end{split}
\end{align}  
where $\bar{\mathcal{F}}_{\mu\rho}\bar{\mathcal{F}}\indices{_\nu^\rho} = \bar{F}_{\mu\rho}\bar{F}\indices{_\nu^\rho}+ \bar{H}_{\mu\rho}\bar{H}\indices{_\nu^\rho}$ and $\bar{\mathcal{F}}_{\rho\sigma}\bar{\mathcal{F}}^{\rho\sigma} = 2\bar{F}_{\mu\nu}\bar{F}^{\mu\nu}= 2\bar{H}_{\mu\nu}\bar{H}^{\mu\nu}$. Note that the effective Maxwell field strength $\mathcal{F}_{\mu\nu}$ satisfies the same background gauge filed \eqref{nbh5} and invariant form \eqref{nbh6} for $Q= Q_1 + Q_2$ and $P = P_1 + P_2$, with a net effective charge $Q_e \equiv \sqrt{\left(Q_1 + Q_2\right)^2 + \left(P_1 + P_2\right)^2}$. For the embedded Reissner-Nordstr\"om black hole, we will always set $Q=P$ to account for both the constraints \eqref{emds6}, which will aid us in obtaining more simplified Seeley-DeWitt coefficient and logarithmic correction results in the subsequent sections.

\subsection{Heat kernel treatment and Seeley-DeWitt coefficient}\label{sdclog}

We now proceed with the similar heat kernel treatment of \cref{EMD1a} to calculate the Seeley-DeWitt coefficient $a_4(x)$ for obtaining logarithmic corrections of embedded black holes in the $U(1)^2$-charged EMD-AdS theory. First, we fluctuate the entire content around the common background $(\bar{g}_{\mu\nu},\, \bar{A}_{1\mu},\, \bar{A}_{2\mu})$ with vanishing dilaton for small fluctuations, 
\begin{align}\label{sdclog1}
	\begin{gathered}
		g_{\mu\nu} = \bar{g}_{\mu\nu} + \sqrt{2}h_{\mu\nu}, \enspace A_\mu = \bar{A}_{1\mu} + \frac{1}{2}a_{1\mu},\enspace A_{2\mu} = \bar{A}_{2\mu} + \frac{1}{2}a_{2\mu},\\
		F_{\mu\nu} = \bar{F}_{\mu\nu}+\frac{1}{2}f_{\mu\nu},\enspace H_{\mu\nu} = \bar{H}_{\mu\nu}+\frac{1}{2}\tilde{f}_{\mu\nu},
	\end{gathered}
\end{align}
where ${f}_{\mu\nu}={D_{[\mu}a_{1\nu]}}$ and $\tilde{f}_{\mu\nu}={D_{[\mu}a_{2\nu]}}$ are strengths associated with the Maxwell fluctuations $a_{1\mu}$ and $a_{2\mu}$, respectively.  Around a vanishing background $\bar{\Phi}=0$, the dilaton acts as its own fluctuation, allowing the coupling functions $f_1(\Phi)=e^{-2\Phi}$ and $f_2(\Phi)=e^{2\Phi}$ to fluctuate perturbatively via the expansion \eqref{embd7}. We then gauge-fix the fluctuated theory by incorporating the gauge-fixing term,
\begin{align}\label{sdclog2}
	-\int \mathrm{d}^4x \sqrt{\det \bar{g}}\left[\Big(D_\mu h^{\mu\rho}-\frac{1}{2}D^\rho h\indices{^\alpha_\alpha}\Big)\Big(D^\nu h_{\nu\rho}-\frac{1}{2}D_\rho h\indices{^\beta_\beta}\Big)  + \frac{1}{2}\Big((D^\mu a_{1\mu})^2+ (D^\mu a_{2\mu})^2\Big) \right],
\end{align}  
and obtain the form of fluctuated action up to total derivatives and quadratic orders as
\begin{subequations}\label{sdclog3}
{\allowdisplaybreaks
\begin{align}\label{sdclog3a}
{\delta^2\mathcal{S}}[h_{\mu\nu}, a_{1\mu}, a_{2\mu}, \Phi]&= \frac{1}{2}\int \mathrm{d}^4x \sqrt{\det \bar{g}}\bigg[h^{\mu\nu}\left(\tilde{\Delta}h\right)_{\mu\nu}+ a_{1\mu}\left( \bar{g}^{\mu\nu} D_\rho D^\rho - R^{\mu\nu} \right)a_{1\nu} \nonumber\\
&\enspace + a_{2\mu}\left( \bar{g}^{\mu\nu} D_\rho D^\rho - R^{\mu\nu} \right)a_{2\nu} + 4\Phi D_\rho D^\rho \Phi - 4 \left( \bar{F}_{\mu\nu}\bar{F}^{\mu\nu}+ \bar{H}_{\mu\nu}\bar{H}^{\mu\nu}\right) \Phi^2 \nonumber \\
&\enspace -8\sqrt{2}\left( \bar{F}_{\mu\nu}\bar{F}\indices{_\alpha^\nu}- \bar{H}_{\mu\nu}\bar{H}\indices{_\alpha^\nu}\right)\Phi h^{\mu\alpha} + 2\sqrt{2}\left( \bar{F}_{\mu\nu}\bar{F}^{\mu\nu} - \bar{H}_{\mu\nu}\bar{H}^{\mu\nu}\right) \Phi h\indices{^\alpha_\alpha} \nonumber \\
&\enspace + 4 \left(\bar{F}_{\mu\nu}f^{\mu\nu}-\bar{H}_{\mu\nu}\tilde{f}^{\mu\nu}\right)\Phi + 4\sqrt{2}h^{\mu\nu}\left(\bar{F}_{\mu\alpha}f\indices{_\nu^\alpha}+ \bar{H}_{\mu\alpha}\tilde{f}\indices{_\nu^\alpha}\right)  \nonumber \\
& \quad - \sqrt{2}h\indices{^\rho_\rho}\left(\bar{F}_{\mu\nu} f^{\mu\nu}+ \bar{H}_{\mu\nu} \tilde{f}^{\mu\nu}\right) + \delta^2\mathscr{L}_{\text{ghost}}\bigg],
\end{align}}
where
{\allowdisplaybreaks
\begin{align}
\left(\tilde{\Delta}h\right)_{\mu\nu} &= D_\rho D^\rho h_{\mu\nu}- \frac{1}{2}\bar{g}_{\mu\nu}\bar{g}^{\alpha\beta}D_\rho D^\rho + 2R_{\mu\alpha\nu\beta}h^{\alpha\beta} -2R_{\mu\rho}h\indices{^\rho_\nu} \nonumber \\
&\quad - \left(R-2\Lambda-\bar{F}_{\rho\sigma}\bar{F}^{\rho\sigma}- \bar{H}_{\rho\sigma}\bar{H}^{\rho\sigma}\right)h_{\mu\nu} - 2R_{\mu\nu}h\indices{^\alpha_\alpha} \nonumber\\
&\quad + \frac{1}{2}\left(R - 2\Lambda-\bar{F}_{\rho\sigma}\bar{F}^{\rho\sigma}- \bar{H}_{\rho\sigma}\bar{H}^{\rho\sigma}\right)\bar{g}_{\mu\nu}h\indices{^\alpha_\alpha} - 4 \left(\bar{F}_{\mu\alpha}\bar{F}_{\nu\beta}+ \bar{H}_{\mu\alpha}\bar{H}_{\nu\beta}\right)h^{\alpha\beta} \nonumber \\
&\quad -8 \left(\bar{F}_{\alpha\mu}\bar{F}^{\alpha\beta}+ \bar{H}_{\alpha\mu}\bar{H}^{\alpha\beta}\right)h_{\beta\nu}+ 4\left(\bar{F}_{\mu\alpha}\bar{F}\indices{_\nu^\alpha}+ \bar{H}_{\mu\alpha}\bar{H}\indices{_\nu^\alpha}\right)h\indices{^\beta_\beta} , \label{sdclog3b}\\[7pt]
\delta^2\mathscr{L}_{\text{ghost}} &= b_{1\mu}\left(\bar{g}^{\mu\nu}D_\rho D^\rho + R^{\mu\nu}\right) c_{1\nu} + b_1 D_\rho D^\rho c_1  -2 b_1 \bar{F}^{\rho\nu} D_\rho c_{1\nu} \nonumber\\
&\quad + b_{2\mu}\left(\bar{g}^{\mu\nu}D_\rho D^\rho + R^{\mu\nu}\right) c_{2\nu} + b_2 D_\rho D^\rho c_2  -2 b_2 \bar{H}^{\rho\nu} D_\rho c_{2\nu}, \label{sdclog3c} 		
\end{align}}
\end{subequations}
with the vector ghosts $(b_{i\mu},\, c_{i\mu})$ and scalar ghosts $(b_{i},\, c_{i})$ associated with the same fluctuated $U(1)$ species $a_{i\mu}$ for $i=1,2$. From this stage, we adjust the quadratic action of the fluctuated $U(1)^2$-charged EMD-AdS theory with similar convention and scalings as we implemented for the $U(1)$-charged case in \cref{sdc}. This also involves the appropriate splitting of the graviton $h_{\mu\nu}$ into its trace $\hat{h}$ and traceless $\hat{h}_{\mu\nu}$ parts, using the equations of motion \eqref{emds7} and embedding conditions \eqref{emds5} at necessary places, and finally expressing the desired kinetic operator $\mathcal{H}$ operating on the fluctuations $\phi_m = \big\lbrace \hat{h}_{\mu\nu}, \hat{h}, a_{1\mu}, a_{2\mu}, \Phi \big\rbrace$ into the following Hermitian and Laplace-type form (excluding ghosts)
\begin{subequations}\label{sdclog4} 
\begin{align}\label{sdclog4a} 
\begin{split}
&\delta^2 \mathcal{S}[\hat{h}_{\mu\nu}, \hat{h}, a_{1\mu}, a_{2\mu}, \Phi] = \frac{1}{2}\int \mathrm{d}^4x \sqrt{\det \bar{g}}\, \phi_m \mathcal{H}^m_n\phi^n, \\[3pt]
& \phi_m \mathcal{H}^m_n\phi^n = \phi_m\Big(D_\rho D^\rho I^{mn} + 2(\omega_\rho D^\rho)^{mn} +  P^{mn}\Big)\phi_n,
\end{split}	 
\end{align}
where the identity or projection operators $I$ characterizing effective degrees of freedom of each fluctuations are expressed as
\begin{align}\label{sdclog4b} 
	\phi_m I^{mn}\phi_n = \hat{h}_{\mu\nu} I^{\hat{h}_{\mu\nu}\hat{h}_{\alpha\beta}} \hat{h}_{\alpha\beta} + \hat{h}\hat{h} + a_{1\mu} \bar{g}^{\mu\nu} a_{1\nu} + a_{2\mu} \bar{g}^{\mu\nu} a_{2\nu} + \Phi\Phi.
\end{align}
The components of matrices $\omega_\rho$ and $P$ controlling all minimal and non-minimal interaction data are obtained as
{\allowdisplaybreaks
\begin{align}
\phi_m (\omega^\rho)^{mn}\phi_n &= \frac{\sqrt{2}}{2} \hat{h}_{\mu\nu}\left(\bar{g}^{\mu\rho}\bar{F}^{\nu\alpha}+\bar{g}^{\nu\rho}\bar{F}^{\mu\alpha}-\bar{g}^{\mu\alpha}\bar{F}^{\nu\rho}-\bar{g}^{\nu\alpha}\bar{F}^{\mu\rho}\right) a_{1\alpha} \nonumber \\
& \quad + \frac{\sqrt{2}}{2} \hat{h}_{\mu\nu}\left(\bar{g}^{\mu\rho}\bar{H}^{\nu\alpha}+\bar{g}^{\nu\rho}\bar{H}^{\mu\alpha}-\bar{g}^{\mu\alpha}\bar{H}^{\nu\rho}-\bar{g}^{\nu\alpha}\bar{H}^{\mu\rho}\right) a_{2\alpha} \nonumber \\
& \quad - \frac{\sqrt{2}}{2} a_{1\alpha}\left(\bar{g}^{\mu\rho}\bar{F}^{\nu\alpha}+\bar{g}^{\nu\rho}\bar{F}^{\mu\alpha}-\bar{g}^{\mu\alpha}\bar{F}^{\nu\rho}-\bar{g}^{\nu\alpha}\bar{F}^{\mu\rho}\right) \hat{h}_{\mu\nu} \nonumber \\
& \quad - \frac{\sqrt{2}}{2} a_{2\alpha}\left(\bar{g}^{\mu\rho}\bar{H}^{\nu\alpha}+\bar{g}^{\nu\rho}\bar{H}^{\mu\alpha}-\bar{g}^{\mu\alpha}\bar{H}^{\nu\rho}-\bar{g}^{\nu\alpha}\bar{H}^{\mu\rho}\right) \hat{h}_{\mu\nu} \nonumber \\
&\quad + a_{1\mu} \bar{F}^{\mu\rho}\Phi -  \Phi\bar{F}^{\mu\rho}a_{1\mu} - a_{2\mu} \bar{H}^{\mu\rho}\Phi +  \Phi\bar{H}^{\mu\rho}a_{2\mu},\label{sdclog4c}  \\[9pt]
\phi_m P^{mn}\phi_n &= \hat{h}_{\mu\nu}\Big( R^{\mu\alpha\nu\beta}+R^{\mu\beta\nu\alpha}-\frac{1}{2}(\bar{g}^{\mu\alpha}R^{\nu\beta}+\bar{g}^{\nu\alpha}R^{\mu\beta}+ \bar{g}^{\mu\beta}R^{\nu\alpha}+\bar{g}^{\nu\beta}R^{\mu\alpha})\nonumber\\
&\enspace -2(\bar{F}^{\mu\alpha}\bar{F}^{\nu\beta}+\bar{F}^{\mu\beta}\bar{F}^{\nu\alpha}) -2(\bar{H}^{\mu\alpha}\bar{H}^{\nu\beta}+\bar{H}^{\mu\beta}\bar{H}^{\nu\alpha}) \nonumber\\
& \enspace - \frac{1}{2}\left(\bar{{F}}_{\rho\sigma}\bar{{F}}^{\rho\sigma}+ \bar{{H}}_{\rho\sigma}\bar{{H}}^{\rho\sigma} - 2\Lambda\right)(\bar{g}^{\mu\alpha} \bar{g}^{\nu\beta} + \bar{g}^{\mu\beta} \bar{g}^{\nu\alpha})\Big) \hat{h}_{\alpha\beta}  \nonumber \\
& \enspace + 2\hat{h}\Lambda\hat{h} - a_{1\mu} R^{\mu\nu}a_{1\nu} - a_{2\mu} R^{\mu\nu}a_{2\nu} - 2i \hat{h}_{\mu\nu} \left(\bar{F}^{\mu\alpha}\bar{F}\indices{^\nu_\alpha}+ \bar{H}^{\mu\alpha}\bar{H}\indices{^\nu_\alpha}\right) \hat{h}  \nonumber \\
&\enspace - 2i \hat{h} \left(\bar{F}^{\mu\alpha}\bar{F}\indices{^\nu_\alpha}+ \bar{H}^{\mu\alpha}\bar{H}\indices{^\nu_\alpha}\right) \hat{h}_{\mu\nu} - \Phi\left( \bar{F}_{\mu\nu}\bar{F}^{\mu\nu}+ \bar{H}_{\mu\nu}\bar{H}^{\mu\nu}\right) \Phi \nonumber\\
&\enspace -2\sqrt{2} \hat{h}_{\mu\nu}\left(\bar{F}^{\mu\alpha}\bar{F}\indices{^\nu_\alpha}- \bar{H}^{\mu\alpha}\bar{H}\indices{^\nu_\alpha}\right) \Phi -2\sqrt{2} \Phi\left(\bar{F}^{\mu\alpha}\bar{F}\indices{^\nu_\alpha}- \bar{H}^{\mu\alpha}\bar{H}\indices{^\nu_\alpha}\right)\hat{h}_{\mu\nu}  \nonumber\\
& \enspace + \frac{\sqrt{2}}{2}\hat{h}_{\mu\nu}\left( D^\mu \bar{F}^{\alpha\nu} + D^\nu \bar{F}^{\alpha\mu} \right)a_{1\alpha} + \frac{\sqrt{2}}{2}a_{1\alpha}\left( D^\mu \bar{F}^{\alpha\nu} + D^\nu \bar{F}^{\alpha\mu} \right)\hat{h}_{\mu\nu}\nonumber \\
& \enspace + \frac{\sqrt{2}}{2}\hat{h}_{\mu\nu}\left( D^\mu \bar{H}^{\alpha\nu} + D^\nu \bar{H}^{\alpha\mu} \right)a_{2\alpha} + \frac{\sqrt{2}}{2}a_{2\alpha}\left( D^\mu \bar{H}^{\alpha\nu} + D^\nu \bar{H}^{\alpha\mu} \right)\hat{h}_{\mu\nu}. \label{sdclog4d} 
\end{align}}
\end{subequations} 
With the help of the above relations, we further determine the important matrices $E$, $E^2$ and $\Omega_{\rho\sigma}\Omega^{\rho\sigma}$ needed in the heat kernel method of \cref{local}. We then calculated the necessary traces for computing the $a_4(x)$ coefficients using the formula \eqref{comp5}. These trace calculations are extremely tedious and complicated compared to the $U(1)$-charged EMD case. We want to refer the readers to \cref{calcul} for some explicit calculation details and relevant on-shell identities used to simplify the traces in $U(1)^2$-charged EMD-AdS theory. Here we quote the calculated trace results in the following simplified forms without mentioning any other intermediate steps and details,
{\allowdisplaybreaks
\begin{align}\label{sdclog5} 
\text{Tr}(I) &= 9 + 1 + 4 + 4 + 1 = 19, \nonumber\\[5pt]
\text{Tr}(E) &= -12\Lambda + 7 \left(\bar{F}_{\mu\nu}\bar{F}^{\mu\nu}+ \bar{H}_{\mu\nu}\bar{H}^{\mu\nu}\right), \nonumber\\[5pt]
\text{Tr}(E^2) &= 3 R_{\mu\nu\rho\sigma}R^{\mu\nu\rho\sigma}-\frac{27}{4}R_{\mu\nu}R^{\mu\nu} + 35\Lambda^2+ \frac{37}{2}\left(\bar{F}_{\mu\nu}\bar{F}^{\mu\nu}\right)^2 + \frac{37}{2} \left(\bar{H}_{\mu\nu}\bar{H}^{\mu\nu}\right)^2 \nonumber\\
&\quad -20\Lambda\bar{F}_{\mu\nu}\bar{F}^{\mu\nu} -20\Lambda \bar{H}_{\mu\nu}\bar{H}^{\mu\nu}+ 3R_{\mu\nu\rho\sigma}\bar{F}^{\mu\nu}\bar{F}^{\rho\sigma}+ 3R_{\mu\nu\rho\sigma}\bar{H}^{\mu\nu}\bar{H}^{\rho\sigma}, \nonumber\\[5pt]
\text{Tr}\left(\Omega_{\rho\sigma}\Omega^{\rho\sigma}\right) &= -8 R_{\mu\nu\rho\sigma}R^{\mu\nu\rho\sigma} + \frac{117}{2}R_{\mu\nu}R^{\mu\nu} -234\Lambda^2 - 111\left(\bar{F}_{\mu\nu}\bar{F}^{\mu\nu}\right)^2 - 111 \left(\bar{H}_{\mu\nu}\bar{H}^{\mu\nu}\right)^2 \nonumber\\
&\quad + 64\Lambda\bar{F}_{\mu\nu}\bar{F}^{\mu\nu} + 64\Lambda \bar{H}_{\mu\nu}\bar{H}^{\mu\nu} -18R_{\mu\nu\rho\sigma}\bar{F}^{\mu\nu}\bar{F}^{\rho\sigma} -18 R_{\mu\nu\rho\sigma}\bar{H}^{\mu\nu}\bar{H}^{\rho\sigma}. 
\end{align}}
The above data excluded the contributions of the ghost part \eqref{sdclog3c}, where each set of ghosts associated with the two Maxwell fluctuations are non-interacting but minimally coupled to the graviton fluctuation. Thus, we can progress like the $U(1)$-charged EMD-AdS case (see \cref{sdc24} onward). As expected, the total ghost trace contributions are found to be exactly twice the results \eqref{sdc28}, i.e.,
{\allowdisplaybreaks
\begin{align}\label{sdclog6}
	\begin{split}
		\text{Tr}(I) &= 2\times (4 + 4 + 1 + 1) = 20, \\[5pt]
		\text{Tr}(E) &= 16\Lambda, \quad \text{Tr}(E^2) = 4 R_{\mu\nu}R^{\mu\nu}, \\[5pt]
		\text{Tr}\left(\Omega_{\rho\sigma}\Omega^{\rho\sigma}\right) &= -4 R_{\mu\nu\rho\sigma}R^{\mu\nu\rho\sigma}.
	\end{split}
\end{align}  }
Finally, we end up calculating the net $a_4(x)$ coefficient by inserting both the trace data \eqref{sdclog5} and \eqref{sdclog6} into the Seeley-DeWitt formula\eqref{comp5}. The ghost traces are associated with $\chi=-1$ to account for the bosonic and scalar ghost fields. This yields,
\begin{align}\label{sdclog7}
(4\pi)^2 {a_4}^{\text{$U(1)^2$-EMD}}(x) &= \frac{209}{180}R_{\mu\nu\rho\sigma}R^{\mu\nu\rho\sigma} - \frac{89}{180}R_{\mu\nu}R^{\mu\nu} - \frac{188}{9}\Lambda^2.
\end{align} 
The same for the $U(1)^2$-charged EMD theory in flat space is obtained by setting $\Lambda=0$, or one can proceed without the cosmological constant $\Lambda$ from the starting point. It is not trivial to simply generalize the $U(1)$-charged EMD result \eqref{sdc29} and derive the ${a_4}^{\text{$U(1)^2$-EMD}}(x)$ coefficient because of the contrary nature of both dilaton coupling functions and the difference in off-shell degrees of freedom after fluctuating the $U(1)^2$-charged EMD-AdS theory \eqref{emds1}. Here one must notice an interesting fact: all the irreducible background invariant terms proportional to $\bar{F}_{\mu\nu}$ and $\bar{H}_{\mu\nu}$ are completely rescinded inside the formula \eqref{comp5}, although they were present in the simplified trace data \eqref{sdclog5}. This indicates the ${a_4}^{\text{$U(1)^2$-EMD}}(x)$ coefficient preserves the electromagnetic duality, just like we have seen for the $U(1)$-charged EMD case (see \cref{sdc29}). This in turn assures that the EM black hole backgrounds are appropriately embedded into the $U(1)^2$-charged EMD theory as well as checks the consistency of the calculated Seeley-DeWitt result \eqref{sdclog7}. 

\subsection{Logarithmic entropy corrections in bosonic $\mathcal{N}=4$ ungauged and gauged supergravity}\label{U2log}
We will now examine the implications of ${a_4}^{\text{$U(1)^2$-EMD}}(x)$ data \eqref{sdclog7} in calculating the logarithmic correction to the entropy of flat and AdS black holes embedded in the $U(1)^2$-charged EMD models \eqref{emds1} intersect with the SO(4) version of $\mathcal{N}=4$ supergravity in 4D \cite{Kallosh:1992ii,Myung:2020dqt,Cvetic:2014vsa}. In particular, the Schwarzschild-AdS, Reissner-Nordstr\"om-AdS and Kerr-AdS black holes, embedded into the $U(1)^2$-charged EMD-AdS theory, represent the background solutions in the bosonic sector of $\mathcal{N}=4$ gauged supergravity. On the other hand, the asymptotically-flat Schwarzschild, Reissner-Nordstr\"om and Kerr black holes, embedded into the $U(1)^2$-charged EMD theory, are the exact bosonic solutions of $\mathcal{N}=4$ ungauged supergravity. The specific choice of embeddings (see \cref{embedd}) suggests that the background setup and thermodynamic behavior of the embedded black are identical to the $U(1)$-charged EMD-AdS and EMD cases we have demonstrated in \cref{EMD1b}. Thus for the present case, we are allowed to proceed with the same treatment, $\mathcal{C}_{\text{local}}$ formulas derived in \cref{nbh11,nbh12,ebh5,ebh11}, and the global $\mathcal{C}_{\text{zm}}$ contributions listed in \Cref{czero}. In addition, it is essential to extract the following central charges and trace anomaly data (as defined in \cref{emd1,emd2})   
\begin{align}\label{sdclog8}
c_{\text{A}} &= \frac{83}{40},\quad a_{\text{A}} = \frac{329}{360},\quad b_{\text{A}} = -\frac{13}{12}.
\end{align}
Then we calculate the logarithmic corrections to the entropy of \textit{non-extremal} Schwarzschild-AdS ($q,p,a= 0$), Reissner-Nordstr\"om-AdS ($q=p,a= 0$) and Kerr-AdS ($q,p= 0$) black holes. The results are
{\allowdisplaybreaks
\begin{align}
\Delta S_{\text{BH}}^{\text{(Sch-AdS)}} &= \bigg[-\frac{599}{180} + \frac{1}{20\left(3 r_+^2+\ell ^2\right)\ell ^2 } \left(343r_+^4 -94\ell ^2 r_+^2  + 83\ell ^4 \right)\bigg]\ln \mathcal{A}_{H}, \label{sdclog9a}\\[5pt]
\Delta S_{\text{BH}}^{\text{(RN-AdS)}}  &= \bigg[-\frac{562}{225} +  \frac{1}{200\pi\ell ^4 r_+} \bigg\lbrace  \frac{1328\pi^2 \ell^4 r_+^2}{\beta} + 1438\pi\ell^2 r_+^3  \nonumber \\
&\qquad + \beta\left(83\ell^4 -1217\ell^2 r_+^2  - 968 r_+^4  \right) \bigg\rbrace\bigg]\ln \mathcal{A}_{H}, \label{sdclog9b}\\[7pt]
\Delta S_{\text{BH}}^{\text{(Kerr-AdS)}}&= \bigg[-\frac{419}{180} + \frac{\beta }{80\pi  \ell ^2 \left(r_+^2 + a^2\right) \left(\ell ^2-a^2\right)r_+ }\bigg\lbrace 343 r_+^6 -260 a^4\ell^2 - 83 a^2\ell^4    \nonumber \\
&\qquad -\left( 94\ell ^2 - 437 a^2\right)r_+^4 +\left(83 \ell ^4 - 686 a^2 \ell ^2  + 260 a^4 \right)r_+^2 \bigg\rbrace\bigg]\ln \mathcal{A}_{H}.\label{sdclog9c}	
\end{align}}
The same for Reissner-Nordstr\"om-AdS and Kerr-AdS black holes in \textit{extremal} limit are obtained as
{\allowdisplaybreaks
\begin{align}
\Delta S_{\text{BH}}^{\text{(ext,RN-AdS)}}  & = -\left[\frac{101}{18} - \frac{47\left(\ell _2^4 + r_0^4\right)}{120 r_0^2 \ell _2^2} \right]\ln \mathcal{A}_{H}, \label{sdclog9d}\\[5pt]
\Delta S_{\text{BH}}^{\text{(ext,Kerr-AdS)}}&= \bigg[-\frac{689}{180} + \frac{1}{40r_0^2 \left(\ell ^2-a^2\right) \left(a^2+r_0^2\right) \left(a^2+6 r_0^2+\ell ^2\right)} \bigg\lbrace 83 a^4\ell^4 + 1029 r_0^8  \nonumber \\
&\qquad  + \left(2152 a^2 -94 \ell^2\right)r_0^6  + \left(144 a^2\ell^2 + 1051 a^4 - 83\ell^4\right)r_0^4 \nonumber \\
&\qquad + \left(332 a^2\ell^4 - 426 a^4\ell^2 + 260 a^6\right)r_0^2\bigg\rbrace\bigg]\ln \mathcal{A}_{H},\label{sdclog9e}
\end{align}
}
where $\ell_2= \ell\sqrt{\frac{\left(a^2 + r_0^2\right)}{\left(\ell^2 + a^2 + 6r_0^2\right)}}$ and $r_0 = \ell\sqrt{\frac{\left(a^2 + q_e^2\right)}{\left(a^2 + \ell^2 + 3r_0^2\right)}}$.
Similarly, in the flat space limit ($\ell \to \infty$), the logarithmic correction formulas for Schwarzschild ($q,p,a= 0$), Reissner-Nordstr\"om ($q=p,a=0$) and Kerr ($q,p=0$) black holes are computed in both \textit{non-extremal} and \textit{extremal} limits of their temperature. The results are expressed in the following simplified forms
\begin{align}
	\Delta S_{\text{BH}}^{\text{(Sch)}} &= \frac{37}{45}\ln \mathcal{A}_{H}, \label{sdclog10a}\\[5pt]
	\Delta S_{\text{BH}}^{\text{(Kerr)}} &= \frac{82}{45}\ln \mathcal{A}_{H}, \label{sdclog10b}\\[5pt]
	\Delta S_{\text{BH}}^{\text{(RN)}} &= \left[\frac{37}{45} + \frac{83\beta q_e^4}{200\pi r_+^5}\right]\ln \mathcal{A}_{H}, \label{sdclog10c}\\[5pt]
	\Delta S_{\text{BH}}^{\text{(ext,Kerr)}} &= \frac{29}{90}\ln \mathcal{A}_{H}, \label{sdclog10d}\\[5pt]
	\Delta S_{\text{BH}}^{\text{(ext,RN)}} &= -\frac{869}{180}\ln \mathcal{A}_{H}, \label{sdclog10e}
\end{align}  
where $\beta = 4\pi r_+^3/\left(r_+^2 - q_e^2\right)$, $r_+ = m + \sqrt{m^2 - q_e^2}$  and $q_e =  \sqrt{2}q = \sqrt{2}p$. The nature of the above quantum corrections is identical to what we found for the black holes for the $U(1)$-charged models in \cref{results}. The relevant coefficients of all the AdS corrections are non-topological. In contrast, the logarithmic corrections for asymptotically-flat black holes are entirely topological and free from any dependence on black hole parameters, except for the non-extremal charged Reissner-Nordstr\"om black hole. All the calculated flat and AdS correction results for the bosonic $\mathcal{N}=4$ ungauged and gauged SO(4) supergravity are the main focus of this paper and novel reports. The extremal Reissner-Nordstr\"om logarithmic correction relation \eqref{sdclog10e} must be in alliance with the appropriate dilaton part in the bosonic contribution of the work \cite{Banerjee:2011pp}, where the relevant computation approach relies on the eigenfunction expansion method that is exclusive for the AdS$_2 \times S^2$ backgrounds.     


\section{Discussion and outlook}\label{diss}

In summary, we have explored logarithmic correction to the entropy flat and AdS black holes embedded in $U(1)$ and $U(1)^2$-charged EMD theories as the most ubiquitous building blocks of compactified or dimensionally-reduced superstring models, supergravity and Kaluza-Klein theories in 4D \cite{Gibbons:1982ih,Gibbons:1987ps,Garfinkle:1990qj,Horowitz:1991cd,Kallosh:1992ii,Khuri:1995xk,Gao:2004tv,Becker:2006dvp,Cvetic:2014vsa,GoulartSantos:2017dun,Guo:2021zed,Zhang:2021edm,Myung:2020dqt}. In particular, we first investigated the three specific cases $\kappa=1$, $\kappa=\sqrt{3}$ and $\kappa=\frac{1}{\sqrt{3}}$ of the $U(1)$ EMD models that are relevant in string theory. Next, as a concrete example, we have calculated the leading quantum corrections for a special case ($\kappa_1, \kappa_2$) $\equiv$ ($1, -1$) of the $U(1)^2$ EMD models that directly intersect with the bosonic sector of a SO(4) version of $\mathcal{N}=4$ supergravity \cite{Kallosh:1992ii,Myung:2020dqt,Cvetic:2014vsa}. For the non-extremal black holes, we cast the standard Euclidean quantum gravity approach developed in \cite{Sen:2013ns}, which has been so successful for asymptotically-flat black holes (e.g., see \cite{Charles:2015nn,Castro:2018tg,Karan:2021teq,Banerjee:2021pdy,Karan:2020sk}) and also extended for the case of asymptotically-AdS black holes in this paper. On the other hand, quantum entropy function (QEF) formalism is a powerful Euclidean gravity avatar, which has been well-investigated for the extremal black holes in the flat spacetimes and provided consistent results matching with the available microscopic counting data. This motivated us to employ the same QEF prescription for analyzing all extremal AdS$_4$ black holes in this paper. 

However, there are reports of a few instances \cite{Liu:2017vll,Jeon:2017ij} where the AdS$_4$ logarithmic corrections obtained from the extremal near-horizon analysis suffer a mismatch with the results of field theory computations. Contrarily in \cite{Liu:2017vbl}, a full geometry treatment over extremal AdS$_4$ black holes exhibited an exact agreement. Later, David et al. in \cite{David:2021eoq} resolved this puzzle and pointed out that the $\mathcal{C}_{\text{local}}$ contributions remain the same for all the treatments with full geometry or near-horizon of extremal black holes (even for AdS backgrounds). It is the $\mathcal{C}_{\text{zm}}$ contributions that explicitly differ the total logarithmic correction results when one proceeds via the different treatments over the two parts of extremal black hole geometry. The whole fact has been verified in \cref{ebh}, where we found identical $\mathcal{C}_{\text{local}}$ formulas via successively proceeding with the QEF prescription (i.e., extremal near-horizon analysis) and setting $\beta \to \infty$ limit over the full geometry relations of finite-temperature. Although we believe QEF formalism is fundamental due to its high conquest, which allows presenting the final relations in \cref{results,U2log} by incorporating zero-mode data from the near-horizon. But, as per the requirement at any point, one can always utilize the full geometry zero-mode data, as discussed in \cref{zeromode}. In the future, it would be interesting to counter the question of whether the degrees of freedom underlying the zero-mode part of quantum entropy for extremal AdS black holes live in near-horizon, full geometry, or somewhere else. This progress may involve finding the correct choice of ensemble and scalings for the extremal AdS black hole backgrounds.

We revisited the feature that only the third-order Seeley-DeWitt coefficient $a_4(x)$ encoding all trace anomaly and central charge data is required for computing the logarithmic corrections in 4D. We computed them by fluctuating the EMD content around Reissner-Nordstr\"om, Kerr and Schwarzschild black holes as the embedded EM backgrounds. The final $a_4(x)$ forms \eqref{sdc29} and \eqref{sdclog7} are managed only in background invariants where all the Maxwell or $U(1)$ gauge field strength terms are canceled out. This proves the heat kernel results are invariant under electromagnetic duality rotation, which justifies the EM embedding and checks the consistency of underlying trace calculations. In fact, the generic ${a_4}^{\text{$U(1)$-EMD}}$ formula \eqref{sdc29} is found to be perfectly in alliance with the result available in \cite{Castro:2018tg} for a pure 4D Kaluza-Klein system having $\kappa= \sqrt{3}$ and $\Lambda=0$. To ensure more accuracy of the delicate Seeley-DeWitt trace computations, we have progressed independently by hand calculation and developing Mathematica algorithms using xAct \cite{Garcia:2002wp} and xPert \cite{Brizuela:2008ra}. All these activities combinedly boosted our confidence in the consistency of the novel $a_4(x)$ and logarithmic correction results reported in this paper. Finally, we want to emphasize that the EM embedding into EMD theories is encountered by constraining the Maxwell background with equal charges (i.e., $Q=P$) for casting a vanishing dilaton background. In the future, we hope to overcome the challenges of looking beyond the EM-embedding or $Q=P$ limit and explore the quantum black hole entropy in all EMD models with a non-vanishing dilaton background.

All the logarithmic corrections in \cref{results,U2log} are obtained by integrating the relevant $a_4(x)$ invariants, including the Euler and Weyl trace anomalies, around the background of concerned EM black hole backgrounds. In this process, we used the prescription of holographic renormalization \cite{Skenderis:2002wp} to regulate the divergences of AdS$_4$ backgrounds. This regularization choice is natural and consistent with the standard Gauss-Bonnet-Chern theorem \cite{Chern:1945wp} (see \cite{David:2021eoq} for more details). As a result, the leading entropy corrections reported in this paper for AdS$_4$ black holes are physically sensible and unambiguous. We have also verified that the integrated AdS$_4$ invariants perfectly matched the known relations in \cite{Sen:2013ns,Charles:2015nn,Karan:2020sk,Karan:2021teq,Bhattacharyya:2012ss} for asymptotically-flat black holes in the limit $\ell \to \infty$. Additionally, the holographic renormalization procedure is always found to provide the correct Euler characteristic value $\chi =2$ via integrating the Euler density $E_4$ around the AdS$_4$ black hole geometries. Similarly, while proceeding via the QEF formalism for extremal black holes, the logarithmic corrections received contributions only from the cut-off independent or finite piece in the AdS$_2$ part of near-horizon. All this guarantees the extremal logarithmic correction results reported in this paper are genuine and robust. 

Finally, we want to comment on the ``universality'' status of the explicit logarithmic corrections and their novel implications in future progress. The coefficient of logarithmic corrections generally depends on both the field content or central charges of theory and geometric parameters of the related black holes. Sometimes they avoid the dependence on black hole parameters, then the logarithmic corrections are recognized as topological or fully universal. Such a universal form of logarithmic corrections is expected since all the available microstate counting and supergravity localization computing examples \cite{Strominger:1996sh,Maldacena:1996gb,Horowitz:1996fn,Emparan:2006it,Mandal:2010cj,Sen:2014aja,Belin:2016knb,Benini:2019dyp,Gang:2019uay,PandoZayas:2020iqr,Liu:2017vbl,Liu:2017vll,Benini:2015eyy} are indeed pure numbers, i.e., topological. In this paper, we found that all the logarithmic correction results for asymptotically-AdS$_4$ black holes are non-topological, while the same is entirely universal or topological for the flat backgrounds except the non-extremal charged Reissner-Nordstr\"om black hole. {Here notice that} the extremal limit is fully ensuring a confirmed topological nature for the logarithmic corrections to the entropy of asymptotically-flat black holes, in contrast to the extremal asymptotically-AdS black holes. One can also realize this via analyzing the AdS/Kaluza-Klein (KK) scale separation conjecture, as stressed in \cite{Tsimpis:2012tu,Gautason:2015tig,Lust:2019zwm}. For the AdS vacua, there is no scale separation between the scales characterizing the AdS background and the internal manifold containing all the KK tower of modes. Thus the AdS backgrounds are not well controllable since there is no low-energy limit in which all KK modes can be decoupled or neglected. Consequently, the nature of related logarithmic corrections for the asymptotically-AdS black holes becomes so robust that even the extremal limit cannot ensure them a topological character. On the other hand, the Minkowski or flat backgrounds are automatically scale separated and well controllable, where KK modes can be safely neglected by setting the compactification radius to be very small. This activity is perfectly consistent with setting the extremal limit on the asymptotically-flat black holes and achieving a guaranteed topological nature of the related {logarithmic correction results}. We can therefore surmise that the ubiquitous non-topological piece in AdS$_4$ logarithmic correction is arising due to the appearance of an additional boundary that is also sensitive to microscopic details. Thus, we conclude that the AdS$_4$ logarithmic corrections in the low-energy $U(1)$ and $U(1)^2$ EMD models encode a lot more information than their flat-space counterparts, serving a much wider ``infrared window into the microstates''.

{The non-topological} logarithmic corrections are very generic and natural in all even-dimensional spacetimes. In odd dimensions, the local contribution trivially vanishes, guaranteeing a topological or universal character of the logarithmic correction. For example, in the one-loop setup of the present paper, the local contribution of logarithmic corrections is determined by the Seeley-DeWitt coefficient $a_D(x)$ for $D$-dimensional spacetimes. In odd $D$-dimensional theories, all the background curvature invariants evaluating $a_D(x)$ vanish due to the lack of diffeomorphism invariant scalar functions connected to the background metric. As a result, the $\mathcal{C}_{\text{local}}$ piece in \eqref{comp1} becomes zero, providing $\Delta S_{\text{BH}}$ that is entirely controlled by the topological zero-mode contributions $\mathcal{C}_{\text{zm}}$. Contrarily, the $a_D(x)$ coefficients (as well as $\mathcal{C}_{\text{local}}$ contributions) in even $D$-dimensions are in principle non-vanishing, hence providing the generic non-topological character to the logarithmic correction in terms of a rather non-trivial function of black hole parameters. This evidently justifies the non-topological nature of logarithmic corrections reported for AdS$_4$ black holes in this paper, where the parent theories are four-dimensional EMD models. Similar nature is also confirmed by the logarithmic corrections calculated for AdS$_4$ black holes embedded in four-dimensional minimal $\mathcal{N}=2$ gauged supergravity \cite{David:2021eoq} using the same one-loop and heat kernel setup as in this paper. 

{It would be fruitful} to test whether the above-mentioned expectation of topological vs. non-topological character of logarithmic correction is fulfilled by the results available via various other computation approaches in different dimensions. We note that the microscopic computations executed in eleven-dimensional supergravity \cite{Benini:2019dyp,Gang:2019uay,PandoZayas:2020iqr,Liu:2017vbl} have confirmed the topological nature of logarithmic entropy corrections. However, the microscopic analysis for the AdS$_4$ black hole embedded in ten-dimensional  theories, such as massive IIA supergravity, exhibits a contrasting character. In this progress, a consistent matching of the Bekenstein-Hawking formula was achieved at the leading order \cite{Azzurli:2017kxo,Hosseini:2017fjo,Benini:2017oxt}, while the subleading logarithmic correction term appears to be topological \cite{Liu:2018bac}. However, the outcome of these supergravity computations should be in agreement with the non-trivial character of $\mathcal{C}_{\text{local}}$ since the parent theory is even-dimensional. To resolve this tension, we believe that the matter multiplets arising from the full KK tower of modes need to be included while embedding the black holes in such higher-dimensional theories, which might remedy the logarithmic correction to be non-topological in ten-dimensional supergravities. 

{Supergravity localization} is another powerful treatment for finding the full quantum black hole entropy. It would be illuminating to compare some recent investigations \cite{Hristov:2018lod,Hristov:2019xku,Hristov:2021zai} in this line with the one-loop computation results achieved in this paper as well as in \cite{David:2021eoq} for AdS black holes embedded in four-dimensional (super-)gravity theories. For example in \cite{Hristov:2021zai}, the logarithmic correction to the entropy of BPS black holes in four-dimensional $\mathcal{N}=2$ gauged supergravity is studied via the localization of QEF. Here the logarithmic correction is interpreted as the Atiyah-Singer index of an appropriate supercharge, where it has been shown that a topological or universal piece is emerging from the Euler term. This clearly contradicts \cite{David:2021eoq} and our results since we expect a non-topological logarithmic correction for AdS black holes embedded in any 4D theory. However, David et al. in \cite{David:2021eoq} already hinted at a possible resolution of this mismatch where one needs to include the contribution from the so-called $\eta$-invariant, not considered in \cite{Hristov:2021zai}, which is a non-topological correction due to the presence of a boundary \cite{Atiyah:1975jf}. In addition, we believe the contributions of auxiliary fields in off-shell supergravity and the full KK tower of modes could also prove handy in rescuing the appropriate character of logarithmic corrections for such supergravity computations.

Arguably, we should not consider the topological or universal nature of logarithmic corrections as an explicit criterion for an effective theory (macroscopic or gravity side) to appear as the low-energy limit of UV-complete counterparts (microscopic or UV side). 
In principle, the universal or topological criteria would strongly constrain any low-energy effective model and its black hole backgrounds. For the specific 4D EMD cases of this paper, we either require the integrated $W_{\mu\nu\rho\sigma}W^{\mu\nu\rho\sigma}$ and $R^2$ as topological or need to set $c_A = b_A =0$. The ungauged supergravities always guarantee a universal logarithmic correction by providing $c_A =0$ via anomaly cancellations between the bosonic and fermionic degrees of freedom in 4D \cite{Charles:2015nn,Karan:2020sk}.\footnote{The asymptotically-flat backgrounds automatically set $b_A =R^2 = 0$.} Similar universal or topological nature is also ensured by the extremal non-rotating and non-extremal uncharged black holes in asymptotically-flat space due to integrating out a vanishing and numerical $W_{\mu\nu\rho\sigma}W^{\mu\nu\rho\sigma}$ contribution, respectively (e.g., see \cite{Karan:2021teq} or use appropriate flat-space limits in the formulas of \cref{holo}). Therefore, when we look beyond the topological or universal limit, for example the AdS$_4$ results in this paper, the non-topological logarithmic corrections appear as a generic probe of whether a low-energy effective theory can admit the UV complete microscopic counterpart. In the future, all these reported logarithmic correction results in the EMD theories and their universality (topological vs. non-topological) status will serve as a strong and wider macroscopic window into the microstates of gravity models and black holes in string theory. 

\acknowledgments

The authors express their gratitude to Suvankar Dutta for his encouragement during the course of this research. SK would like to thank Binata Panda, Shamik Banerjee, Abhishek Chowdhury and Arnab Priya Saha for various important discussions and insights. SK would like to acknowledge IIT(ISM) Dhanbad for their hospitality during the final stage of this research. This work of SK is supported by IISER Bhopal.

\appendix

\section{Heat kernel trace calculations and on-shell identities}\label{calcul}
The useful on-shell equations of motion for the EM backgrounds embedded into $U(1)^2$-charged EMD-AdS theory or bosonic $\mathcal{N}=4$ supergravity \eqref{emds1} are listed as
{
\allowdisplaybreaks
	\begin{align}
\bar{F}_{\mu\rho}{\bar{F_\nu}}^\rho + \bar{H}_{\mu\rho}\bar{H}\indices{_\nu^\rho} &= \frac{1}{2}R_{\mu\nu} -\frac{1}{8}\bar{g}_{\mu\nu}R +\frac{1}{4}\bar{g}_{\mu\nu}\left(\bar{F}_{\rho\sigma}\bar{F}^{\rho\sigma}+ \bar{H}_{\rho\sigma}\bar{H}^{\rho\sigma}\right), \label{calcul1a}\\[4pt]
\bar{H}_{\mu\nu}\bar{H}^{\mu\nu} &= \bar{F}_{\mu\nu}\bar{F}^{\mu\nu},\enspace R= 4\Lambda,\label{calcul1b}\\[4pt]
D_\mu \bar{F}^{\mu\nu} &=0, \enspace D_\mu \bar{H}^{\mu\nu} =0, \label{calcul1c}\\[4pt]
D_{[\mu}\bar{F}_{\nu\rho]} &=0,\enspace D_{[\mu}\bar{H}_{\nu\rho]}=0,\label{calcul1d}\\[4pt]
\left(\bar{F}_{\mu\nu}\bar{F}^{\mu\nu}\right)^2 &= \left(\bar{H}_{\mu\nu}\bar{H}^{\mu\nu}\right)^2 = \bar{F}_{\mu\nu}\bar{F}^{\mu\nu}\bar{H}_{\rho\sigma}\bar{H}^{\rho\sigma}= \bar{F}_{\mu\nu}\bar{H}^{\mu\nu}\bar{F}_{\rho\sigma}\bar{H}^{\rho\sigma}.\label{calcul1e}
\end{align}}
With the help of the above evolution equations and the gravitational Bianchi $R_{\mu[\nu\rho\sigma]}=0$, we derive the following induced on-shell identities
{
\allowdisplaybreaks
\begin{align}
R_{\mu\rho\nu\sigma}R^{\mu\nu\rho\sigma}&=\frac{1}{2}R_{\mu\nu\rho\sigma}R^{\mu\nu\rho\sigma}, \label{calcul2a}\\[3pt]
{R}_{\mu\nu}\bar{F}^{\mu\rho}\bar{F}\indices{^\nu_\rho}&= {R}_{\mu\nu}\bar{H}^{\mu\rho}\bar{H}\indices{^\nu_\rho}= \frac{1}{4}R_{\mu\nu}R^{\mu\nu}-\Lambda^2+ \frac{1}{2}\Lambda\left(\bar{F}_{\mu\nu}\bar{F}^{\mu\nu}+ \bar{H}_{\mu\nu}\bar{H}^{\mu\nu}\right), \label{calcul2b}\\[3pt]
R_{\mu\rho\nu\sigma}\bar{F}^{\mu\nu}\bar{F}^{\rho\sigma}&= R_{\mu\rho\nu\sigma}\bar{H}^{\mu\nu}\bar{H}^{\rho\sigma} = \frac{1}{2}R_{\mu\nu\rho\sigma}\bar{F}^{\mu\nu}\bar{F}^{\rho\sigma}=\frac{1}{2}R_{\mu\nu\rho\sigma}\bar{H}^{\mu\nu}\bar{H}^{\rho\sigma}, \label{calcul2c}\\[3pt]
\left(D_\rho \bar{F}_{\mu\nu}\right)\left(D^\rho \bar{F}^{\mu\nu}\right) &= 2\left(D_\mu \bar{F}\indices{_\rho^\nu}\right)\left(D_\nu \bar{F}^{\rho\mu}\right) \nonumber\\
&= R_{\mu\nu\rho\sigma}\bar{F}^{\mu\nu}\bar{F}^{\rho\sigma}-\frac{1}{2}R_{\mu\nu}R^{\mu\nu}+ 2\Lambda^2- \Lambda\left(\bar{F}_{\mu\nu}\bar{F}^{\mu\nu}+ \bar{H}_{\mu\nu}\bar{H}^{\mu\nu}\right),\label{calcul2d}\\[3pt]
\left(D_\rho \bar{H}_{\mu\nu}\right)\left(D^\rho \bar{H}^{\mu\nu}\right) &= 2\left(D_\mu \bar{H}\indices{_\rho^\nu}\right)\left(D_\nu \bar{H}^{\rho\mu}\right) \nonumber\\
&= R_{\mu\nu\rho\sigma}\bar{H}^{\mu\nu}\bar{H}^{\rho\sigma}-\frac{1}{2}R_{\mu\nu}R^{\mu\nu}+ 2\Lambda^2- \Lambda\left(\bar{F}_{\mu\nu}\bar{F}^{\mu\nu}+ \bar{H}_{\mu\nu}\bar{H}^{\mu\nu}\right), \label{calcul2e}\\
\bar{F}^{\mu\rho}\bar{F}\indices{^\nu_\rho} \bar{F}_{\mu\sigma}\bar{F}\indices{_\nu^\sigma} &= \bar{H}^{\mu\rho}\bar{H}\indices{^\nu_\rho} \bar{H}_{\mu\sigma}\bar{H}\indices{_\nu^\sigma}\nonumber\\
& = \bar{F}^{\mu\rho}\bar{H}\indices{^\nu_\rho} \bar{H}_{\mu\sigma}\bar{F}\indices{_\nu^\sigma}\nonumber\\
&= \frac{1}{16}R_{\mu\nu}R^{\mu\nu}-\frac{1}{4}\Lambda^2  + \frac{1}{8}\left(\bar{F}_{\mu\nu}\bar{F}^{\mu\nu}\right)^2+ \frac{1}{8}\left(\bar{H}_{\mu\nu}\bar{H}^{\mu\nu}\right)^2,\label{calcul2f}			
\end{align}
}
where the derivative identities are structured up to total derivatives using all the Maxwell and Maxwell-Bianchi equations as
\begin{align}\label{calcul3}
\left(D_\rho \bar{F}_{\mu\nu}\right)\left(D^\rho \bar{F}^{\mu\nu}\right) &= 2\bar{F}\indices{^\mu_\nu} D_\rho D_\mu \bar{F}^{\nu\rho}= 2\bar{F}\indices{^\mu_\nu} [D_\rho,D_\mu] \bar{F}^{\nu\rho},
\end{align}
also utilizing the covariant derivative commutation acting on a rank-2 tensor,
\begin{align}\label{calcul4}
[D_\rho,D_\sigma]\bar{F}_{\mu\nu} &= R\indices{_\mu^\alpha_{\rho\sigma}}\bar{F}_{\alpha\nu} + R\indices{_\nu^\alpha_{\rho\sigma}}\bar{F}_{\mu\alpha}.
\end{align}
Note that the same steps \eqref{calcul3} and relation \eqref{calcul4} also hold for the other background $U(1)$ field strength $\bar{H}_{\mu\nu}$. Next, we need to derive the matrices $E$ and $\Omega_{\rho\sigma}$ by utilizing the $P$ and $\omega_\rho$ data from \cref{sdclog4b,sdclog4c,sdclog4d} into the formulas \eqref{comp4c} and \eqref{comp4d}, respectively. All the valid components of the $E$ are obtained and simplified as
{\allowdisplaybreaks
\begin{align}\label{calcul5}
E\indices{^{\hat{h}_{\mu\nu}}^{\hat{h}_{\alpha\beta}}} &= P\indices{^{\hat{h}_{\mu\nu}}^{\hat{h}_{\alpha\beta}}} - \left(\omega^\rho\right)\indices{^{\hat{h}_{\mu\nu}}^{a_{1\sigma}}}\left(\omega_\rho\right)\indices{_{a_{1\sigma}}^{\hat{h}_{\alpha\beta}}}- \left(\omega^\rho\right)\indices{^{\hat{h}_{\mu\nu}}^{a_{2\sigma}}}\left(\omega_\rho\right)\indices{_{a_{2\sigma}}^{\hat{h}_{\alpha\beta}}}= R^{\mu\alpha\nu\beta} + R^{\mu\beta\nu\alpha},\\
E\indices{^{\hat{h}}^{\hat{h}}} &=  P\indices{^{\hat{h}}^{\hat{h}}} = 2\Lambda,\\
E\indices{^{a_{1\alpha}}^{a_{1\beta}}} &= P\indices{^{a_{1\alpha}}^{a_{1\beta}}} - \left(\omega^\rho\right)\indices{^{a_{1\alpha}}^{\hat{h}_{\mu\nu}}}\left(\omega_\rho\right)\indices{_{\hat{h}_{\mu\nu}}^{a_{1\beta}}}- \left(\omega^\rho\right)\indices{^{a_{1\alpha}}^{\Phi}}\left(\omega_\rho\right)\indices{_{\Phi}^{a_{1\beta}}}\nonumber\\
&= -R^{\alpha\beta} + 3\bar{F}^{\alpha\rho}\bar{F}\indices{^\beta_\rho} + \bar{g}^{\alpha\beta}\bar{F}_{\mu\nu}\bar{F}^{\mu\nu},\\[2pt]
E\indices{^{a_{2\alpha}}^{a_{2\beta}}} &= P\indices{^{a_{2\alpha}}^{a_{2\beta}}} - \left(\omega^\rho\right)\indices{^{a_{2\alpha}}^{\hat{h}_{\mu\nu}}}\left(\omega_\rho\right)\indices{_{\hat{h}_{\mu\nu}}^{a_{2\beta}}}- \left(\omega^\rho\right)\indices{^{a_{2\alpha}}^{\Phi}}\left(\omega_\rho\right)\indices{_{\Phi}^{a_{2\beta}}}\nonumber\\
&= -R^{\alpha\beta} + 3\bar{H}^{\alpha\rho}\bar{H}\indices{^\beta_\rho} + \bar{g}^{\alpha\beta}\bar{H}_{\mu\nu}\bar{H}^{\mu\nu},\\[3pt]
E\indices{^\Phi^\Phi} &= P\indices{^\Phi^\Phi} - \left(\omega^\rho\right)\indices{^\Phi^{a_{1\alpha}}}\left(\omega_\rho\right)\indices{_{a_{1\alpha}}^\Phi}- \left(\omega^\rho\right)\indices{^\Phi^{a_{2\alpha}}}\left(\omega_\rho\right)\indices{_{a_{2\alpha}}^\Phi} =0,\\[2pt]
E\indices{^{a_{1\alpha}}^{a_{2\beta}}} &=  - \left(\omega^\rho\right)\indices{^{a_{1\alpha}}^{\hat{h}_{\mu\nu}}}\left(\omega_\rho\right)\indices{_{\hat{h}_{\mu\nu}}^{a_{2\beta}}}- \left(\omega^\rho\right)\indices{^{a_{1\alpha}}^{\Phi}}\left(\omega_\rho\right)\indices{_{\Phi}^{a_{2\beta}}} =  \bar{F}^{\alpha\rho}\bar{H}\indices{^\beta_\rho} + \bar{g}^{\alpha\beta}\bar{F}_{\mu\nu}\bar{H}^{\mu\nu},\\
E\indices{^{a_{2\alpha}}^{a_{1\beta}}} &=  - \left(\omega^\rho\right)\indices{^{a_{2\alpha}}^{\hat{h}_{\mu\nu}}}\left(\omega_\rho\right)\indices{_{\hat{h}_{\mu\nu}}^{a_{1\beta}}}- \left(\omega^\rho\right)\indices{^{a_{2\alpha}}^{\Phi}}\left(\omega_\rho\right)\indices{_{\Phi}^{a_{1\beta}}} =  \bar{H}^{\alpha\rho}\bar{F}\indices{^\beta_\rho} + \bar{g}^{\alpha\beta}\bar{H}_{\mu\nu}\bar{F}^{\mu\nu},\\
E\indices{^{\hat{h}_{\mu\nu}}^{\hat{h}}} &= P\indices{^{\hat{h}_{\mu\nu}}^{\hat{h}}} = -2i\left(\bar{F}^{\mu\alpha}\bar{F}\indices{^\nu_\alpha}+ \bar{H}^{\mu\alpha}\bar{H}\indices{^\nu_\alpha}\right),\\
E\indices{^{\hat{h}}^{\hat{h}_{\mu\nu}}} &= P\indices{^{\hat{h}}^{\hat{h}_{\mu\nu}}} = -2i\left(\bar{F}^{\mu\alpha}\bar{F}\indices{^\nu_\alpha}+ \bar{H}^{\mu\alpha}\bar{H}\indices{^\nu_\alpha}\right),\\
E\indices{^{\hat{h}_{\mu\nu}}^{a_{1\alpha}}} &= P\indices{^{\hat{h}_{\mu\nu}}^{a_{1\alpha}}} - \left(D_\rho\omega^\rho\right)\indices{^{\hat{h}_{\mu\nu}}^{a_{1\alpha}}}= -\frac{\sqrt{2}}{2}\left(D^\mu\bar{F}^{\nu\alpha}+ D^\nu \bar{F}^{\mu\alpha}\right),\\
E\indices{^{a_{1\alpha}}^{\hat{h}_{\mu\nu}}} &= P\indices{^{a_{1\alpha}}^{\hat{h}_{\mu\nu}}} - \left(D_\rho\omega^\rho\right)\indices{^{a_{1\alpha}}^{\hat{h}_{\mu\nu}}} = -\frac{\sqrt{2}}{2}\left(D^\mu\bar{F}^{\nu\alpha}+ D^\nu \bar{F}^{\mu\alpha}\right),\\[3pt]
E\indices{^{\hat{h}_{\mu\nu}}^{a_{2\alpha}}} &= P\indices{^{\hat{h}_{\mu\nu}}^{a_{2\alpha}}} - \left(D_\rho\omega^\rho\right)\indices{^{\hat{h}_{\mu\nu}}^{a_{2\alpha}}}= -\frac{\sqrt{2}}{2}\left(D^\mu\bar{H}^{\nu\alpha}+ D^\nu \bar{H}^{\mu\alpha}\right),\\
E\indices{^{a_{2\alpha}}^{\hat{h}_{\mu\nu}}} &= P\indices{^{a_{2\alpha}}^{\hat{h}_{\mu\nu}}} - \left(D_\rho\omega^\rho\right)\indices{^{a_{2\alpha}}^{\hat{h}_{\mu\nu}}} = -\frac{\sqrt{2}}{2}\left(D^\mu\bar{H}^{\nu\alpha}+ D^\nu \bar{H}^{\mu\alpha}\right),\\
E\indices{^{\hat{h}_{\mu\nu}}^{\Phi}} &= P\indices{^{\hat{h}_{\mu\nu}}^{\Phi}}  - \left(\omega^\rho\right)\indices{^{\hat{h}_{\mu\nu}}^{a_{1\sigma}}}\left(\omega_\rho\right)\indices{_{a_{1\sigma}}^{\Phi}}- \left(\omega^\rho\right)\indices{^{\hat{h}_{\mu\nu}}^{a_{2\sigma}}}\left(\omega_\rho\right)\indices{_{a_{2\sigma}}^{\Phi}} =0,\\[2pt]
E\indices{^{\Phi}^{\hat{h}_{\mu\nu}}} &= P\indices{^{\Phi}^{\hat{h}_{\mu\nu}}}  - \left(\omega^\rho\right)\indices{^\Phi^{a_{1\sigma}}}\left(\omega_\rho\right)\indices{_{a_{1\sigma}}^{\hat{h}_{\mu\nu}}}- \left(\omega^\rho\right)\indices{^\Phi^{a_{2\sigma}}}\left(\omega_\rho\right)\indices{_{a_{2\sigma}}^{\hat{h}_{\mu\nu}}} =0,\\[2pt]
E\indices{^{a_{1\alpha}}^{\Phi}} &=  - \left(D_\rho\omega^\rho\right)\indices{^{a_{1\alpha}}^{\Phi}} =0,\\[2pt] 
E\indices{^{\Phi}^{a_{1\alpha}}} &=  - \left(D_\rho\omega^\rho\right)\indices{^{\Phi}^{a_{1\alpha}}} =0,\\[2pt]
E\indices{^{a_{2\alpha}}^{\Phi}} &=  - \left(D_\rho\omega^\rho\right)\indices{^{a_{2\alpha}}^{\Phi}} =0, \\[2pt]
E\indices{^{\Phi}^{a_{2\alpha}}} &=  - \left(D_\rho\omega^\rho\right)\indices{^{\Phi}^{a_{2\alpha}}} =0.
\end{align}
}
On the other hand, from the curvature commutation operation $\phi_m \left[\mathcal{D}_\rho,\mathcal{D}_\sigma\right]\phi^m$ over the fluctuations $\phi_m = \big\lbrace \hat{h}_{\mu\nu}, \hat{h}, a_{1\mu}, a_{2\mu}, \Phi \big\rbrace$, we derive the following components of $\Omega_{\rho\sigma}$
{\allowdisplaybreaks
\begin{align}
\left(\Omega_{\rho\sigma}\right)\indices{^{\hat{h}_{\mu\nu}}^{\hat{h}_{\alpha\beta}}} &= \frac{1}{2}\left(\bar{g}^{\mu\alpha}R\indices{^\nu^\beta_\rho_\sigma}+ \bar{g}^{\mu\beta}R\indices{^\nu^\alpha_\rho_\sigma} +\bar{g}^{\nu\alpha}R\indices{^\mu^\beta_\rho_\sigma}+\bar{g}^{\nu\beta}R\indices{^\mu^\alpha_\rho_\sigma} \right) \nonumber\\
&\qquad\quad + \left( \left(\omega_\rho\right)\indices{^{\hat{h}_{\mu\nu}}^{a_{1\theta}}}\left(\omega_\sigma\right)\indices{_{a_{1\theta}}^{\hat{h}_{\alpha\beta}}}+ \left(\omega_\rho\right)\indices{^{\hat{h}_{\mu\nu}}^{a_{2\theta}}}\left(\omega_\sigma\right)\indices{_{a_{2\theta}}^{\hat{h}_{\alpha\beta}}} - \left(\rho \leftrightarrow \sigma\right)\right),      \\[3pt]
\left(\Omega_{\rho\sigma}\right)\indices{^{a_{1\alpha}}^{a_{1\beta}}} &=  R\indices{^\alpha^\beta_\rho_\sigma} + \left(\left(\omega_\rho\right)\indices{^{a_{1\alpha}}^{\hat{h}_{\mu\nu}}}\left(\omega_\sigma\right)\indices{_{\hat{h}_{\mu\nu}}^{a_{1\beta}}} + \left(\omega_\rho\right)\indices{^{a_{1\alpha}}^{\Phi}}\left(\omega_\sigma\right)\indices{_{\Phi}^{a_{1\beta}}}- \left(\rho \leftrightarrow \sigma\right)\right),  \\[3pt]
\left(\Omega_{\rho\sigma}\right)\indices{^{a_{2\alpha}}^{a_{2\beta}}} &=  R\indices{^\alpha^\beta_\rho_\sigma} + \left(\left(\omega_\rho\right)\indices{^{a_{2\alpha}}^{\hat{h}_{\mu\nu}}}\left(\omega_\sigma\right)\indices{_{\hat{h}_{\mu\nu}}^{a_{2\beta}}} + \left(\omega_\rho\right)\indices{^{a_{2\alpha}}^{\Phi}}\left(\omega_\sigma\right)\indices{_{\Phi}^{a_{2\beta}}}- \left(\rho \leftrightarrow \sigma\right)\right),\\[3pt]
\left(\Omega_{\rho\sigma}\right)\indices{^\Phi^\Phi} &= \left(\omega_\rho\right)\indices{^\Phi^{a_{1\alpha}}}\left(\omega_\sigma\right)\indices{_{a_{1\alpha}}^\Phi} + \left(\omega_\rho\right)\indices{^\Phi^{a_{2\alpha}}}\left(\omega_\sigma\right)\indices{_{a_{2\alpha}}^\Phi}- \left(\rho \leftrightarrow \sigma\right),\\[3pt]
\left(\Omega_{\rho\sigma}\right)\indices{^{\hat{h}_{\mu\nu}}^{a_{1\alpha}}} &=  \left(D_\rho\omega_\sigma\right)\indices{^{\hat{h}_{\mu\nu}}^{a_{1\alpha}}}- \left(\rho \leftrightarrow \sigma\right), \\[3pt]
\left(\Omega_{\rho\sigma}\right)\indices{^{\hat{h}_{\mu\nu}}^{a_{2\alpha}}} &=  \left(D_\rho\omega_\sigma\right)\indices{^{\hat{h}_{\mu\nu}}^{a_{2\alpha}}}- \left(\rho \leftrightarrow \sigma\right), \\[3pt]
\left(\Omega_{\rho\sigma}\right)\indices{^{\hat{h}_{\mu\nu}}^{\Phi}} &=  \left(\omega_\rho\right)\indices{^{\hat{h}_{\mu\nu}}^{a_{1\alpha}}}\left(\omega_\sigma\right)\indices{_{a_{1\alpha}}^{\Phi}} + \left(\omega_\rho\right)\indices{^{\hat{h}_{\mu\nu}}^{a_{2\alpha}}}\left(\omega_\sigma\right)\indices{_{a_{2\alpha}}^{\Phi}} -\left(\rho \leftrightarrow \sigma\right),\\[3pt]
\left(\Omega_{\rho\sigma}\right)\indices{^{a_{1\alpha}}^{\hat{h}_{\mu\nu}}} &=  \left(D_\rho\omega_\sigma\right)\indices{^{a_{1\alpha}}^{\hat{h}_{\mu\nu}}}-\left(\rho \leftrightarrow \sigma\right),\\[3pt]
\left(\Omega_{\rho\sigma}\right)\indices{^{a_{1\alpha}}^{a_{2\beta}}} &= \left(\omega_\rho\right)\indices{^{a_{1\alpha}}^{\hat{h}_{\mu\nu}}}\left(\omega_\sigma\right)\indices{_{\hat{h}_{\mu\nu}}^{a_{2\beta}}} + \left(\omega_\rho\right)\indices{^{a_{1\alpha}}^{\Phi}}\left(\omega_\sigma\right)\indices{_{\Phi}^{a_{2\beta}}}- \left(\rho \leftrightarrow \sigma\right),  \\[3pt]
\left(\Omega_{\rho\sigma}\right)\indices{^{a_{1\alpha}}^{\Phi}} &=  \left(D_\rho\omega_\sigma\right)\indices{^{a_{1\alpha}}^{\Phi}}- \left(\rho \leftrightarrow \sigma\right),\\
\left(\Omega_{\rho\sigma}\right)\indices{^{a_{2\alpha}}^{\hat{h}_{\mu\nu}}} &=  \left(D_\rho\omega_\sigma\right)\indices{^{a_{2\alpha}}^{\hat{h}_{\mu\nu}}}-\left(\rho \leftrightarrow \sigma\right),\\[3pt]
\left(\Omega_{\rho\sigma}\right)\indices{^{a_{2\alpha}}^{a_{1\beta}}} &= \left(\omega_\rho\right)\indices{^{a_{2\alpha}}^{\hat{h}_{\mu\nu}}}\left(\omega_\sigma\right)\indices{_{\hat{h}_{\mu\nu}}^{a_{1\beta}}} + \left(\omega_\rho\right)\indices{^{a_{2\alpha}}^{\Phi}}\left(\omega_\sigma\right)\indices{_{\Phi}^{a_{1\beta}}}- \left(\rho \leftrightarrow \sigma\right),  \\[3pt]
\left(\Omega_{\rho\sigma}\right)\indices{^{a_{2\alpha}}^{\Phi}} &=  \left(D_\rho\omega_\sigma\right)\indices{^{a_{2\alpha}}^{\Phi}}- \left(\rho \leftrightarrow \sigma\right),\\[3pt] 
\left(\Omega_{\rho\sigma}\right)\indices{^{\Phi}^{\hat{h}_{\mu\nu}}} &=  \left(\omega_\rho\right)\indices{^\Phi^{a_{1\sigma}}}\left(\omega_\sigma\right)\indices{_{a_{1\sigma}}^{\hat{h}_{\mu\nu}}} + \left(\omega_\rho\right)\indices{^\Phi^{a_{2\sigma}}}\left(\omega_\sigma\right)\indices{_{a_{2\sigma}}^{\hat{h}_{\mu\nu}}}- \left(\rho \leftrightarrow \sigma\right),\\[3pt]
\left(\Omega_{\rho\sigma}\right)\indices{^{\Phi}^{a_{1\alpha}}} &=  \left(D_\rho\omega_\sigma\right)\indices{^{\Phi}^{a_{1\alpha}}} - \left(\rho \leftrightarrow \sigma\right),\\[3pt] 
\left(\Omega_{\rho\sigma}\right)\indices{^{\Phi}^{a_{2\alpha}}} &=  \left(D_\rho\omega_\sigma\right)\indices{^{\Phi}^{a_{2\alpha}}} - \left(\rho \leftrightarrow \sigma\right),    
\end{align}}
where one needs to use the covariant derivative commutation relations,
{\allowdisplaybreaks
\begin{align}
	\phi_m [D_\rho,D_\sigma] \phi^m &= \hat{h}_{\mu\nu} [D_\rho,D_\sigma] \hat{h}^{\mu\nu}+\hat{h} [D_\rho,D_\sigma] \hat{h} + a_{1\alpha} [D_\rho,D_\sigma] {a_1}^{\alpha}\nonumber \\
	&\qquad + a_{2\alpha} [D_\rho,D_\sigma] {a_2}^{\alpha}+\Phi [D_\rho,D_\sigma] \Phi\nonumber\\[5pt]
	&=\frac{1}{2} \hat{h}_{\mu\nu}\bigg(\bar{g}^{\mu\alpha} R\indices{^{\nu\beta}_{\rho\sigma}} +\bar{g}^{\mu\beta}R\indices{^{\nu\alpha}_{\rho\sigma}}+\bar{g}^{\nu\alpha}R\indices{^{\mu\beta}_{\rho\sigma}}+\bar{g}^{\nu\beta}R\indices{^{\mu\alpha}_{\rho\sigma}}\bigg)\hat{h}_{\alpha\beta} \nonumber \\
	&\qquad\qquad\quad + a_{1\alpha} R\indices{^\alpha^\beta_\rho_\sigma} a_{1\beta} + a_{2\alpha} R\indices{^\alpha^\beta_\rho_\sigma} a_{2\beta},
\end{align}}
followed by the appropriate form of the gauge-coupling $\omega_\rho$,
{\allowdisplaybreaks
	\begin{align}
		\left(\omega^\rho\right)\indices{^{h_{\mu\nu}}^{a_{1\alpha}}} &=-\left(\omega^\rho\right)\indices{^{a_{1\alpha}}^{h_{\mu\nu}}}= \frac{\sqrt{2}}{2} \left(\bar{g}^{\mu\rho}\bar{F}^{\nu\alpha}+\bar{g}^{\nu\rho}\bar{F}^{\mu\alpha}-\bar{g}^{\mu\alpha}\bar{F}^{\nu\rho}-\bar{g}^{\nu\alpha}\bar{F}^{\mu\rho}\right),\\
		\left(\omega^\rho\right)\indices{^{h_{\mu\nu}}^{a_{2\alpha}}} &=-\left(\omega^\rho\right)\indices{^{a_{2\alpha}}^{h_{\mu\nu}}}= \frac{\sqrt{2}}{2} \left(\bar{g}^{\mu\rho}\bar{H}^{\nu\alpha}+\bar{g}^{\nu\rho}\bar{H}^{\mu\alpha}-\bar{g}^{\mu\alpha}\bar{H}^{\nu\rho}-\bar{g}^{\nu\alpha}\bar{H}^{\mu\rho}\right),\\[3pt]
		\left(\omega^\rho\right)\indices{^{a_{1\alpha}}^{\Phi}} &=-\left(\omega^\rho\right)\indices{^{\Phi}^{a_{1\alpha}}}= \bar{F}^{\mu\rho}, \enspace \left(\omega^\rho\right)\indices{^{a_{2\alpha}}^{\Phi}} =-\left(\omega^\rho\right)\indices{^{\Phi}^{a_{2\alpha}}}= \bar{H}^{\mu\rho}.
\end{align}}  
Next, with the help of all non-vanishing $E$ and $\Omega_{\rho\sigma}$ components, we successively define $\text{Tr}(E)$, $\text{Tr}(E^2)$ and $\text{Tr}\left(\Omega_{\rho\sigma}\Omega^{\rho\sigma}\right)$ as follows 
{\allowdisplaybreaks
\begin{align}\label{calcul6}
\text{Tr}(E) &= \text{Tr}\bigg[E\indices{^{\hat{h}_{\mu\nu}}_{\hat{h}_{\alpha\beta}}}+ E\indices{^{\hat{h}}_{\hat{h}}} + E\indices{^{a_{1\alpha}}_{a_{1\beta}}} + E\indices{^{a_{2\alpha}}_{a_{2\beta}}} \bigg],\\[10pt]
\text{Tr}\left(E^2\right) &= \text{Tr}\bigg[E\indices{^{\hat{h}_{\mu\nu}}_{\hat{h}_{\theta\phi}}}E\indices{^{\hat{h}_{\theta\phi}}_{\hat{h}_{\alpha\beta}}}+ E\indices{^{\hat{h}}_{\hat{h}}}E\indices{^{\hat{h}}_{\hat{h}}}+ E\indices{^{a_{1\alpha}}_{a_{1\gamma}}}E\indices{^{a_{1\gamma}}_{a_{1\beta}}} \nonumber\\
&\qquad\enspace + E\indices{^{a_{2\alpha}}_{a_{2\gamma}}}E\indices{^{a_{2\gamma}}_{a_{2\beta}}} + E\indices{^{a_{1\alpha}}_{a_{2\gamma}}}E\indices{^{a_{2\gamma}}_{a_{1\beta}}}+ E\indices{^{a_{2\alpha}}_{a_{1\gamma}}}E\indices{^{a_{1\gamma}}_{a_{2\beta}}}  \nonumber\\
&\qquad\enspace +  E\indices{^{\hat{h}_{\mu\nu}}_{\hat{h}}}E\indices{^{\hat{h}}_{\hat{h}_{\alpha\beta}}}+ E\indices{^{\hat{h}}_{\hat{h}_{\mu\nu}}}E\indices{^{\hat{h}_{\mu\nu}}_{\hat{h}}} +  E\indices{^{\hat{h}_{\mu\nu}}_{a_{1\theta}}}E\indices{^{a_{1\theta}}_{\hat{h}_{\alpha\beta}}} \nonumber \\
& \qquad\enspace + E\indices{^{a_{1\alpha}}_{\hat{h}_{\mu\nu}}}E\indices{^{\hat{h}_{\mu\nu}}_{a_{1\beta}}}+  E\indices{^{\hat{h}_{\mu\nu}}_{a_{2\theta}}}E\indices{^{a_{2\theta}}_{\hat{h}_{\alpha\beta}}}+ E\indices{^{a_{2\alpha}}_{\hat{h}_{\mu\nu}}}E\indices{^{\hat{h}_{\mu\nu}}_{a_{2\beta}}} \bigg],\\[5pt]
\text{Tr}\left(\Omega_{\rho\sigma}\Omega^{\rho\sigma}\right) &= \text{Tr}\bigg[{\left(\Omega_{\rho\sigma}\right)}\indices{^{\hat{h}_{\mu\nu}}_{\hat{h}_{\theta\phi}}}{\left(\Omega^{\rho\sigma}\right)}\indices{^{\hat{h}_{\theta\phi}}_{\hat{h}_{\alpha\beta}}} + {\left(\Omega_{\rho\sigma}\right)}\indices{^{a_{1\alpha}}_{a_{1\gamma}}}{\left(\Omega^{\rho\sigma}\right)}\indices{^{a_{1\gamma}}_{a_{1\beta}}}\nonumber \\
&\qquad\enspace + {\left(\Omega_{\rho\sigma}\right)}\indices{^{a_{2\alpha}}_{a_{2\gamma}}}{\left(\Omega^{\rho\sigma}\right)}\indices{^{a_{2\gamma}}_{a_{2\beta}}}+ {\left(\Omega_{\rho\sigma}\right)}\indices{^{\Phi}_{\Phi}}{\left(\Omega^{\rho\sigma}\right)}\indices{^{\Phi}_{\Phi}} \nonumber\\
&\qquad\enspace  + {\left(\Omega_{\rho\sigma}\right)}\indices{^{a_{1\alpha}}_{a_{2\gamma}}}{\left(\Omega^{\rho\sigma}\right)}\indices{^{a_{2\gamma}}_{a_{1\beta}}}+ {\left(\Omega_{\rho\sigma}\right)}\indices{^{a_{2\alpha}}_{a_{1\gamma}}}{\left(\Omega^{\rho\sigma}\right)}\indices{^{a_{1\gamma}}_{a_{2\beta}}} \nonumber \\
&\qquad\enspace +  {\left(\Omega_{\rho\sigma}\right)}\indices{^{\hat{h}_{\mu\nu}}_{a_{1\theta}}}{\left(\Omega^{\rho\sigma}\right)}\indices{^{a_{1\theta}}_{\hat{h}_{\alpha\beta}}}+ {\left(\Omega_{\rho\sigma}\right)}\indices{^{a_{1\alpha}}_{\hat{h}_{\mu\nu}}}{\left(\Omega^{\rho\sigma}\right)}\indices{^{\hat{h}_{\mu\nu}}_{a_{1\beta}}}\nonumber\\
&\qquad\enspace +  {\left(\Omega_{\rho\sigma}\right)}\indices{^{\hat{h}_{\mu\nu}}_{a_{2\theta}}}{\left(\Omega^{\rho\sigma}\right)}\indices{^{a_{2\theta}}_{\hat{h}_{\alpha\beta}}} +{\left(\Omega_{\rho\sigma}\right)}\indices{^{a_{2\alpha}}_{\hat{h}_{\mu\nu}}}{\left(\Omega^{\rho\sigma}\right)}\indices{^{\hat{h}_{\mu\nu}}_{a_{2\beta}}}  \nonumber\\
&\qquad\enspace +  {\left(\Omega_{\rho\sigma}\right)}\indices{^{\hat{h}_{\mu\nu}}_{\Phi}}{\left(\Omega^{\rho\sigma}\right)}\indices{^{\Phi}_{\hat{h}_{\alpha\beta}}} +{\left(\Omega_{\rho\sigma}\right)}\indices{^{\Phi}_{\hat{h}_{\mu\nu}}}{\left(\Omega^{\rho\sigma}\right)}\indices{^{\hat{h}_{\mu\nu}}_{\Phi}}  \nonumber\\
&\qquad\enspace +{\left(\Omega_{\rho\sigma}\right)}\indices{^{a_{1\alpha}}_{\Phi}}{\left(\Omega^{\rho\sigma}\right)}\indices{^{\Phi}_{a_{1\beta}}} +  {\left(\Omega_{\rho\sigma}\right)}\indices{^{\Phi}_{a_{1\alpha}}}{\left(\Omega^{\rho\sigma}\right)}\indices{^{a_{1\alpha}}_{\Phi}} \nonumber \\
& \qquad\enspace +{\left(\Omega_{\rho\sigma}\right)}\indices{^{a_{2\alpha}}_{\Phi}}{\left(\Omega^{\rho\sigma}\right)}\indices{^{\Phi}_{a_{2\beta}}} +  {\left(\Omega_{\rho\sigma}\right)}\indices{^{\Phi}_{a_{2\alpha}}}{\left(\Omega^{\rho\sigma}\right)}\indices{^{a_{2\alpha}}_{\Phi}} \bigg].
\end{align}}   
To execute the above traces, we pursued the following explicit treatments and steps
{\allowdisplaybreaks
\begin{align}\label{calcul44}
	\begin{split}
		A\indices{^{m}_{n}} &= A^{mp}I_{pn},\\[2pt]
		\mathrm{tr}(A) &= A\indices{^{m}_{m}} = A^{m p}I_{p m},\\[2pt]
		\mathrm{tr}(A^2) &= A\indices{^{m}_{n}}A\indices{^{n}_{m}}= \left(A^{m p}I_{p n}\right)\left(A^{nq}I_{q m}\right),\\[2pt]
		\mathrm{tr}(AB) &= A\indices{^{m}_{n}}B\indices{^{n}_{m}}= \left(A^{m p}I_{p n}\right)\left(B^{n q}I_{q m}\right)=\mathrm{tr}(BA),
	\end{split}
\end{align}}
where $A^{\tilde{\xi}_m \tilde{\xi}_n}$ and $B^{\tilde{\xi}_m \tilde{\xi}_n}$ are any two arbitrary matrix components associated with the $U(1)^2$ EMD fluctuations $\phi_m = \big\lbrace \hat{h}_{\mu\nu}, \hat{h}, a_{1\mu}, a_{2\mu}, \Phi \big\rbrace$. In this process, we require to utilize the following projection operators
{\allowdisplaybreaks
\begin{align}
I^{\hat{h}\hat{h}} &= I\indices{^{\Phi}^{\Phi}} = 1,\\
I\indices{^{a_{1\mu}}^{a_{1\nu}}} &= I\indices{^{a_{2\mu}}^{a_{2\nu}}} = \bar{g}^{\mu\nu},\\	
I^{\hat{h}_{\mu\nu}\hat{h}_{\alpha\beta}} &= \frac{1}{2}\left(\bar{g}^{\mu\alpha} \bar{g}^{\nu\beta} + \bar{g}^{\mu\beta} \bar{g}^{\nu\alpha}-\frac{1}{2}\bar{g}^{\mu\nu}\bar{g}^{\alpha\beta}\right).
\end{align}}
With all that mentioned above, we computed the following trace results
{\allowdisplaybreaks
\begin{align}
\text{Tr}\left(E\right) &= 7\left(\bar{F}_{\mu\nu}\bar{F}^{\mu\nu}+ \bar{H}_{\mu\nu}\bar{H}^{\mu\nu}\right) -3R, \\[10pt]
\text{Tr}\left(E^2\right) &= 2 R_{\mu\nu\rho\sigma}R^{\mu\nu\rho\sigma} +2R_{\mu\rho\nu\sigma}R^{\mu\nu\rho\sigma} + 4 \Lambda^2 + \frac{1}{4}R^2 -2R\bar{F}_{\mu\nu}\bar{F}^{\mu\nu} \nonumber\\[2pt]
&\quad  -2R\bar{H}_{\mu\nu}\bar{H}^{\mu\nu} -6{R}_{\mu\nu}\bar{F}^{\mu\rho}\bar{F}\indices{^\nu_\rho} -6{R}_{\mu\nu}\bar{H}^{\mu\rho}\bar{H}\indices{^\nu_\rho} + 12\left(\bar{F}_{\mu\nu}\bar{F}^{\mu\nu}\right)^2  \nonumber\\[2pt]
&\quad + 12\left(\bar{H}_{\mu\nu}\bar{H}^{\mu\nu}\right)^2 + 4\bar{F}_{\mu\nu}\bar{F}^{\mu\nu}\bar{H}_{\rho\sigma}\bar{H}^{\rho\sigma} + 12\bar{F}_{\mu\nu}\bar{H}^{\mu\nu}\bar{F}_{\rho\sigma}\bar{H}^{\rho\sigma} \nonumber \\[2pt]
&\quad + \bar{F}^{\mu\rho}\bar{F}\indices{^\nu_\rho} \bar{F}_{\mu\sigma}\bar{F}\indices{_\nu^\sigma} + \bar{H}^{\mu\rho}\bar{H}\indices{^\nu_\rho} \bar{H}_{\mu\sigma}\bar{H}\indices{_\nu^\sigma} - 14\bar{F}^{\mu\rho}\bar{H}\indices{^\nu_\rho} \bar{H}_{\mu\sigma}\bar{F}\indices{_\nu^\sigma}   \nonumber\\[2pt]
&\quad + 2\left(D_\rho \bar{F}_{\mu\nu}\right)\left(D^\rho \bar{F}^{\mu\nu}\right) + 2\left(D_\rho \bar{H}_{\mu\nu}\right)\left(D^\rho \bar{H}^{\mu\nu}\right)\nonumber \\[2pt]
&\quad + 2\left(D_\mu \bar{F}\indices{_\rho^\nu}\right)\left(D_\nu \bar{F}^{\rho\mu}\right)+ 2\left(D_\mu \bar{H}\indices{_\rho^\nu}\right)\left(D_\nu \bar{H}^{\rho\mu}\right), \\[15pt]
\text{Tr}\left(\Omega_{\rho\sigma}\Omega^{\rho\sigma}\right) &= -8R_{\mu\nu\rho\sigma}R^{\mu\nu\rho\sigma}  +4 R_{\mu\rho\nu\sigma}\bar{F}^{\mu\nu}\bar{F}^{\rho\sigma} +4 R_{\mu\rho\nu\sigma}\bar{H}^{\mu\nu}\bar{H}^{\rho\sigma} + 4R\bar{F}_{\mu\nu}\bar{F}^{\mu\nu}  \nonumber\\[2pt]
&\quad + 4R \bar{H}_{\mu\nu}\bar{H}^{\mu\nu}+ 8{R}_{\mu\nu}\bar{F}^{\mu\rho}\bar{F}\indices{^\nu_\rho}  + 8{R}_{\mu\nu}\bar{H}^{\mu\rho}\bar{H}\indices{^\nu_\rho} - 70\left(\bar{F}_{\mu\nu}\bar{F}^{\mu\nu}\right)^2 \nonumber \\[2pt]
&\quad -70\left(\bar{H}_{\mu\nu}\bar{H}^{\mu\nu}\right)^2 - 116\bar{F}_{\mu\nu}\bar{F}^{\mu\nu}\bar{H}_{\rho\sigma}\bar{H}^{\rho\sigma}- 104 \bar{F}_{\mu\nu}\bar{H}^{\mu\nu}\bar{F}_{\rho\sigma}\bar{H}^{\rho\sigma} \nonumber\\[2pt]
&\quad  + 82\bar{F}^{\mu\rho}\bar{F}\indices{^\nu_\rho} \bar{F}_{\mu\sigma}\bar{F}\indices{_\nu^\sigma} + 82 \bar{H}^{\mu\rho}\bar{H}\indices{^\nu_\rho} \bar{H}_{\mu\sigma}\bar{H}\indices{_\nu^\sigma} + 388 \bar{F}^{\mu\rho}\bar{H}\indices{^\nu_\rho} \bar{H}_{\mu\sigma}\bar{F}\indices{_\nu^\sigma}  \nonumber\\[2pt]
&\quad  - 24\left(D_\rho \bar{F}_{\mu\nu}\right)\left(D^\rho \bar{F}^{\mu\nu}\right) - 24\left(D_\rho \bar{H}_{\mu\nu}\right)\left(D^\rho \bar{H}^{\mu\nu}\right)\nonumber\\[2pt]
&\quad + 8\left(D_\mu \bar{F}\indices{_\rho^\nu}\right)\left(D_\nu \bar{F}^{\rho\mu}\right)+ 8\left(D_\mu \bar{H}\indices{_\rho^\nu}\right)\left(D_\nu \bar{H}^{\rho\mu}\right). 
\end{align}}
Finally, one needs to simplify the above traces using the on-shell identities of \cref{calcul1a,calcul1b,calcul1c,calcul1d,calcul1e,calcul2a,calcul2b,calcul2c,calcul2d,calcul2e,calcul2f}, leading to the explicit irreducible results written in \cref{sdclog5}. 

\section{Holographic renormalization for AdS$_4$ black holes}\label{holo}

In this section, we want to provide some essential details of the holographic boundary counterterm \eqref{nbh8} and the related regularization process in obtaining the curvature invariant integration results \eqref{nbh9}. 
 
With the Euclidean continuation ($t \to -i\tau$), the boundary geometry of AdS$_4$ background \eqref{nbh1} at $r = r_c$ is given by   
\begin{align}\label{holo1}
	\mathrm{d}s^2 = \gamma_{\mu\nu}\mathrm{d}y^\mu \mathrm{d}y^\nu &= - \frac{\Delta _r}{{\rho_c}^2} \left( i\mathrm{d}\tau + \frac{a \sin^2 \theta }{\Xi} \mathrm{d}\phi \right)^2  + \frac{{\rho_c}^2}{\Delta_\theta} \mathrm{d}\theta^2  + \frac{\Delta_\theta \sin^2 \theta}{{\rho_c}^2} \left(ia\,\mathrm{d}\tau + \frac{{r_c}^2+a^2}{\Xi} \mathrm{d}\phi \right)^2,
\end{align}
where the boundary parameters $\Delta_{r_c}$, $\Delta_\theta$, $\rho_c$, and $\Xi$ are defined as
\begin{align}\label{holo2}
	\begin{gathered}
		\Delta_{r_c} = ({r_c}^2+a^2) \bigg(1+\frac{{r_c}^2}{\ell^2}\bigg) - 2m {r_c} + q^2 + p^2,\\
		\Delta_\theta = 1 -\frac{a^2}{\ell^2} \cos^2 \theta, \quad {\rho_c}^2 = {r_c}^2 + a^2 \cos^2 \theta, \quad \Xi = 1 -\frac{a^2}{\ell^2}.
	\end{gathered}
\end{align} 
Then, we can write the following forms for the determinant of metric $\gamma_{\mu\nu}$ describing the boundary geometry and related Ricci scalar $\mathcal{R}$
\begin{align}\label{holo3}
\begin{split}
\det \gamma &= \frac{1}{\Xi^2}\left[\left({r_c}^2 + a^2\right)\left(\frac{{r_c}^2}{\ell^2}+1\right)-2mr_c + q^2+ p^2\right]\left({r_c}^2 + a^2 \cos^2 \theta\right) \sin^2\theta,\\[3pt]
\mathcal{R} &= \frac{2}{\left({r_c}^2 + a^2 \cos^2 \theta\right)^3}\bigg[{r_c}^4 + \left({r_c}^2-2m{r_c} + q^2+ p^2\right)a^2\cos^2 \theta \\
&\qquad\qquad\qquad\qquad +\frac{a^2}{\ell^2}\Big({r_c}^2\left(1-5\cos^2\theta\right)-3a^2\cos^4\theta\Big)\left({r_c}^2 + a^2 \cos^2 \theta\right)\bigg].
\end{split}
\end{align} 
With the above boundary setup, we can now express and expand the holographic boundary counterterm as
\begin{align}
	\mathcal{C}_{\text{HCT}} &= \int_{\partial(\text{AdS$_4$})} \mathrm{d}^3y \sqrt{\det\gamma} \left(c_1 + c_2 \mathcal{R}\right) \label{holo5} \\[2pt]
	& = \int_{0}^{\beta}\mathrm{d}\tau \int_{0}^{2\pi}\mathrm{d}\phi \int_{0}^{\pi}\mathrm{d}\theta \sqrt{\det\gamma} \left(c_1 + c_2 \mathcal{R}\right) \label{holo6} \\[2pt]
	&= \frac{4\pi\beta}{\Xi\ell}\bigg[c_1 {r_c}^3 + \frac{1}{6\ell^2}\Big(c_1\left(4a^2 + 3\ell^2\right)\ell^2 + 4c_2\left(3\ell^2 - 2a^2\right)\Big)r_c - c_1m\ell^2\bigg] + \mathcal{O}\left({r_c}^{-1}\right).\label{holo4}
\end{align}
Throughout this paper, we have added the above form of boundary term to the local part of logarithmic correction (i.e., the bulk contribution) and then extracted a finite and renormalized $\mathcal{C}_{\text{local}}$ contribution to AdS$_4$ black hole entropy in the limit $r_c \to \infty$,
{
	\renewcommand{\arraystretch}{2.0}
	\begin{table}[t]
		\centering
		\begin{tabular}{|>{\centering}p{2.2in}||>{\centering}p{.7in}>{\centering}p{.7in}|}
			\hline
			\textbf{Integrated AdS$\bm{_4}$ Invariants} & {$\bm{c_1}$} & {$\bm{c_2}$} \tabularnewline \hline \hline
			$\frac{1}{16\pi^2}\int \mathrm{d}^4x \sqrt{\det\bar{g}}\, R_{\mu\nu\rho\sigma}R^{\mu\nu\rho\sigma}$ & $-\dfrac{8}{\ell^3}$ & $\dfrac{2}{\ell}$ \tabularnewline \hline
			$\frac{1}{16\pi^2}\int \mathrm{d}^4x \sqrt{\det\bar{g}}\, R_{\mu\nu}R^{\mu\nu}$ & $-\dfrac{12}{\ell^3}$ & $\dfrac{3}{\ell}$ \tabularnewline \hline
			$\frac{1}{16\pi^2}\int \mathrm{d}^4x \sqrt{\det\bar{g}}\,R^2$ & $-\dfrac{48}{\ell^3}$ & $\dfrac{12}{\ell}$ \tabularnewline \hline
			$\frac{1}{16\pi^2}\int \mathrm{d}^4x \sqrt{\det\bar{g}}\, W_{\mu\nu\rho\sigma}W^{\mu\nu\rho\sigma}$ & $0$ & $0$ \tabularnewline \hline
			$\frac{1}{16\pi^2}\int \mathrm{d}^4x \sqrt{\det\bar{g}}\,E_4$ & $-\dfrac{8}{\ell^3}$ & $\dfrac{2}{\ell}$ \tabularnewline \hline
		\end{tabular}
		\caption{Holographic renormalization data $(c_1, c_2)$ for different integrated background invariants around the AdS$_4$ black holes for finite $\mathcal{C}_{\text{local}}$ contributions.}\label{holograph}
	\end{table}
}
\begin{align}\label{holo7}
	\mathcal{C}_{\text{local}} = \lim_{r_c \to \infty}\left[\frac{1}{16\pi^2}\int_0^\beta \mathrm{d}\tau \int_{r_+}^{r_c} \mathrm{d}r \int_0^{\pi} \mathrm{d}\theta \int_0^{2\pi} \mathrm{d}\phi\, \sqrt{\det\bar{g}}\,a_4(x) + \mathcal{C}_{\text{HCT}}\right].
\end{align}
During this holographic renormalization process, the boundary counterterm \eqref{holo4} is found to be canceling the only and explicit bulk divergent terms involving ${r_c}^3$ and $r_c$ for the appropriate choice of $c_1$ and $c_2$ coefficients in the limit $r_c \to \infty$. For our specific aim of EMD-AdS theory, we have always used the typical $a_4(x)$ form \eqref{emd1} or the $\mathcal{C}_{\text{local}}$ formula \eqref{emd4}. Hence, it is also convenient to find the regulated results of integrated $W_{\mu\nu\rho\sigma}W^{\mu\nu\rho\sigma}$, $E_4$ and $R^2$ invariants via the same renormalization process \eqref{holo7}. The relevant relations are presented in \eqref{nbh9} for a generic charged and rotating background, where the associated $c_1$ and $c_2$ values are listed in \Cref{holograph}. These data are common for the integrations over all Kerr-AdS ($q,p =0$), Reissner-Nordstr\"om-AdS ($a=0$), and Schwarzschild-AdS ($q,p,a =0$) backgrounds. Note that there is no such divergence due to infinite AdS boundary while integrating the curvature invariants around asymptotically-flat black holes ($\ell \to \infty$), and thus, we do not need any renormalization for them.

Finally, we want to mention the explicit forms of the integrated curvature invariants that are regulated via the mentioned holographic renormalization process and directly associated with the $\mathcal{C}_{\text{local}}$ formulas derived in \cref{nbh11,ebh5}. The contributions of Weyl tensor square $W^2$ $(=W_{\mu\nu\rho\sigma}W^{\mu\nu\rho\sigma})$ and Ricci scalar square $R^2$ are listed as
{\allowdisplaybreaks
\begin{align}
&\frac{1}{16\pi^2}\int_{\text{Sch-AdS}}\hspace{-0.3in} \mathrm{d}^4x \sqrt{\det\bar{g}}\, R^2 = \frac{24 r_{+}^{2}\left(\ell^{2}-r_{+}^{2}\right)}{\ell^{2}\left(\ell^{2}+3 r_{+}^{2}\right)}, \label{holo8a}\\[3pt]
& \frac{1}{16\pi^2}\int_{\text{Sch-AdS}}\hspace{-0.3in} \mathrm{d}^4x \sqrt{\det\bar{g}}\, W^2 =\frac{4\left(\ell^{2}+r_{+}^{2}\right)^{2}}{\ell^{2}\left(\ell^{2}+3 r_{+}^{2}\right)},\label{holo8b}\\[3pt]
&\frac{1}{16\pi^2}\int_{\text{RN-AdS}}\hspace{-0.28in} \mathrm{d}^4x \sqrt{\det\bar{g}}\, R^2 = -\frac{24 r_{+}^{2}}{\ell^{2}} + \frac{12 r_{+}\left(r_{+}^{2}+\ell^{2}\right)}{\pi \ell^{4}} \beta, \label{holo8c}\\[3pt]
& \frac{1}{16\pi^2}\int_{\text{RN-AdS}}\hspace{-0.28in} \mathrm{d}^4x \sqrt{\det\bar{g}}\, W^2 =\frac{4}{5} -\frac{28 r_{+}^{2}}{5 \ell^{2}} + \frac{32 \pi r_{+}}{5 \beta}+\frac{ 2\left(\ell^{4}+\ell^{2} r_{+}^{2}+4 r_{+}^{4}\right)}{ 5 \pi \ell^{4} r_{+}} \beta,\label{holo8d}\\[3pt]
&\lim_{\beta \to \infty} \frac{1}{16\pi^2}\int_{\text{RN-AdS}}\hspace{-0.28in} \mathrm{d}^4x \sqrt{\det\bar{g}}\, R^2 = 4-\frac{2 r_0^2}{\ell _2^2}-\frac{2 \ell _2^2}{r_0^2}, \label{holo8e}\\[3pt]
&\lim_{\beta \to \infty} \frac{1}{16\pi^2}\int_{\text{RN-AdS}}\hspace{-0.28in} \mathrm{d}^4x \sqrt{\det\bar{g}}\, W^2 = -\frac{2 \left(\ell _2^2-r_0^2\right){}^2}{3 r_0^2 \ell _2^2},\label{holo8f}\\[3pt]
&\frac{1}{16\pi^2}\int_{\text{Kerr-AdS}}\hspace{-0.32in} \mathrm{d}^4x \sqrt{\det\bar{g}}\, R^2 = \frac{6 \beta  \left(a^2+r_+^2\right) \left(\ell ^2-r_+^2\right)}{\pi  r_+ \ell ^2 \left(\ell^2 - a^2\right)}, \label{holo8g}\\[3pt]
& \frac{1}{16\pi^2}\int_{\text{Kerr-AdS}}\hspace{-0.32in} \mathrm{d}^4x \sqrt{\det\bar{g}}\, W^2 =\frac{\beta  \left(r_+^2-a^2\right) \left(r_+^2+\ell ^2\right)^2}{\pi  r_+ \ell ^2 \left(\ell^2 - a^2\right) \left(a^2+r_+^2\right)},\label{holo8h}\\[3pt]
&\lim_{\beta \to \infty} \frac{1}{16\pi^2}\int_{\text{Kerr-AdS}}\hspace{-0.32in} \mathrm{d}^4x \sqrt{\det\bar{g}}\, R^2 = \frac{12 \left(a^2+r_0^2\right) \left(a^2+3 r_0^2-\ell ^2\right)}{\left(a^2-\ell ^2\right) \left(a^2+6 r_0^2+\ell^2\right)}, \label{holo8i}\\[3pt]
& \lim_{\beta \to \infty} \frac{1}{16\pi^2}\int_{\text{Kerr-AdS}}\hspace{-0.32in} \mathrm{d}^4x \sqrt{\det\bar{g}}\, W^2 = \frac{2 \left(r_0^2+\ell ^2\right) \Big(a^4 \left(\ell ^2-3 r_0^2\right)+4 a^2 r_0^2 \left(r_0^2+\ell
	^2\right)-r_0^4 \ell ^2+3 r_0^6\Big)}{r_0^2 \ell ^2 \left(a^2+r_0^2\right) \left(a^2+6 r_0^2+\ell ^2\right)}. \label{holo8j}
\end{align}
}
However, the contribution of integrated Euler density $E_4$ is the same for all the cases,
\begin{align}
\frac{1}{16\pi^2}\int_{\text{Sch/RN/Kerr-AdS}}\hspace{-0.5in} \mathrm{d}^4x \sqrt{\det\bar{g}}\, E_4 &= 4.
\end{align}
This is justified because the integration of $E_4$ over all the extremal and non-extremal black hole backgrounds defines the 4D Euler characteristics \eqref{nbh10}, which is a fixed number, i.e., $\chi=2$.

\section{Curvature invariants of AdS$_4$ extremal-near-horizon background}\label{enhci}
In this section, we list out the explicit forms of the curvature invariants for extremal near-horizon (ENH) backgrounds. 
These are crucial when turn to our analysis via the quantum entropy function formalism for computing $\mathcal{C}_{\text{local}}$ contributions to extremal black hole entropy (see \cref{ebh}). The four-derivative invariants $R^2$, $E_4$ and $W^2=W_{\mu\nu\rho\sigma}W^{\mu\nu\rho\sigma}$ associated with the ENH background \eqref{nhe} are expressed as 
{\allowdisplaybreaks  
\begin{align}
	R^2 & = \frac{144}{\ell^4},\\[5pt]
	E_4 & = \frac{8 }{\ell ^4 \left(a^2 \cos ^2\theta+r_0^2\right)^6}\bigg[a^4 \cos ^4\theta \Big(3a^4\left( a^{4} \cos ^{4}\theta+6 a^{2} r_0^2 \cos ^{2}\theta + 15 r_0^4 \right)\cos ^4\theta \nonumber\\
	&\quad + 5 \big\lbrace a^4 \left(10 r_0^2 \ell ^2+7 r_0^4+\ell ^4\right)+2 a^2 \left(5 r_0^2 \ell ^4+16 r_0^4 \ell ^2+9 r_0^6\right)+ r_0^4\left(7 \ell ^4+18 r_0^2 \ell ^2+18 r_0^4\right) \big\rbrace \nonumber\\
	&\quad -6 a^2 r_0^2 \big\lbrace a^4+2 a^2 \left(2 r_0^2+\ell ^2\right)+4 r_0^2 \ell ^2-6 r_0^4+\ell ^4\big\rbrace\cos ^2\theta \Big) -2 a^2 r_0^2  \Big(a^4 \left(22 r_0^2 \ell ^2+4 r_0^4+19 \ell ^4\right) \nonumber\\
	&\quad +2 a^2 r_0^2 \left(11 \ell ^4+7 r_0^2 \ell ^2-3 r_0^4\right)+ 2r_0^4 \left(2 \ell^4 -3 r_0^2 \ell ^2-9 r_0^4\right)\Big)\cos ^2\theta \nonumber\\
	&\quad -r_0^4 \Big(a^4 \left(r_0^4 -2 r_0^2 \ell ^2 -5 \ell ^4\right)+a^2 r_0^2\left(8 r_0^2 \ell ^2+6 r_0^4-2 \ell ^4\right)+r_0^4 \ell ^2 \left(6 r_0^2+\ell ^2\right)\Big)  \bigg],\\[10pt]
	W^2 &= \frac{48}{\ell ^4 \left(a^2 \cos ^2\theta+r_0^2\right)^6} \bigg[r_0^4 \left(r_0^4-a^2 \ell ^2\right)^2 + a^8\ell^4\cos ^4\theta - a^2r_0^2\Big\lbrace a^4 \left(a^2+2 r_0^2+\ell ^2\right)^2\cos^4\theta \nonumber\\
	&\quad  - a^2\Big(16 r_0^4 \left(a^2+\ell ^2\right)+8 a^2 \ell ^2 \left(a^2+\ell ^2\right)+2 r_0^2 \left(3 a^4+13 a^2 \ell ^2+3 \ell ^4\right)+9 r_0^6\Big)\cos^2\theta \nonumber\\
	& +6 a^4 \ell ^4-4 r_0^6 \left(a^2+\ell ^2\right)+8 a^2 r_0^2 \ell ^2 \left(a^2+\ell ^2\right)+r_0^4 \left(a^4+6 a^2 \ell ^2+\ell ^4\right)-6 r_0^8\Big\rbrace\cos^2\theta \bigg].	
\end{align}}
As mentioned in \cref{ebh}, integration of the above near-horizon extreme invariants are found to be exactly in the same form as the full geometry extreme results \eqref{ebh4} but differ by a factor of $-2\pi$ due to the removal of divergence in the AdS$_2$ part of ENH geometry,
\begin{align}
	\begin{split}
		\lim_{\beta \to \infty} \int_{\text{full geometry}}\hspace{-0.3in} \mathrm{d}^4x \sqrt{\det\bar{g}}\, R^2 &= -2\pi \int_{\text{ENH}} \mathrm{d}\theta\,\mathrm{d}\tilde{\phi}\,\mathcal{G}(\theta)\, R^2, \\[5pt]
		\lim_{\beta \to \infty} \int_{\text{full geometry}}\hspace{-0.3in} \mathrm{d}^4x \sqrt{\det\bar{g}}\, E_4 &= -2\pi \int_{\text{ENH}} \mathrm{d}\theta\,\mathrm{d}\tilde{\phi}\,\mathcal{G}(\theta)\, E_4, \\[5pt]
		\lim_{\beta \to \infty} \int_{\text{full geometry}}\hspace{-0.3in} \mathrm{d}^4x \sqrt{\det\bar{g}}\, W^2 &= -2\pi \int_{\text{ENH}} \mathrm{d}\theta\,\mathrm{d}\tilde{\phi}\,\mathcal{G}(\theta)\, W^2,
	\end{split}
\end{align} 
where the ENH function $\mathcal{G}(\theta) = \frac{\ell ^2 \ell _2^2 }{(\ell ^2-a^2)}\left(a^2 \cos^2+r_0^2\right) \sin \theta$ must be operated in the range $0 \leq \theta \leq \pi$ and $0 \leq \tilde{\phi} \leq 2\pi$. Here note that, we have obtained exactly similar relations also in the flat-space limit $\ell\to \infty$.

\section{Logarithmic correction formulas for arbitrary dilaton coupling \boldmath $\kappa$}\label{genlog}
This section presents the general logarithmic correction formulas for each black hole embedded in the $U(1)$-charged EMD-AdS and EMD models with an arbitrary dilaton coupling constant $\kappa$. For that, one can utilize the trace anomaly or central charge data \eqref{emd3} in the $\mathcal{C}_{\text{local}}$ formulas derived in \cref{nbh11,nbh12,ebh5,ebh11}. In addition, the $\mathcal{C}_{\text{zm}}$ data recorded in \Cref{czero} is found to be useful. In terms of a general $\kappa$ parameter, the logarithmic correction formulas for the extremal and non-extremal Schwarzschild-AdS, Reissner-Nordstr\"om-AdS, and Kerr-AdS black holes are obtained as
{\allowdisplaybreaks
\begin{align}
\Delta S_{\text{BH}}^{\text{(Sch-AdS)}} &= \Bigg[- \frac{139}{36} + \kappa^2 -\frac{\kappa^4}{6}  + \frac{1}{\left(3 r_+^2+\ell ^2\right)\ell ^2 } \bigg\lbrace\frac{35}{4}r_+^4 + \left(\frac{\kappa^4}{2}-3\kappa^2 + 5\right)\ell ^2 r_+^2 \nonumber \\
&\qquad + \left(\frac{\kappa^4}{6}- \kappa^2+ \frac{55}{12}\right)\ell ^4 \bigg\rbrace\Bigg]\ln \mathcal{A}_{H}, \label{genlog1a}\\[5pt]
\Delta S_{\text{BH}}^{\text{(RN-AdS)}}  &= \Bigg[- \frac{53}{18} + \frac{4}{5}\kappa^2 -\frac{2}{15}\kappa^4   - \frac{2}{5\pi\ell ^4 r_+} \bigg\lbrace  \left(\kappa^4 - 6\kappa^2 + \frac{45}{8}\right)\pi\ell^2 r_+^3 \nonumber \\
& \qquad - 8\pi^2\left(\frac{\kappa^4}{12}-\frac{\kappa^2}{2}+\frac{55}{24}\right)\frac{ \ell^4 r_+^2}{\beta} -\frac{\beta}{2} \bigg(\left(\frac{\kappa^4}{2}- 3\kappa^2 - \frac{65}{8}\right)\ell^2 r_+^2 \nonumber\\
&\qquad    +  \left(\frac{3}{4}\kappa^4 - \frac{9}{2}\kappa^2 - \frac{5}{4}\right)r_+^4  + \left(\frac{\kappa^4}{12}- \frac{\kappa^2}{2}+ \frac{55}{24}\right)\ell^4\bigg) \bigg\rbrace\Bigg]\ln \mathcal{A}_{H}, \label{genlog1b}\\[5pt]
\Delta S_{\text{BH}}^{\text{(Kerr-AdS)}} &= \Bigg[ - \frac{103}{36} + \kappa^2 - \frac{\kappa^4}{6} + \frac{\beta }{2\pi  \ell ^2 \left(r_+^2 + a^2\right) \left(\ell ^2-a^2\right)r_+ }\bigg\lbrace  \nonumber\\
&\qquad \left(\frac{\kappa^4}{12}- \frac{\kappa^2}{2}- \frac{25}{12}\right) a^4\ell^2 - \left(\frac{\kappa^4}{12}-\frac{\kappa^2}{2}+\frac{55}{24}\right) a^2\ell^4 + \frac{35}{8}r_+^6    \nonumber\\
&\qquad   +\bigg( \left(\frac{\kappa^4}{4}- \frac{3}{2}\kappa^2 + \frac{5}{2}\right)\ell ^2 -\left(\frac{\kappa^4}{4}- \frac{3}{2}\kappa^2 - \frac{15}{8}\right)a^2\bigg)r_+^4  \nonumber\\
&\quad  +\bigg(\left(\frac{\kappa^4}{12}-\frac{\kappa^2}{2}+\frac{55}{24}\right) \ell ^4 -\frac{35}{4}a^2 \ell ^2  -\left(\frac{\kappa^4}{12}-\frac{\kappa^2}{2}-\frac{25}{12}\right)a^4 \bigg)r_+^2 \bigg\rbrace \Bigg]\ln \mathcal{A}_{H}, \label{genlog1c} \\[5pt] 
\Delta S_{\text{BH}}^{\text{(ext,RN-AdS)}}  & = \Bigg[-\frac{163}{36}+ \frac{\kappa^2}{2}- \frac{\kappa^4}{12}-\left(\frac{\kappa^4}{24}- \frac{\kappa^2}{4} + \frac{5}{12}\right) \frac{\left(\ell _2^4 + r_0^4\right)}{ r_0^2 \ell _2^2}\Bigg]\ln \mathcal{A}_{H},\label{genlog1d} \\[5pt]
\Delta S_{\text{BH}}^{\text{(ext,Kerr-AdS)}} &= \Bigg[-\frac{157}{36} + \kappa^2 - \frac{\kappa^4}{6} + \frac{1}{r_0^2 \left(\ell ^2-a^2\right) \left(a^2+r_0^2\right) \left(a^2+6 r_0^2+\ell ^2\right)} \bigg\lbrace\nonumber\\
&\qquad \left(\frac{\kappa^4}{12}-\frac{\kappa^2}{2}+\frac{55}{24}\right) a^4\ell^4 +\frac{105}{8}r_0^8 + \bigg(\left(\frac{95}{4}+ \frac{3}{2}\kappa^2 - \frac{\kappa^4}{4}\right)a^2  \nonumber \\
&\quad  + \left(\frac{\kappa^4}{4}-\frac{3}{2}\kappa^2 + \frac{5}{2}\right)\ell^2\bigg)r_0^6 + \bigg(\left(\frac{5}{6}\kappa^4 - 5 \kappa^2 + \frac{85}{6} \right)a^2\ell^2  \nonumber\\
& \quad  - \left(\frac{2}{3}\kappa^4 - 4\kappa^2 - \frac{85}{24}\right)a^4 - \left(\frac{\kappa^4}{12}-\frac{\kappa^2}{2}+\frac{55}{24}\right)\ell^4\bigg)r_0^4 \nonumber \\
&\quad + \bigg(\left(\frac{\kappa^4}{3}- 2\kappa^2 +\frac{55}{6}\right) a^2\ell^4 - \left(\frac{\kappa^4}{12}- \frac{\kappa^2}{2} + \frac{20}{3}\right)a^4\ell^2 \nonumber\\
&\qquad - \left(\frac{\kappa^4}{12}- \frac{\kappa^2}{2}- \frac{25}{12}\right)a^6\bigg)r_0^2\bigg\rbrace\Bigg]\ln \mathcal{A}_{H}. \label{genlog1e} 
\end{align}}
In the flat-space limit ($\ell \to \infty$), the same formulas for the extremal and non-extremal Schwarzschild, Reissner-Nordstr\"om and Kerr black holes can be written as 
{\allowdisplaybreaks
\begin{align}
	\Delta S_{\text{BH}}^{\text{(Sch)}} &= \frac{13}{18}\ln \mathcal{A}_{H}, \label{genlog2a}\\[5pt]
	\Delta S_{\text{BH}}^{\text{(Kerr)}} &= \frac{31}{18}\ln \mathcal{A}_{H}, \label{genlog2b}\\[5pt]
	\Delta S_{\text{BH}}^{\text{(RN)}} &= \left[\frac{13}{18} + \left(\frac{\kappa^4}{60}- \frac{\kappa^2}{10} + \frac{11}{24}\right)\frac{\beta q_e^4}{\pi r_+^5}\right]\ln \mathcal{A}_{H}, \label{genlog2c}\\[5pt]
	\Delta S_{\text{BH}}^{\text{(ext,Kerr)}} &= \frac{2}{9}\ln \mathcal{A}_{H}, \label{genlog2d}\\[5pt]
	\Delta S_{\text{BH}}^{\text{(ext,RN)}} &= \left[\kappa^2 - \frac{\kappa^4}{6}- \frac{193}{36}\right]\ln \mathcal{A}_{H}, \label{genlog2e}
\end{align} } 
where the symbols and notations carry the same notion as depicted in \cref{nbh,ebh}. The above formulas appear to be crucial in understanding the quantum black holes in a wide range of EMD models characterizing the 4D description of different higher-dimensional gravity theories. In the case of EMD-AdS and EMD theories embedded into the low-energy string theory or supergravity models, one can set the appropriate dilaton coupling constant values, i.e., $\kappa=1$, $\kappa=\sqrt{3}$, and $\kappa=\frac{1}{\sqrt{3}}$ into the relations to obtain the results mentioned in \cref{results}.


\end{document}